\documentclass[twocolumn]{aastex62}
\usepackage{graphicx}
\usepackage{subfigure}
\usepackage{multirow}
\usepackage{CJK}
\usepackage{mathptmx} 
\usepackage{hyperref}

\hypersetup{linkcolor=blue,citecolor=blue,filecolor=cyan,urlcolor=blue}

\makeatletter 
  \newcommand{\nfcaption}{\def\@captype{figure}\caption} 
\makeatother

\newcommand{\survey}{Surveys of Clumps, Cores, and Condensations in Cygnus X}
\newcommand{\surveyShowAb}{Surveys of {\bf C}lumps, Cor{\bf E}s, and Co{\bf N}den{\bf S}ations in Cygn{\bf US}-X}
\newcommand{\surveyAb}{CENSUS}

\newcommand{\hii}{H{\scriptsize\ II}}
\newcommand{\degree}{$^\circ$} 
\newcommand{\msun}{$M_{\odot}$}

\newcommand{\um}{$\rm{\mu m}$}
\newcommand{\kms}{$\rm km\cdot s^{-1}$}

\newcommand{\ColThreshold}{$3.5\times10^{22} \rm cm^{-2}$}
\newcommand{\massThresholdCompleteness}{23.3 \msun} 
\newcommand{\massThresholdMDC}{35 \msun} 

\newcommand{\numField}{40}
\newcommand{\numTrueField}{39}
\newcommand{\numNewField}{14}
\newcommand{\numMDC}{151}
\newcommand{\numMDCOrig}{33}
\newcommand{\numMDCSpitzer}{78}
\newcommand{\numMDCMSX}{40}

\newcommand{\numMDCIRB}{15}
\newcommand{\numMDCIRQ}{136}
\newcommand{\numMDCStarless}{26}  
\newcommand{\numCoreNew}{963}
\newcommand{\numCoreGood}{496}
\newcommand{\numCoreBig}{205}
\newcommand{\numCoreMMII}{10}
\newcommand{\numCoreUCHII}{8}

\newcommand{\IRThresholdSpitzer}{23}
\newcommand{\IRThresholdMSX}{17}

\newcommand{\listDegradedMDC}{N32, N58, N69, NW02, S07, S10, S15, S18, and S26} 

\newcommand{\motte}{\hyperlink{M07}{M07}}

\received{- -, --}
\revised{- -, --}
\accepted{- -, --}
\submitjournal{ApJS}

\begin{document} 
\begin{CJK*}{UTF8}{gbsn}

\title{\survey:\\ I. a New Catalog of $\sim$0.1 pc Massive Dense Cores}

\author[0000-0002-6368-7570]{Yue Cao (曹越)}
\affiliation{School of Astronomy and Space Science, Nanjing University, 163 Xianlin Avenue, Nanjing 210023, P.R.China}
\affiliation{Key Laboratory of Modern Astronomy and Astrophysics (Nanjing University), Ministry of Education, Nanjing 210023, P.R.China}

\author[0000-0002-5093-5088]{Keping Qiu}
\affiliation{School of Astronomy and Space Science, Nanjing University, 163 Xianlin Avenue, Nanjing 210023, P.R.China}
\affiliation{Key Laboratory of Modern Astronomy and Astrophysics (Nanjing University), Ministry of Education, Nanjing 210023, P.R.China}

\author[0000-0003-2384-6589]{Qizhou Zhang}
\affiliation{Harvard-Smithsonian Center for Astrophysics, 60 Garden St., MS 42, Cambridge, MA 02138 USA}

\author[0000-0001-6630-0944]{Yuwei Wang}
\affiliation{School of Astronomy and Space Science, Nanjing University, 163 Xianlin Avenue, Nanjing 210023, P.R.China}
\affiliation{Key Laboratory of Modern Astronomy and Astrophysics (Nanjing University), Ministry of Education, Nanjing 210023, P.R.China}

\author[0000-0002-3286-5469]{Bo Hu}
\affiliation{School of Astronomy and Space Science, Nanjing University, 163 Xianlin Avenue, Nanjing 210023, P.R.China}
\affiliation{Key Laboratory of Modern Astronomy and Astrophysics (Nanjing University), Ministry of Education, Nanjing 210023, P.R.China}

\author[0000-0002-4774-2998]{Junhao Liu}
\affiliation{School of Astronomy and Space Science, Nanjing University, 163 Xianlin Avenue, Nanjing 210023, P.R.China}
\affiliation{Key Laboratory of Modern Astronomy and Astrophysics (Nanjing University), Ministry of Education, Nanjing 210023, P.R.China}

\correspondingauthor{Keping Qiu}
\email{kpqiu@nju.edu.cn}

\begin{abstract} 

Using infrared to (sub)millimeter data from \emph{Spitzer}, \emph{Herschel}, the James Clerk Maxwell Telescope, and the IRAM 30-m telescope, we conducted an unbiased survey of the massive dense cores (MDCs) in the Cygnus X molecular cloud complex, aimed at characterizing the physical conditions of high-mass star formation (HMSF) at $\sim$0.1 pc scales. We created 5\degree$\times$6\degree\ images of the 70--1200 \um\ dust continuum, gas column density, and dust temperature of Cygnus X. A spatial relation between the dense regions ($A_{\rm v}\ge15$) and the developed \hii\ regions was found, indicating the impact of the latter on the global structures of Cygnus X. With a 35-\msun\ mass threshold implied by HMSF signposts, we identified \numMDC\ MDCs with sizes of $\sim$0.1 pc, masses of 35--1762 \msun, and temperatures of 8--35 K. Our MDC sample is statistically complete in Cygnus X and is three times larger than that in \citet{2007A&A...476.1243M}. The MDCs were classified into IR-bright/IR-quiet ones based on their mid-infrared fluxes and a large ``IR-quiet'' proportion (90\%) was found in our sample. Two possible scenarios were proposed to interpret this: accelerated HMSF and the incapability of HMSF of the IR-quiet MDCs. We also found \numMDCStarless\ starless MDCs by their lack of compact emissions at 21--70 \um\ wavelengths, of which the most massive ones are probably the best candidates of initial HMSF sites in Cygnus X.
\end{abstract}


\section{Introduction} \label{sec:intro}

High-mass stars ($M_*>8$ \msun) play an important role in physical and chemical evolution of interstellar medium and energy budget of galaxies. Studies on their formation have made significant progress over the past decades \citep[see][]{2002ASPC..267..267T,2007prpl.conf..181H,2007ARA&A..45..481Z,2014prpl.conf..149T,2017arXiv170600118M,2017arXiv171205281S,2018IAUS..332..139T}. Observational studies have confirmed that high-mass stars form in massive dense cores (MDCs), which are $\sim$0.1 pc dense entities in molecular clouds with masses of tens to hundreds of solar masses \citep[e.g.,][]{2007A&A...476.1243M,2010A&A...518L..78B,2011A&A...527A.135C,2014ASSP...36..271H,2017A&A...602A..77T,2018ApJ...865..110C}. The physical conditions of high-mass star formation (HMSF) are closely related to the formation and dynamical evolution of MDCs. 

One of the key arguments in HMSF is whether high-mass stars form as a scaled-up version of the classic low-mass star formation paradigm \citep{1987ARA&A..25...23S}. Two theoretical models have been developed in this context: the core accretion model \citep{2002Natur.416...59M,2007ApJ...656..959K} and the competitive accretion model \citep{2001MNRAS.323..785B,2006MNRAS.370..488B}. In the core accretion scenario, high-mass stars form through the monolithic collapse of high-mass analogs of low-mass prestellar cores \citep{2002Natur.416...59M,2007ApJ...656..959K}. While in the competitive accretion scenario, ``seeds'' of high-mass stars accrete and grow competitively in massive cores or clumps \citep{2001MNRAS.323..785B,2006MNRAS.370..488B,2009ApJ...696..268Z,2012MNRAS.420.1457H}. Whether high-mass prestellar cores exist, is then an essential observable distinction of the two models. So far there have been quite a number of millimeter/submillimeter surveys and case studies targeting massive clumps and cores \citep{2007A&A...476.1243M,2010A&A...524A..18B,2012ApJ...754...87W,2013A&A...558A.125D,2014ApJ...796L...2C,2015ApJ...804..141Z,2017A&A...602A..77T,2018ApJ...853..160C}, and infrared-dark clouds (IRDCs) \citep{2006ApJ...641..389R,2012ApJ...754....5B,2013ApJ...779...96T,2014A&A...561A..83P,2014MNRAS.439.3275W,2015ApJ...805..171L,2016ApJ...833..209O,2017ApJ...834..193K}, yet no bona fide high-mass prestellar cores have been identified. 

In this paper we focus on Cygnus X, one of the most massive and active high-mass star-forming regions in our Galaxy. Lying at a distance of 1.4 kpc \citep{2012A&A...539A..79R}, it harbors numerous \hii\ regions \citep{1991A&A...241..551W}, OB associations \citep{2001A&A...371..675U}, and high-mass star-forming sites, including W75N, DR21, DR21(OH), and S106 \citep{2002A&A...384..225S,2004ApJ...601..952S}. \citet{2007A&A...476.1243M} (hereafter \motte) carried out a dust continuum survey covering a 3-square-degree area in Cygnus X, and detected a total of 129 dense cores, of which 42 cores are considered as probable precursors of high-mass stars (i.e., MDCs). Here we take the advantage of a newly available large data set from \emph{Spitzer}, \emph{Herschel}, and the James Clerk Maxwell Telescope (JCMT), and perform a detailed multi-wavelength survey of the MDCs in Cygnus X with a spatial coverage one order of magnitude larger than that of \motte.

The distance of Cygnus X has been debated over decades. Due to its location on the tangent point of the Galaxy rotation, Cygnus X has been once considered as a superposition of various objects \citep[e.g.,][]{1969A&A.....1..270D,1988A&A...191..313P,2001A&A...371..675U}. However, recent CO surveys toward Cygnus X suggested that it is more likely to be located at a coherent distance \citep[e.g.,][]{2006A&A...458..855S,2012A&A...541A..79G}; \citet{2012A&A...539A..79R} measured the parallaxes of five maser sites in Cygnus X and concluded that Cygnus X North is a physically connected region at $\sim1.4$ kpc; \citet{2013ApJ...769...15X} provided a comprehensive result of parallax measurement of the Local Arm, of which eight out of nine maser sites in Cygnus X have distances within 1.3--1.6 kpc, except AFGL 2591 (3.3 kpc). We therefore will assume that most of the objects in Cygnus X share a coherent distance of 1.4 kpc, except a few regions, e.g., AFGL 2591 (see Sect. \ref{subsec:field}). It should be noticed that \motte\ adopted a distance value of 1.7 kpc in their analysis, which was derived by the $\rm ^{13}CO$ study of \citet{2006A&A...458..855S}. Since this value has been updated to 1.4 kpc, we will modify their results to match the new distance before any comparisons are made, and the corrected results will hereafter be referred to as ``\motte'' unless otherwise specified. 

As the first paper of our project, \surveyShowAb\ (\surveyAb\footnote{\surveyAb\ project (PI: Keping Qiu) is dedicated to a systematic study of the 0.01--10 pc hierarchical cloud structures and high-mass star formation in the Cygnus X complex. The project consists of several parallel surveys involving the SMA, JVLA, JCMT, Tianma 65m radio telescope (Shanghai 65 m Radio Telescope), as well as case studies with ALMA, NOEMA, and CARMA.}), we describe our data used in Sect. \ref{sec:obs}. In Sect. \ref{sec:res} we discuss our data processing, including generating large maps of Cygnus X, MDC sample selection, SED fitting, classification of MDCs, and analysis on HMSF signposts. We discuss these main results in Sect. \ref{sec:disc}, including the formation and HMSF potential of large-scale cloud structures, review of the MDC sample and comparison with \motte, infrared properties of the MDCs and the interpretations of HMSF, and the starless MDCs in Cygnus X. Our conclusions are summarized in Sect. \ref{sec:conc}.

\section{Data} \label{sec:obs}

\subsection{Ground-based Observations} \label{subsec:obs_continuum}

The published 1.2 mm continuum images from the MAMBO and MAMBO-2 cameras of the 30-m telescope of Institut de Radioastronomie Millim\'etrique (IRAM) in \motte\ were used for our analysis. The 3 $\rm deg^2$ spatial coverage of these images are shown in Figure \ref{fig:panorama}. The beam size is 11\arcsec\ and the mean $1\sigma$ rms level is 12.9 ${\rm mJy\cdot beam^{-1}}$ (see Table \ref{tab:beam}). Detailed observational parameters and data processing can be found in \motte. 

The JCMT SCUBA-2 continuum data at 450 \um\ and 850 \um\ toward Cygnus X were obtained from the Canadian Astronomy Data Centre\footnote{\url{http://www.cadc-ccda.hia-iha.nrc-cnrc.gc.ca/en/}}, which have not been published before. The observations were made in scan mode during 2011--2015 and the average integration time of each map is $\sim$0.5 h. The FWHM beam sizes are 7.9\arcsec\ and 13\arcsec\, and the $1\sigma$ rms levels are 340 $\rm{mJy\ beam^{-1}}$ and 34 $\rm{mJy\ beam^{-1}}$ for 450 \um\ and 850 \um, respectively. See Figure \ref{fig:panorama} for the coverage of the images. The related program IDs are M11BEC30, M12BU30, M13AD03, M13AU03, M13BU29, M15AI133, and M15AI144. 

\subsection{Space Telescope Observations} \label{subsec:obs_IR}

Continuum images in the wavelengths of \emph{Herschel} PACS (70 and 160 \um) and SPIRE (250, 350, and 500 \um) were obtained from the Herschel Science Archive\footnote{\url{http://archives.esac.esa.int/hsa/whsa/}}. These data are from the HOBYS\footnote{Herschel imaging survey of OB Young Stellar objects, \url{http://www.herschel.fr/cea/hobys/en/index.php}} project \citep{2010A&A...518L..77M}, the Hi-GAL\footnote{The Herschel Infrared Galactic Plane Survey, \url{https://hi-gal.ifsi-roma.inaf.it/higal/}} project \citep{2010PASP..122..314M}, and open-time observations toward Cygnus X by \citet{2011hers.prop.1667H}. See the corresponding references for their observational configurations and data processing. The beam sizes are 5.2\arcsec, 12\arcsec, 17.6\arcsec, 23.9\arcsec, and 35.2\arcsec, and the $1\sigma$ rms levels are $\rm4.0\ mJy\cdot(3.2\arcsec\ pixel)^{-1}$, $\rm2.9\ mJy\cdot(3.2\arcsec\ pixel)^{-1}$, $\rm41\ mJy\cdot beam_{250\mu m}^{-1}$, $\rm50\ mJy\cdot beam_{350\mu m}^{-1}$, and $\rm30\ mJy\cdot beam_{500\mu m}^{-1}$, respectively in the order of increasing wavelengths (Table \ref{tab:beam}). 

The \emph{Spitzer} IRAC 8-\um\ and MIPS 24-\um\ maps from the Spitzer Legacy Survey of the Cygnus X Complex\footnote{\url{https://www.cfa.harvard.edu/cygnusX/}} \citep{2010AAS...21541401K} were obtained from the NASA/IPAC Infrared Science Archive\footnote{IRSA is chartered to curate the science products of NASA's infrared and submillimeter missions, providing access to more than 20 billion astronomical measurements, including all-sky coverage in 24 bands. \url{http://irsa.ipac.caltech.edu/frontpage/}}. Both maps cover the entire Cygnus X region (see Figure \ref{fig:S8} and \ref{fig:S24}). The resolutions are 2\arcsec\ and 6\arcsec, and the 1$\sigma$ rms noise levels are 47.9 and 5.2 $\rm{MJy\cdot sr^{-1}}$ respectively for 8 \um\ and 24 \um. 


\startlongtable
\begin{deluxetable}{c|c|ccc}
\tabletypesize{\footnotesize}
\tablecaption{Main Parameters of the Data Used \label{tab:beam}}  
\tablehead{\colhead{Telescopes}&\colhead{Instruments}&\colhead{Bands}&\colhead{$FWHM$}&\colhead{$1\sigma$ rms}\\
		   \colhead{}&\colhead{}&\colhead{(\um)}&\colhead{(\arcsec)}&\colhead{($\rm{MJy\cdot sr^{-1}}$)} }
\startdata
\multirow{2}{*}{\emph{Spitzer}} & IRAC & 8 & 1.98 & 47.9 \\
\cline{2-5} & MIPS & 24 & 6 & 5.2 \\
\hline
\multirow{5}{*}{\emph{Herschel}} & \multirow{2}{*}{PACS} & 70 & 9.2 & 16.5 \\
\cline{3-5} & & 160 & 12.6 & 12.0 \\
\cline{2-5} & \multirow{3}{*}{SPIRE} & 250 & 18.4 & 4.97 \\
\cline{3-5} & & 350 & 25.2 & 3.29\\
\cline{3-5} & & 500 & 36.7 & 0.91 \\
\hline
\multirow{2}{*}{JCMT} & \multirow{2}{*}{SCUBA-2} & 450 & 7.9 & 204 \\
\cline{3-5} & & 850 & 13 & 7.6 \\
\hline 
IRAM & MAMBO/MAMBO-2 & 1200 & 11 & 4.0 \\
\enddata
\end{deluxetable}

\section{Analysis} \label{sec:res}

\subsection{Mosaic Maps of Cygnus X}\label{subsec:map}

We generated mosaic images of the whole Cygnus X region in the bands of \emph{Herschel} (70, 160, 250, 350, and 500 \um), JCMT (450 and 850 \um), and the IRAM 30 m telescope (1.2 mm) using the \emph{Montage} package\footnote{\url{http://montage.ipac.caltech.edu/}} \citep{2010arXiv1005.4454J} (see Figures \ref{fig:H70} to \ref{fig:I1200}). The procedures include re-projection, re-gridding, and merging of the original data in each band. For the \emph{Herschel} data, the zero-level offset correction was conducted with the Planck (217, 353, 545, and 857 GHz) and IRAS (100 \um) data following the procedure of \citet{2010A&A...518L..88B}. All the mosaic maps are 5\degree$\times$6\degree\ in size with resolutions presented in Table \ref{tab:beam}. An RGB color image made out of three \emph{Herschel} mosaics was also created (see Figure \ref{fig:rgb}). 

High-resolution column density map (Figure \ref{fig:panorama}) and temperature map (Figure \ref{fig:T}) of Cygnus X were generated from the 160, 250, 350, and 500 \um\ mosaic images using the \emph{hirescoldens} command in the \emph{getsources}\footnote{\url{http://www.herschel.fr/cea/gouldbelt/en/getsources/}} (v2.180813) procedure \citep{2012A&A...542A..81M}. This command decomposed the continuum images into single-spatial-scale components and pixel-by-pixel fitted the SEDs at each scale using a modified blackbody model fully described in Sect. \ref{subsec:extraction}. The column density map was then constructed by combining the column density ``components'' at each scale, and its final resolution was set by that of the second shorted wavelength (18.4\arcsec @250 \um). The temperature map is nevertheless not addable and was generated using the same SED fitting routine with the images convolved to the lowest resolution (36.7\arcsec @500 \um). The 70 \um\ data were not used since the emission traces warm dust and a single-temperature model is no longer applicable \citep[see][]{2010A&A...518L..77M}. For parameter settings, we used a dust-opacity spectral index $\beta=2$ and a gas-to-dust ratio of 100. We conservatively estimated the relative flux uncertainties in each band as 0.2. The uncertainty maps of column density and temperature (Figure \ref{fig:Ne2N} and \ref{fig:T_err}) were generated by \emph{hirescoldens}, which considered both the flux uncertainties and the fitting errors.

\subsection{Complete Sample of Massive Dense Cores}\label{subsec:sample}

In principle, our target sample should contain all the MDCs in Cygnus X which are massive enough to be able to form high-mass stars. We will describe the procedure for generating our MDC sample in this subsection. 

\subsubsection{Selection of High-column-density Fields}\label{subsec:field}

High-mass stars form preferentially in high-column-density regions, both proposed by theoretical works \citep{2008Natur.451.1082K} and verified by observations \citep[e.g.,][]{2007A&A...476.1243M,2010ApJ...723L...7K}. To find regions in Cygnus X with potential of HMSF, i.e., regions where MDCs reside, we selected \numField\ rectangular fields on the column density map that cover regions where $N_{\rm H_2}\ge$\ColThreshold\ (see Figure \ref{fig:panorama} and \ref{fig:field}). Informations of the fields are listed in Table \ref{tab:field}. The column density threshold \citep[\ColThreshold, corresponding to $A_{\rm v}\gtrsim15$,][]{2009MNRAS.400.2050G} guarantees that all the cores with masses above certain value can be selected in our sample. Specifically, assuming that cores have 2D Gaussian profiles in the column density map with FWHM sizes $d_{\rm FWHM}$ and peak column densities $N_{\rm H_2,p}$, their masses can be estimated as:

\begin{equation}\label{eq:core_mass}
M=\frac{\pi}{\rm 4ln2}\mu_{\rm H_2}m_{\rm H}N_{\rm H_2,p}D^2d_{\rm FWHM}^2,
\end{equation}
where $\mu_{\rm H_2}=2.8$ is the mean molecular weight per hydrogen molecule, $m_{\rm H}$ is the mass of hydrogen atom, and $D$ is the source distance. Cores with $N_{\rm H_2,p}\ge$\ColThreshold\ and typical sizes of 0.1 pc, or 0.15 pc convolved with the 18\arcsec\ beam, will have masses above \massThresholdCompleteness. In other words, our core sample to be extracted in the selected fields will be \emph{statistically complete for the cores with $M\ge$\massThresholdCompleteness}. 

To rule out the regions of which the distances are not coherent with 1.4 kpc, we used the JVLA K-band observation of ammonia inversion lines toward the fields in Cygnus X (proposal ID: VLA/17A-107, PI: Keping Qiu, Zhang et al. in prep.) and found that the systematic velocities of the field C01 and part of the field C08 are -41 and -65 \kms, respectively, which correspond to BeSSeL\footnote{Bar and Spiral Structure Legacy survey, \url{https://www3.mpifr-bonn.mpg.de/staff/abrunthaler/BeSSeL/index.shtml}} kinematic distances of 6.1 and 9.1 kpc, respectively. We excluded \emph{the whole C01 field} and \emph{part of the C08 field}, as well as \emph{AFGL 2591 in field S29} from the regions for core extraction. See Figure \ref{fig:panorama} and the corresponding sub-figures in Figure \ref{fig:field} for the excluded regions. 

To select MDCs in Cygnus X, \citet{2007A&A...476.1243M} used a criterion of $A_{\rm V}\ge15$ mag and selected $\sim3$ deg$^2$ high-extinction regions (see the white dashed polygons in Figure \ref{fig:panorama}). Compared with their coverage, our new fields have a total area of 2.2 deg$^2$ but are focused on denser regions. In addition, \numNewField\ out of \numTrueField\ fields are newly added compared with \motte. These fields have peak column densities comparable to those of the regions in \motte\ but were not covered in their IRAM observations. This indicates that our sample of MDCs represents great improvements on statistical completeness.

\subsubsection{Core Extraction and SED Fitting}\label{subsec:extraction}

\startlongtable
\begin{deluxetable}{ccc}
\tabletypesize{\footnotesize}
\tablecaption{Percentages of the Cores \label{tab:percent}}  
\tablehead{\colhead{Band}&\colhead{Flux\tablenotemark{a}}&\colhead{SED fitting\tablenotemark{b}} }
\startdata
$\emph{Herschel}\ 70\  {\rm\mu m}$&	54.1\%&	0.0\%\\
$\emph{Herschel}\ 160\ {\rm\mu m}$&	68.3\%&	66.7\%\\
$\emph{Herschel}\ 250\ {\rm\mu m}$&	64.3\%&	61.6\%\\
$\emph{Herschel}\ 350\ {\rm\mu m}$&	62.2\%&	57.1\%\\
$\bf JCMT\ 450\ {\rm\mu m}$&	22.8\%&	12.2\%\\
$\emph{Herschel}\ 500\ {\rm\mu m}$&	56.0\%&	49.4\%\\
$\bf JCMT\ 850\ {\rm\mu m}$&	47.9\%&	46.7\%\\
$\bf IRAM\ 1.2\ {\rm mm}$&	21.8\%&	21.5\%\\

\enddata
\tablenotetext{a}{Percentages of cores that \emph{getsources} successfully extracted their fluxes in this band. }
\tablenotetext{b}{Percentages of cores of which the fluxes in this band were used for SED fitting. The \emph{Herschel} 70 \um\ fluxes were not used intentionally (see Sect. \ref{subsec:extraction}). }
\end{deluxetable}

Cores in each field are extracted using \emph{getsources} \citep{2012A&A...542A..81M,2017A&A...607A..64M}, which is a multi-scale, multi-wavelength source extraction algorithm. In spirit, it consists of two main procedures: detection and measurement. Detection of compact sources is based on the detection images in each band, which are generated by decomposed, cleaned and background-removed single-scale images. The measurement procedure then measures the properties of sources (sizes, positions, and fluxes in each band). We used the 70--1200 \um\ continuum images of the fields as inputs of \emph{getsources} and a \emph{deconvolved} maximum core size of 35\arcsec\ (0.23 pc) for background removal (the actual scale for background removal in each band is this deconvolved size convolved with the corresponding beam). Data in 70 \um\ were not used to constrain source positions since their emission peaks can deviate significantly from density peaks. After running through the \numTrueField\ fields, \numCoreNew\ cores were identified by \emph{getsources}. Table \ref{tab:percent} lists the percentages of cores that \emph{getsources} have successfully measured their fluxes in each band. As can be seen, at least nearly half of the cores do not have well-defined fluxes in each band, indicating that this core catalog needs re-selection in order to truly reflect the $\sim0.1$ pc entities in Cygnus X.

To create such a ``robust'' core catalog, we used the following four criteria: (1) having fluxes in at least three bands; (2) having SEDs that can be fitted to obtain masses and temperatures (see below); (3) relative mass error in SED fitting $\Delta M/M<0.8$; (4) relative temperature error in SED fitting $\Delta T/T<0.5$. The SED fitting procedure are described as follows. 

We used the \emph{curve\_fit} function in the Python package \emph{scipy}\footnote{\url{http://www.scipy.org/}} to fit the SEDs of cores with the modified blackbody model

\begin{equation}\label{eq:SED}
F_{\nu} = \frac{\kappa_{\nu}B_{\nu}(T)M}{D^2},
\end{equation}
where $B_{\nu}(T)$ is the Planck function and $\kappa_{\nu}$ is the dust mass opacity, which is evaluated following the HOBYS consortium: $\kappa_{\nu}=\kappa_0 (\nu/\nu_0)^\beta$, where $\kappa_0=0.1\ \rm cm^2g^{-1}$, $\nu_0=1$ THz, and $\beta=2$. We used the fluxes at $\lambda\ge160$ \um\ for the fitting and the data points. During the fitting, we found that some fluxes significantly deviated from the model (see Figure \ref{fig:SED_example}). These fluxes were poorly extracted by \emph{getsources} due to the low resolutions or poor data qualities in some bands (e.g., \emph{Herschel} 500 \um\ and JCMT 450 \um). We removed them from our SED fitting by eye-inspection and the fractions of the cores of which the fluxes were used in SED fitting are listed in Table \ref{tab:percent}. 

With the criteria mentioned above, we selected a total of \numCoreGood\ ``robust'' cores, which we considered to reliably reflect the $\sim0.1$ pc entities in Cygnus X. Furthermore, \numCoreBig\ out of the \emph{robust} cores have $M\ge$\massThresholdCompleteness\ and are thus statistically complete (\emph{complete} cores, see Sect. \ref{subsec:field}). See Figure \ref{fig:venn} for an illustration of the relative relations of these core sets. 

Once the masses and temperatures of cores were determined, physical parameters such as integrated far-IR luminosity, column density, and volume density can also be obtained as

\begin{equation}\label{eq:L}
L_{\rm FIR} = 4\pi D^2\int_0^{+\infty} F_{\nu}\rm{d}\nu,
\end{equation}

\begin{equation}\label{eq:N}
N_{\rm{H_2}}=\frac{4}{\pi}\frac{M}{\mu_{\rm H_2}m_{\rm H}FWHM_{\rm dec}^2},
\end{equation}

\begin{equation}\label{eq:n}
n_{\rm{H_2}}=\frac{6}{\pi}\frac{M}{\mu_{\rm H_2}m_{\rm H}FWHM_{\rm dec}^3},
\end{equation}
where $FWHM_{\rm dec}$ is the deconvolved FWHM size at 160 \um, since 160 \um\ is the highest-resolution band that can reflect optically thin dust emissions.

\subsubsection{Signposts of High-mass Star Formation and Sample of MDCs}\label{subsec:mdc}

\startlongtable
\begin{deluxetable*}{c|c|cccccc}
\tabletypesize{\normalsize}
\tablecaption{Physical Properties of the MDCs in Cygnus X \label{tab:MDC_overview}}  
\tablehead{ 
    \colhead{Samples} & \colhead{} & \colhead{$FWHM$ size} & \colhead{$T$} & \colhead{$M$} & \colhead{$N_{\rm{H_2}}$} & \colhead{$n_{\rm{H_2}}$} & \colhead{$L_{\rm FIR}$} \\ 
    \colhead{} & \colhead{} & \colhead{(pc)} & \colhead{(K)} & \colhead{(\msun)} & \colhead{($10^{22} \rm{cm^{-2}}$)} & \colhead{($10^5 \rm{cm^{-3}}$)} & \colhead{($L_{\odot}$)} }
\startdata
\multirow{3}{*}{All}&	min&	0.09&	8.4&	35&	2.5&	0.4&	2\\
&	max&	0.31&	34.7&	1762&	1263.3&	1715.0&	$2.04\times10^{4}$\\
&	mean&	0.14&	15.2&	131&	83.7&	58.8&	631\\
\hline
IR-bright&	mean&	0.15&	22.9&	398&	147.6&	67.1&	$4.47\times10^{3}$\\
\hline
IR-quiet but not starless&	mean&	0.14&	15.0&	109&	82.1&	63.6&	252\\
\hline
Starless&	mean&	0.13&	11.8&	71&	22.2&	9.6&	22\\

\enddata
\end{deluxetable*}

In our definition, MDCs are cores that are sufficiently massive to have the potential of HMSF. To find them in our \emph{complete} core sample, we took the simple idea that MDCs should have masses exceeding some lower limit, if a constant core-to-star efficiency is assumed \citep{2007A&A...462L..17A}. To set such limit, we used two observational signposts of HMSF: class II methanol masers and ultra-compact \hii\ (UC\hii) regions. 

Discovered by \citet{1991ApJ...380L..75M}, class II methanol masers were found to be exclusively associated with high-mass star forming regions \citep{2003A&A...403.1095M,2008A&A...485..729X,2013MNRAS.430..808G,2013MNRAS.435..524B}. We cross-correlated our \emph{complete} cores with the published catalogs in \citet{2003A&AT...22....1M,2005A&A...432..737P}; and \citet{2016ApJ...833...18H} using a positional tolerance of 10\arcsec\ (0.07 pc) and found that \numCoreMMII\ cores are associated with class II maser sites (see Table \ref{tab:MDC_sign}). 

UC\hii\ regions are a product of newly formed (or forming) high-mass stars on their ambient materials and are bright in radio emissions \citep{1990A&ARv...2...79C}. We used the 3-GHz continuum maps from the VLA Sky Survey (VLASS\footnote{\url{https://science.nrao.edu/science/surveys/vlass}}) to search for compact radio sources that are associated with our \emph{complete} cores within 10\arcsec. These maps cover the whole Cygnus X region with a resolution of 2.5\arcsec\ and a $1\sigma$ rms level of 0.1 $\rm mJy\cdot beam^{-1}$. To rule out possible radio emissions from low-mass star forming regions \citep[i.e., radio jets or ionized accretion shocks, see][]{1998ApJ...509..802C,2018A&ARv..26....3A}, we only selected radio sources with $F_{\rm 3GHz}\ge0.01$ Jy. The flux limit was set using typical physical values of UC\hii\ regions in \citet{2005IAUS..227..111K} (see their Table 3). Consequently, a total of \numCoreUCHII\ cores were found to be associated with bright compact radio sources (see Table \ref{tab:MDC_sign}). 

The lower limit of mass for MDCs was then determined as the minimal mass of cores associated with either of the two HMSF signposts, which is \massThresholdMDC\ (see Figure \ref{fig:IR_class}). Thus \numMDC\ out of \numCoreBig\ \emph{complete} cores were defined as MDCs. We name these MDCs in the format of ``field name-number'', where the number is the ranking by mass in the fields. See Table \ref{tab:MDC_overview} for the overview properties of the MDCs, and Table \ref{tab:MDC_property} for their detailed physical properties. See also Figure \ref{fig:MDC_table} for a visualization of the MDCs, and Figure \ref{fig:MDC} for their maps in each band.


\subsection{Coincidence of MDCs with Mid-infrared Sources and Infrared Classification}\label{subsec:IR_class}

To better understand the infrared properties of MDCs, we cross-correlated our MDCs with archival mid-infrared sources. The Cygnus-X Archive catalog\footnote{\url{https://irsa.ipac.caltech.edu/cgi-bin/Gator/nph-dd}} from the \emph{Spitzer} Cygnus-X Legacy Survey \citep{2010AAS...21541401K} was used preferentially, and the Midcourse Space Experiment (\emph{MSX}) Point Source Catalog\footnote{\url{https://irsa.ipac.caltech.edu/cgi-bin/Gator/nph-dd}} was used in the cases where the \emph{Spitzer} images are saturated. Using a positional tolerance of 10\arcsec, we found that \numMDCSpitzer\ and \numMDCMSX\ MDCs are associated with \emph{Spitzer} and \emph{MSX} sources, respectively. The mid-infrared flux of each MDC was then determined as the sum of the fluxes of all associated infrared sources. If an MDC does not have any associated infrared sources but shows certain compact emissions in \emph{Spitzer} 24 \um\ or \emph{MSX} 21 \um\ maps, its mid-infrared flux was obtained by manual aperture photometry. See Table \ref{tab:MDC_IR} for the results. 

\begin{figure}[htb!]
\epsscale{1.2}\plotone{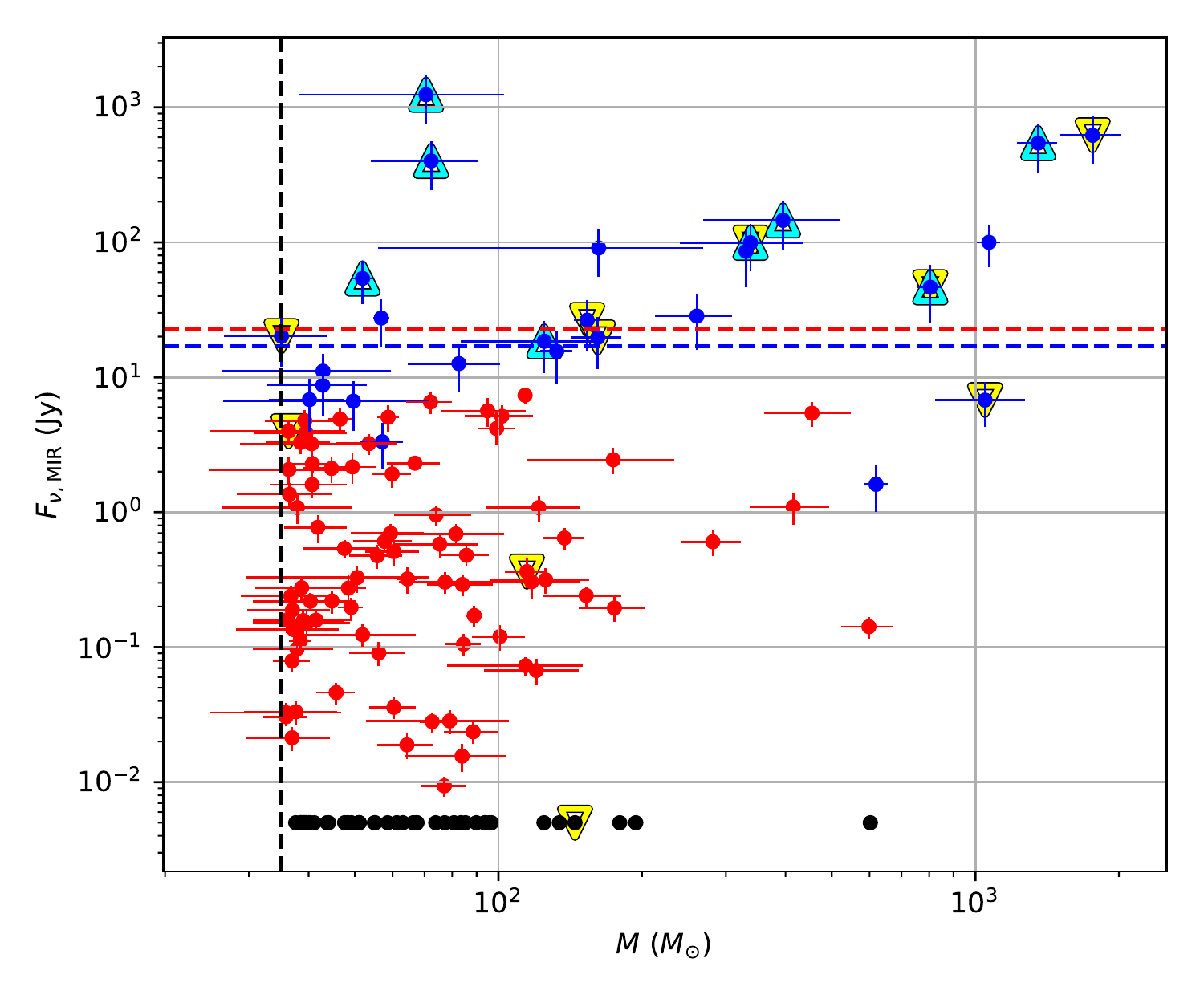}
\caption{Mid-IR flux (red for \emph{Spitzer} 24 \um, blue for \emph{MSX} 21 \um, and black for the 5 mJy detection limit of \emph{Spitzer} at 24 \um) versus mass diagram of the MDCs in Cygnus X. MDCs associated with UC\hii\ regions and class II methanol masers are attached with cyan and yellow triangles, respectively. Black vertical dashed line shows the 35-\msun\ mass threshold for defining MDCs (see Sect. \ref{subsec:mdc}). Red and blue dashed lines are the 23-Jy (\emph{Spitzer} 24 \um) and 17-Jy (\emph{MSX} 21 \um) thresholds separating IR-bright/IR-quiet MDCs (see Sect. \ref{subsec:IR_class}). \label{fig:IR_class}}
\end{figure}

We classified the MDCs into IR-bright/IR-quiet ones following the method of \motte. In our case, the mid-IR flux of a B3-type stellar embryo at a distance of 1.4 kpc is $F_{Spitzer\ 24\rm \mu m}=$\IRThresholdSpitzer\ Jy, or $F_{MSX\ 21\rm \mu m}=$\IRThresholdMSX\ Jy. MDCs are defined as IR-bright if their mid-IR fluxes exceed this threshold, which indicates that they should already host at least one high-mass ($\ge8$ \msun) stellar embryo, or as IR-quiet if it is on the contrary (see Figure \ref{fig:IR_class}). Of the \numMDC\ MDCs, \numMDCIRB\ and \numMDCIRQ\ MDCs were classified as IR-bright and IR-quiet, respectively. See Table \ref{tab:MDC_property} for the detailed list, Figure \ref{fig:number_in_fields} for the numbers of different MDCs in each field and Figure \ref{fig:histIR} for the distributions of physical properties. The surprisingly large proportion of IR-quiet MDCs (90\%) is our main deviation from the results of \motte\ (40\%). We will discuss this in Sect. \ref{subsec:degeneracy}.

High-mass prestellar cores are small (0.01--0.1pc), dense ($10^5$--$10^7\ {\rm cm^{-3}}$) entities with no hydrostatic protostellar objects in their centers \citep{2002Natur.416...59M}. Since our resolutions can hardly resolve them, we chose to search for MDCs in their earliest evolutionary stages, in which high-mass prestellar cores are most likely to reside \citep[see][]{2017arXiv170600118M}. We define an MDC as ``starless'' if it has neither mid-IR (\emph{Spitzer} 24 \um\ or \emph{MSX} 21 \um) sources nor compact 70 \um\ emissions, and \numMDCStarless\ MDCs were found to meet these criteria. Caveats should be made that further observational analyses are still needed to determine whether star-forming activities are absent from these MDCs (i.e., truly ``starless''). We found that the candidate of starless MDC N69 (near N68-4 in our catalog) identified in \motte\ by its possible lack of SiO emissions shows emission peak in the \emph{Herschel} 70 \um\ data (see the corresponding panels in Figure \ref{fig:MDC}), indicating that it is not truly starless. We will discuss these starless MDCs in Sect. \ref{subsec:prestellar}.

\section{Discussion} \label{sec:disc}

\subsection{Dense Regions in Cygnus X} \label{subsec:high_density}

\subsubsection{Spatial Correlation with Developed \hii\ Regions}\label{subsec:HII}

To investigate the spatial distributions of the dense ($N_{\rm H_2}\ge$\ColThreshold) regions and (developed) \hii\ regions in Cygnus X, we selected \hii\ regions in the WISE Catalog of Galactic \hii\ Regions (V2.0)\footnote{\url{http://astro.phys.wvu.edu/wise/}} \citep{2014ApJS..212....1A} with the following criteria: (1) 303\degree.7$\le$ra$\le$310\degree.5 and 37\degree.2$\le$dec$\le$43\degree.2; (2) radii $r\ge10$ pc; (3) line-of-sight velocity range $-15\ {\rm km\cdot s^{-1}}\le v_{\rm lsr}\le20\ \rm km\cdot s^{-1}$ from the CO survey results of \citet{2012A&A...541A..79G}. A total of 7 \hii\ regions were found to meet these criteria (see Table \ref{tab:HII_region}). As can be seen in Figure \ref{fig:panorama}, our fields, which cover all the dense regions in Cygnus X, locate preferentially on the edges of or in the spaces between the \hii\ regions. To further analyze the impact of the \hii\ regions on cloud structures, we calculated their radial profiles of column density and temperature (Figure \ref{fig:HII_profile}). Out of the 7 \hii\ regions, 5 show increasing radial density profiles within their radii, indicating their ``cavity-like'' density structures. In addition, 6 \hii\ regions have increasing radial temperature profiles within their radii. This seems counter-intuitive since the inner temperature should be higher in a central-heating scenario. However, the central hotness actually can not be reflected in our continuum data since our data traces the molecular components, which have been ionized and depleted in the central regions. Instead, the low temperatures probably come from the cold molecular material along the line-of-sight that do not belong to the \hii\ region. On the contrary, molecular gas on the edge of the \hii\ region suffer less from depletion and are more likely to be heated by the central high-mass stars. These increasing temperature profiles confirm the impact of the \hii\ regions on surrounding materials and indicate that the positional correlation mentioned before should have a causal relation. For the dense regions located on the edges of the \hii\ regions, their formation can be explained by the ``collect and collapse'' process, which was firstly proposed by \citet{1977ApJ...214..725E} and supported by follow-up studies \citep[e.g.,][]{1998ASPC..148..150E,2005A&A...433..565D,2007A&A...472..835Z}. In this view, cloud materials are compressed after the passage of shock fronts driven by OB stars, and dense structures (and probably MDCs) are formed due to gravitational instability. On the other hand, the dense regions located in the space between \hii\ regions can not be affected by these shocks. A more reasonable explanation for the positional correlation of these regions would be that the formation of dense regions can not happen within the \hii\ regions due to significant feedbacks. Detailed quantitative analyses are needed to fully explain this positional correlation but it is beyond the scope of this paper.

\startlongtable
\begin{deluxetable}{ccccc}
\tabletypesize{\footnotesize}
\tablecaption{Developed \hii\ Regions in Cygnus X \label{tab:HII_region}}  
\tablehead{ 
    \colhead{Name} & \colhead{$\rm RA_{J2000}$} & \colhead{$\rm DEC_{J2000}$} & \colhead{$R$} & \colhead{$v_{\rm lsr}$} \\
    \colhead{} & \colhead{(h m s)} & \colhead{(\degree\ \arcmin\ \arcsec)} & \colhead{(pc)} & \colhead{($\rm km\cdot s^{-1}$)}  } 
\startdata
G076.951+01.718&	20:19:20.1&	+39:11:15&	12.1&	6.7\\
G077.402+00.841&	20:24:22.4&	+39:03:28&	26.5&	-0.9\\
G078.698+01.902&	20:23:43.8&	+40:43:38&	19.8&	-3.7\\
G078.714+03.115&	20:18:28.8&	+41:25:45&	22.5&	0.9\\
G079.915-00.504&	20:37:43.7&	+40:17:30&	25.6&	1.6\\
G080.362+01.212&	20:31:51.5&	+41:40:40&	18.2&	-12.8\\
G081.920+00.138&	20:41:31.8&	+42:16:24&	14.7&	10.5\\
 
\enddata
\end{deluxetable}

\subsubsection{Potential of High-mass Star Formation of the Fields}\label{subsec:Field_HMSF} 

In order to assess the HMSF potential of the fields, we calculated their $M(r)-r$ relations by drawing contours on the column density maps of the fields and measuring mass and size for each contour. The contour levels are exponentially distributed from the background levels to the maxima of the fields, and the sizes are defined as the effective radius: $r=(A/\pi)^{1/2}$, where $A$ is the area of the contour region. The results are shown in Figure \ref{fig:M-r}. As can be seen, most of the fields have $M(r)-r$ relations that can be well fitted by a power-law: $M(r)\propto r^\alpha$. The mean index value of the fields is 1.4. In addition, almost all the fields (with only one exception S01S) meet the HMSF criterion $M(r)\ge870\ M_{\odot}(r/\rm{pc})^{1.33}$ proposed by \citet{2010ApJ...723L...7K}, which supports the notion that most of the fields have potential to form high-mass stars. Interestingly, we found that no obvious systematic deviations of the $M(r)-r$ relation have been found among the fields with and without known HMSF signposts. This indicates that the fields without HMSF signposts are still highly likely to form high-mass stars in the future. 

\subsection{A New Complete Sample of MDCs in Cygnus X} \label{subsec:disc_MDC}

Our new MDC sample contains \numMDC\ cores in Cygnus X with $M\ge$\massThresholdMDC. These MDCs have an average size of $\sim$0.1 pc, median mass of 60.4 \msun, average temperature of 15.2 K, and a median density of $1.7\times10^6\ \rm cm^{-3}$. Compared with cores in nearby low-mass star forming regions \citep[e.g.,][]{1998A&A...336..150M,1999MNRAS.305..143W}, our MDCs have roughly the same sizes but are 1--2 orders of magnitude more massive (and thus denser), which shows the intrinsic HMSF nature of the MDCs. In this subsection, we will provide a detailed comparison between our MDC catalog and the catalog created by \motte\ (Sect. \ref{subsec:M07_compare}) and analyze the properties of MDCs and their constraints on HMSF (Sect. \ref{subsec:degeneracy} and Sect. \ref{subsec:prestellar})

\subsubsection{Comparison with the MDC Sample of M07} \label{subsec:M07_compare}

Compared with \motte, our MDC sample was created using new data (especially the \emph{Herschel} data) and more advanced extraction techniques in a larger spatial coverage. It is necessary and interesting to make a detailed comparison between the two samples.

Using their own source-extraction technique and the \emph{Gaussclumps} program \citep{1990ApJ...356..513S,1998A&A...329..249K}, \motte\ identified and characterized a total of 129 cores on their 3-$\rm deg^2$ 1.2-mm continuum maps of the IRAM 30 m telescope, which focused on the high-extinction ($A_{\rm v}\ge15$) regions in Cygnus X. They further found 42 probable precursors of high-mass stars (i.e., MDCs'') out of their core sample by identifying HMSF signposts (embedded \hii\ regions, high infrared luminosity, and strong SiO line emissions). Here we cross-correlated our cores with those of \motte\ allowing up to 10\arcsec\ (0.07 pc) offsets. The results are shown in Table \ref{tab:MDC_property} and Figure \ref{fig:venn}. For the 42 MDCs in \motte, \numMDCOrig\ (79\%) were found to be associated with our MDCs, 6 were ``degraded'' to be associated to our cores (because of their low masses or poor definitions), and 3 do not have any associations. In the following three paragraphs, we will respectively discuss the unique MDCs in \motte, the MDCs in common, and our new MDCs that were not reported in \motte. 

A total of 9 MDCs in \motte\ were no longer included in our MDC sample (see Figure \ref{fig:venn}). These MDCs are \listDegradedMDC\ in \motte's nomenclature. NW02, S07, S10, and S15 are at the low-mass end of \motte's MDC sample. Indeed, these MDCs have masses in \motte\ all below 30 \msun\ and should not to be included in our MDC sample. S26 is actually AFGL 2591 and has been excluded from our analysis (see Sect. \ref{sec:intro}). N58 is surrounded by the shell of an \hii\ region IRAS 20375+4109 \citep{2002A&A...390..337K}, and shows no peak but a shell structure in the 70--850 \um\ continuum and column-density maps. Its peak emission in the 1.2-mm map, by which \motte\ identified this core, is possibly the result of the contamination of free-free emissions. In our source extraction procedure, N32 was merged into our MDC W75N-1 and N69 was splitted into two cores due to its elongated shape, of which one becomes our MDC N68-4 and the other becomes a low-mass core. Finally, S18 has position and size in \motte\ that can not fit well with our \emph{Herschel} and column-density maps, and is thus unreliable. It was substituted by our MDC S106-1 of which the position and size are consistent with our data. To summarize, owing to the improved data quality and extraction algorithm, we found that these 9 cores identified in \motte\ are no longer qualified as MDCs. 

By comparing the 33 common MDCs in the two samples (Figure \ref{fig:venn}), we found that our MDCs have similar sizes as those of \motte's MDCs but are $\sim$3.5 times more massive on average (compared with \motte). This deviation is due to our larger extracted fluxes ($\sim1.4$ times larger in $F_{\nu}$ at 1.2 mm), smaller $\kappa_{\nu}$ (e.g., $\kappa_{\nu, 1.2\rm mm}=0.00625\ \rm cm^2g^{-1}$ in our case versus 0.01 $\rm cm^2g^{-1}$ in \motte), and colder temperature (average temperature of $\sim15$ K in our case versus a fixed temperature of 20 K in \motte). While the first reason comes from the systematic deviation of the different extraction techniques, the latter two reasons, which jointly increase the mass evaluation by a factor of $\sim$2, truly reflect that the masses of MDCs were probably underestimated in \motte. This indicates that MDCs in Cygnus X may be more massive than we expected before. 

With our larger data coverage, a great number of MDCs (118) were found that are not in \motte's MDC sample, making our MDC sample three times larger than theirs. Interestingly, these MDCs have physical properties that systematically deviate from those of the MDCs associated with \motte's MDCs: they are generally less massive, colder, and less luminous (see Figure \ref{fig:histProp}). We think this result can be attributed to the relatively high sensitivity and quality of the \emph{Herschel} data (see Table \ref{tab:beam}). With the help of the powerful \emph{Herschel} equipments, less massive dense-gas entities can be identified. Thus compared with \motte, our sample is much more improved in statistical completeness, especially for the most quiescent MDCs.   

\subsubsection{Interpretations of the Large proportion of IR-quiet MDCs} \label{subsec:degeneracy}

One unique feature of our MDC sample is that IR-quiet MDCs take up a surprisingly large proportion (90\%), compared with other studies using the same techniques, e.g., \motte\ (40\%) and \citet{2017A&A...602A..77T} (63\%). We here propose two interpretations to explain this phenomenon.  

\begin{figure*}[htb!]
\epsscale{0.8}\plotone{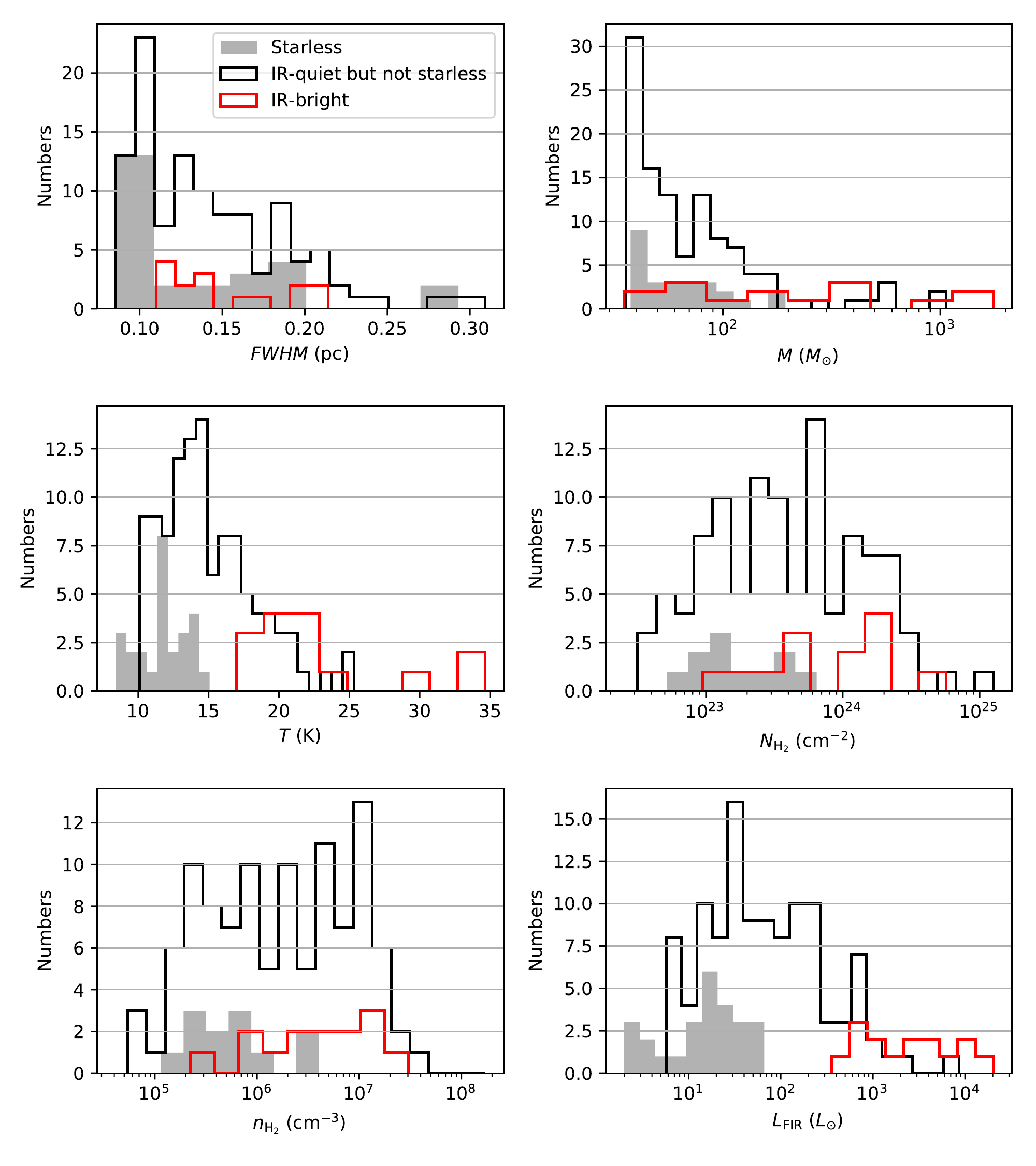}
\caption{Distributions of the \emph{convolved} FWHM sizes (at \emph{Herschel} 160 \um), masses, dust temperatures, $\rm H_2$ column and volume densities, and far-IR luminosities of the MDCs in different infrared classes.  \label{fig:histIR}}
\end{figure*}

\begin{figure*}[htb!]
\epsscale{1.15}\plotone{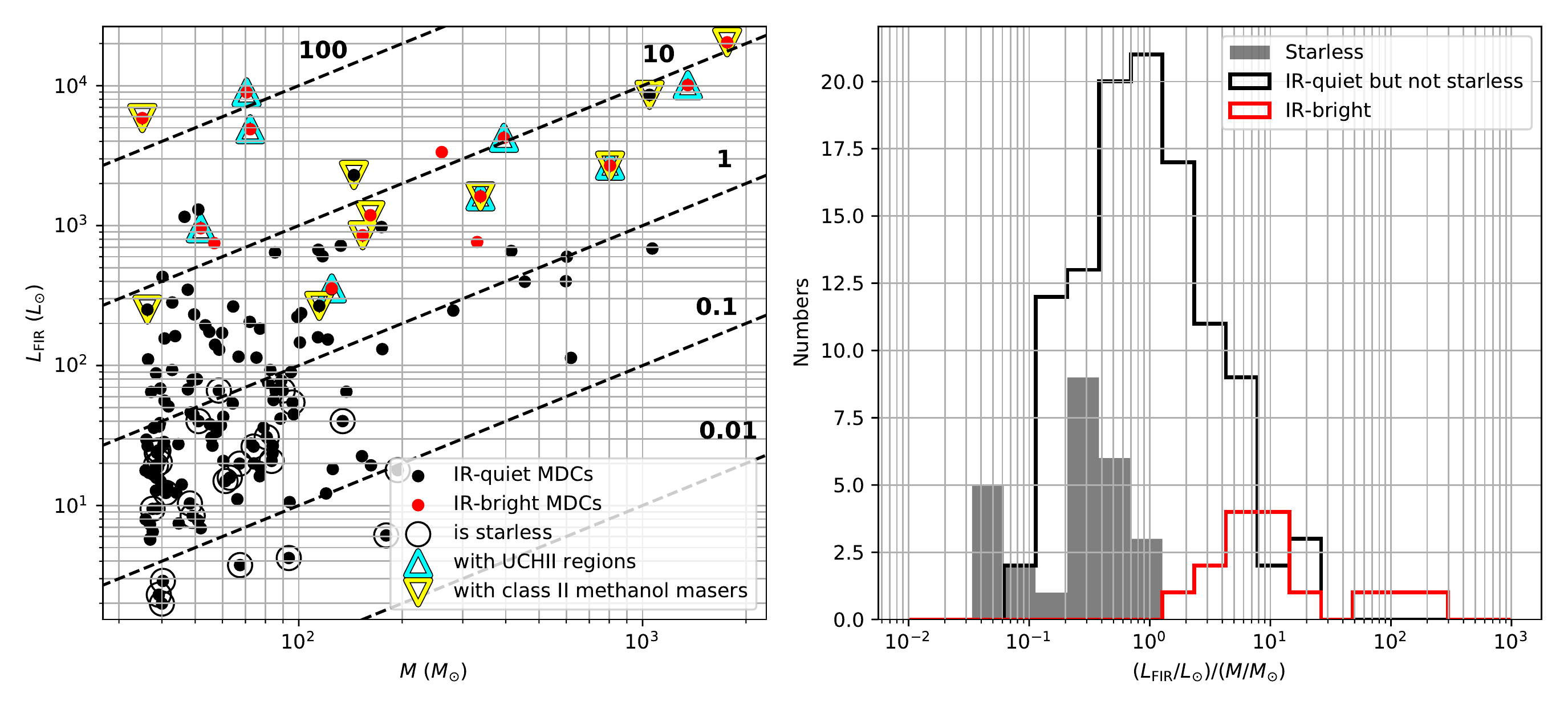}
\caption{\emph{Left}: luminosity--mass diagram of the MDCs in Cygnus X. Dashed lines are the $L/M=const.$ curves with the constant values near them. \emph{Right}: distributions of the $L/M$ ratios of different kinds of MDCs. It is clear that $L/M$ ratios increase along the sequence: starless--IR-quiet but not starless--IR-bright.  \label{fig:M-L}}
\end{figure*}

One is to assume that all the MDCs (including the IR-quiet ones) are on the evolutionary track of \emph{HMSF}, as was done in \motte\ and \citet{2017A&A...602A..77T}. This interpretation can explain the discrepancies of physical properties of MDCs in different infrared classes (see Table \ref{tab:MDC_overview}, Figure \ref{fig:histIR}, and Figure \ref{fig:M-L}): the increasing average temperature and luminosity of the MDCs in starless, IR-quiet but not starless, and IR-bright classes are a consequence of the heating from growing (high-mass) protostars, and the increasing average mass is due to the accretion of MDCs. If we assume that the star formation rate of Cygnus X has remained constant over the past 1--2 Myr, the lifetimes of the MDCs in different stages can be estimated by the number-counting method. For better comparison, we adopted the statistics of OB stars in Cygnus X from \motte\ ($n_{\rm OB}=2600\pm1000$, $t_{\rm OB}=2\pm1$ Myr). An average MDC-to-high-mass-star production ratio of 2 was assumed, which was implied by the observational results of \citet{2010A&A...524A..18B}. In result, the lifetimes of IR-quiet and IR-bright stages are 210$\pm$131 kyr and 23.1$\pm$15.5 kyr, respectively. The lifetime of IR-quiet stage is even comparable with the whole HMSF timescale, of order $\sim100$ kyr, proposed by some theoretical and simulation studies \citep{1999astro.ph..9473M,2000ApJ...534..976N,2001A&A...373..190B,2002Natur.416...59M}. This indicates that HMSF probably needs to go through a relatively long and quiescent low- to intermediate-mass stage before reaching the more dynamic high-mass stage. This accelerated HMSF scenario was also reported by \citep{2005ApJ...625..864Z,2015ApJ...804..141Z}, who suggested an increasing accretion ratio with protostellar mass based on the CO observations of outflows in HMSF sites, and proposed by theoretical works \citep{1941MNRAS.101..227H,1952MNRAS.112..195B,2000A&A...359.1025N,2002Natur.416...59M,2003ApJ...599.1196K}. 




An alternative interpretation of the large proportion of IR-quiet MDCs is to assume that not all the IR-quiet MDCs are able to form high-mass stars. In this sense, the lower mass, luminosity, and mid-IR flux of the IR-quiet MDCs are due to their intrinsic nature of forming low- to intermediate-mass stars and the true timescale of IR-quiet stage is perhaps much shorter than that of the first scenario. This idea was also supported by \citet{2017MNRAS.466.3682B}, who analyzed the effect of source distance on the identification of HMSF candidates by synthesizing ``observational'' images of ``moved'' star-forming clouds. They found that more ``false positives'' will be identified if low-mass star-forming regions are moved to larger distances, probably because more inter-core emissions will be counted into those of cores. The proportion of IR-quiet MDCs that can form high-mass stars is unclear and only high-resolution studies can provide some hints. For instance, \citet{2010A&A...524A..18B} observed six massive IR-quiet MDCs in Cygnus X using the IRAM Plateau de Bure interferometer with resolutions as high as 1700 AU, and found that five of them have massive fragments that are probably precursors of high-mass stars. \citet{2013A&A...558A.125D} studied the CO outflows of the same six MDCs and found that 8 out of 9 high-mass cores (except N53 MM2) are driving outflows stronger than typical low-mass ones. However, their sample is small and biased towards the most massive IR-quiet MDCs.

Both of the interpretations will provide strong constraints on the early evolution of HMSF once validated but unfortunately, we can not yet unravel them based on our data. We expect that this so-called ``mass--evolution degeneracy'' can be disentangled by comprehensive high-resolution surveys of MDCs using interferometers such as SMA, NOEMA, and ALMA.  

\subsubsection{Quest for High-mass Prestellar Cores in Cygnus X} \label{subsec:prestellar}

\begin{figure}[htb!]
\epsscale{1.2}\plotone{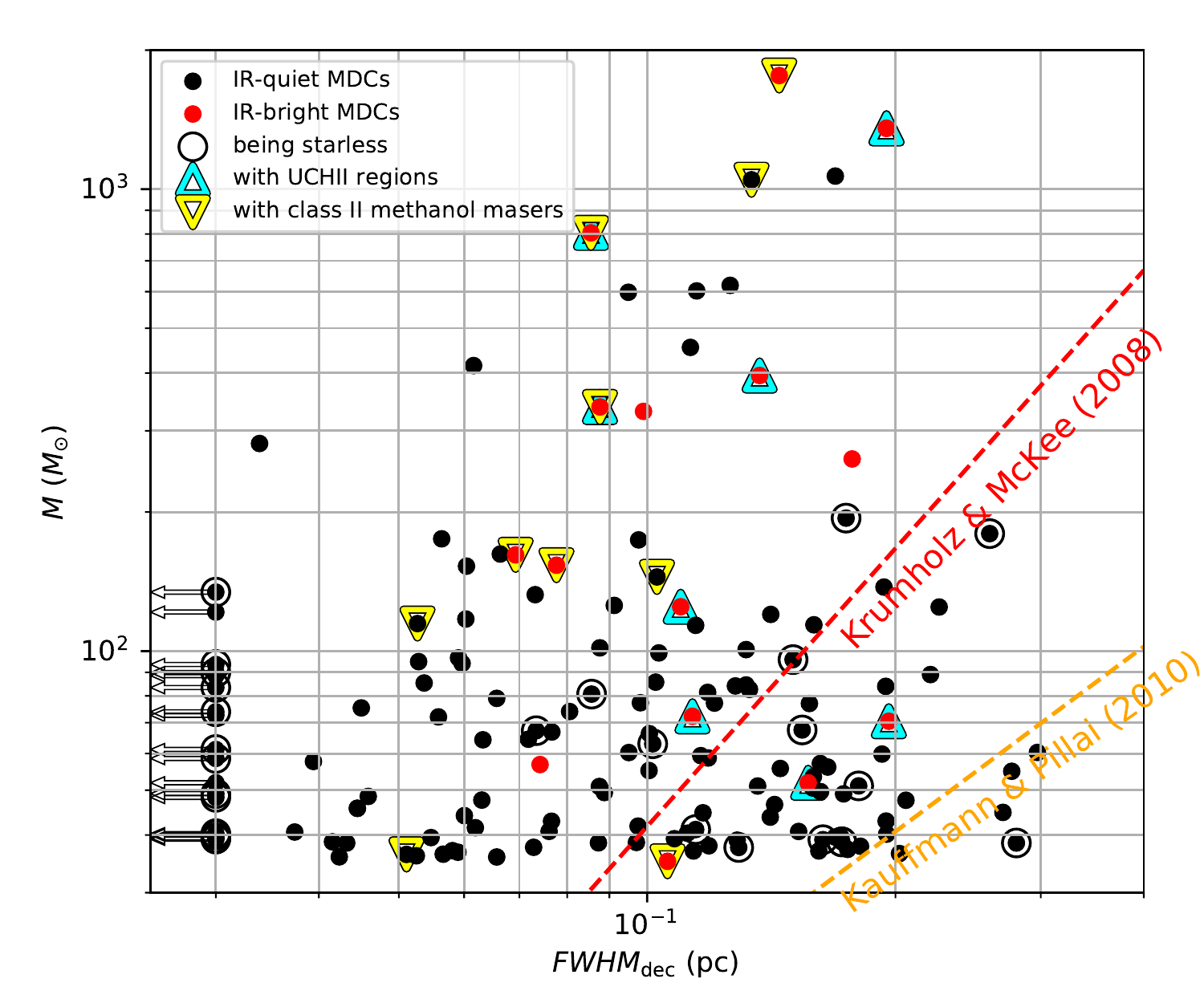}
\caption{Mass--size diagram of the MDCs in Cygnus X. Sizes are presented as deconvolved FWHM sizes at \emph{Herschel} 160 \um. MDCs with sizes smaller than the \emph{Herschel} beam at 160 \um\ are attached with left arrows. The infrared classes and the signposts of HMSF of the MDCs are explained in the legend. Two HMSF thresholds are shown as dashed lines: column density $=1\rm\ g\ cm^{-2}$ proposed by \citet{2008Natur.451.1082K} (red), and $M(r)=870\ M_{\odot}(r/\rm pc)^{1.33}$ proposed by \citet{2010ApJ...723L...7K} (orange). \label{fig:M-size}}
\end{figure}

We found a total of \numMDCStarless\ starless MDCs out of the \numMDC\ MDCs by their absence of compact emissions in the \emph{Herschel} 70 \um\ maps. The average size, mass, and density of these starless MDCs are 0.16 pc, 70 \msun, and $7\times10^5\ \rm cm^{-3}$, respectively. With the caveat of the mass--evolutionary degeneracy discussed in Sect. \ref{subsec:degeneracy}, whether high-mass prestellar cores exist within these starless MDCs remains unclear. However, with the simple idea that more denser cores are more likely to form high-mass stars, we propose that the 7 starless MDCs of which densities higher than the 1 $\rm g\ cm^{-2}$ threshold of HMSF proposed by \citet{2008Natur.451.1082K} (see Figure \ref{fig:M-size}) are among the most probable HMSF candidates. These MDCs are N68-5, S43-3, W75N-2, S106W2-6, S11-1, DR21-8, and S106-1, sorted by decreasing densities. It would be interesting to investigate the detailed properties of these MDCs using interferometers to reveal the early stages of HMSF. 

\section{Summary} \label{sec:conc}

\begin{figure*}[htb!]
\epsscale{1}\plotone{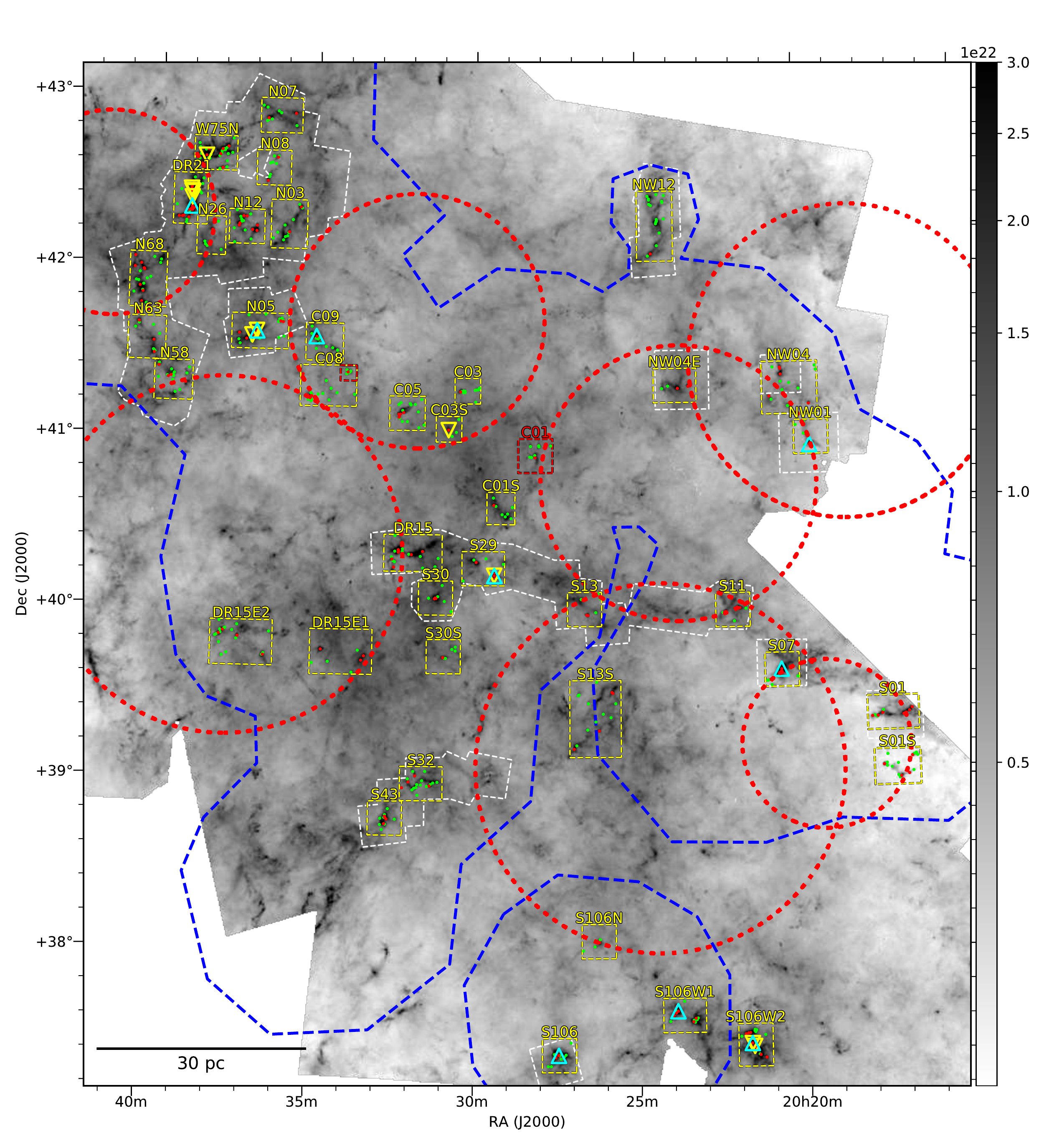}
\caption{High-resolution (18\arcsec) $N_{\rm H_2}$ map of Cygnus X derived from \emph{Herschel} images, units are in $\rm cm^{-2}$. Yellow dashed rectangles represent the fields considered in this paper, which cover dense regions ($N_{\rm H_2}\ge$\ColThreshold) in Cygnus X. Regions in red dashed rectangles are excluded from our analysis due to their inconsistent distances (see Sect. \ref{subsec:field}). Blue and white dashed polygons outline the coverages of the JCMT and IRAM 30 m telescope images, respectively. Green and red dots mark the positions of our cores and MDCs, respectively, extracted by \emph{getsources} (see Sect. \ref{subsec:extraction}). UC\hii\ regions and sites of class II methanol masers are marked as cyan and yellow triangles, respectively (see Sect. \ref{subsec:mdc}). Developed \hii\ regions in Cygnus X are represented as dotted red circles (see Sect. \ref{subsec:HII}). \label{fig:panorama}}
\end{figure*}

\begin{figure}[htb!]
\epsscale{1.2}\plotone{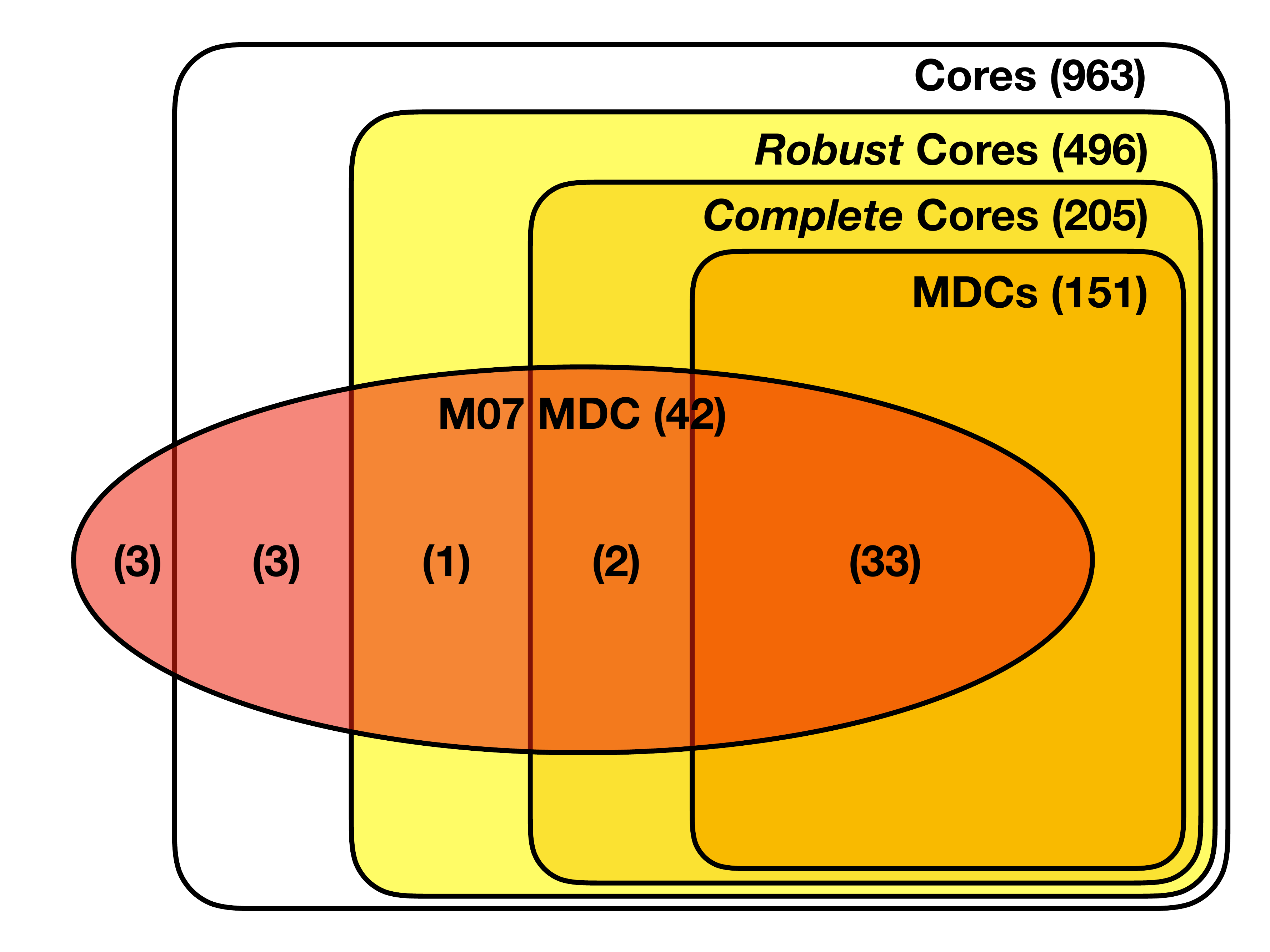}
\caption{Venn diagram of the core/MDC samples in this paper and in \motte. Sizes of the samples and the intersections are shown in brackets.\label{fig:venn}}
\end{figure}

\begin{figure}[htb!]
\epsscale{.9}\plotone{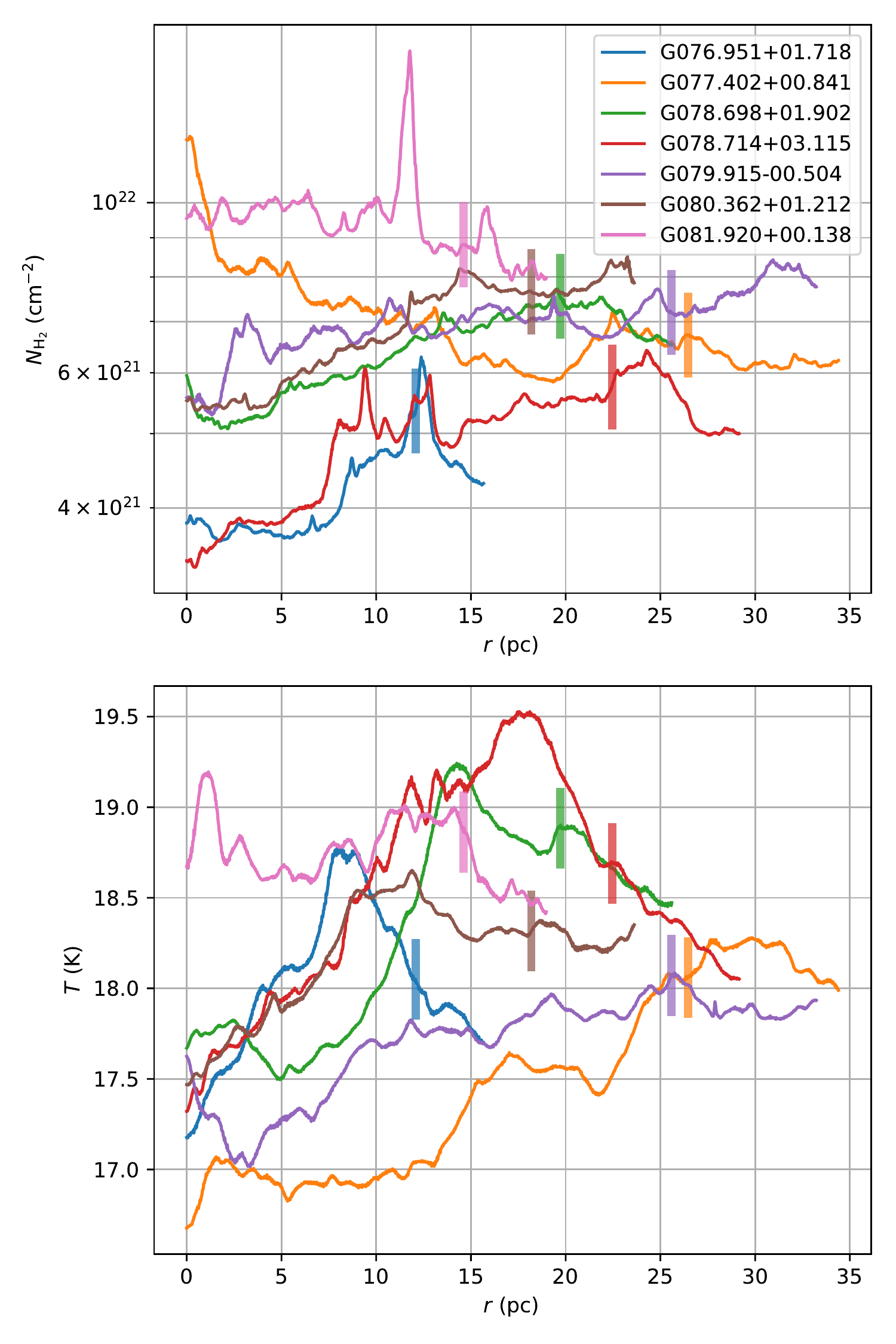}
\caption{Radial profiles of column density (up) and temperature (bottom) of the 7 developed \hii\ regions in Cygnus X (see Sect. \ref{subsec:HII}). Radial distance range is from the centers of the \hii\ regions to 1.3 times of their radii (marked by short vertical segments). \label{fig:HII_profile}}
\end{figure}

\begin{figure}[htb!]
\epsscale{1.2}\plotone{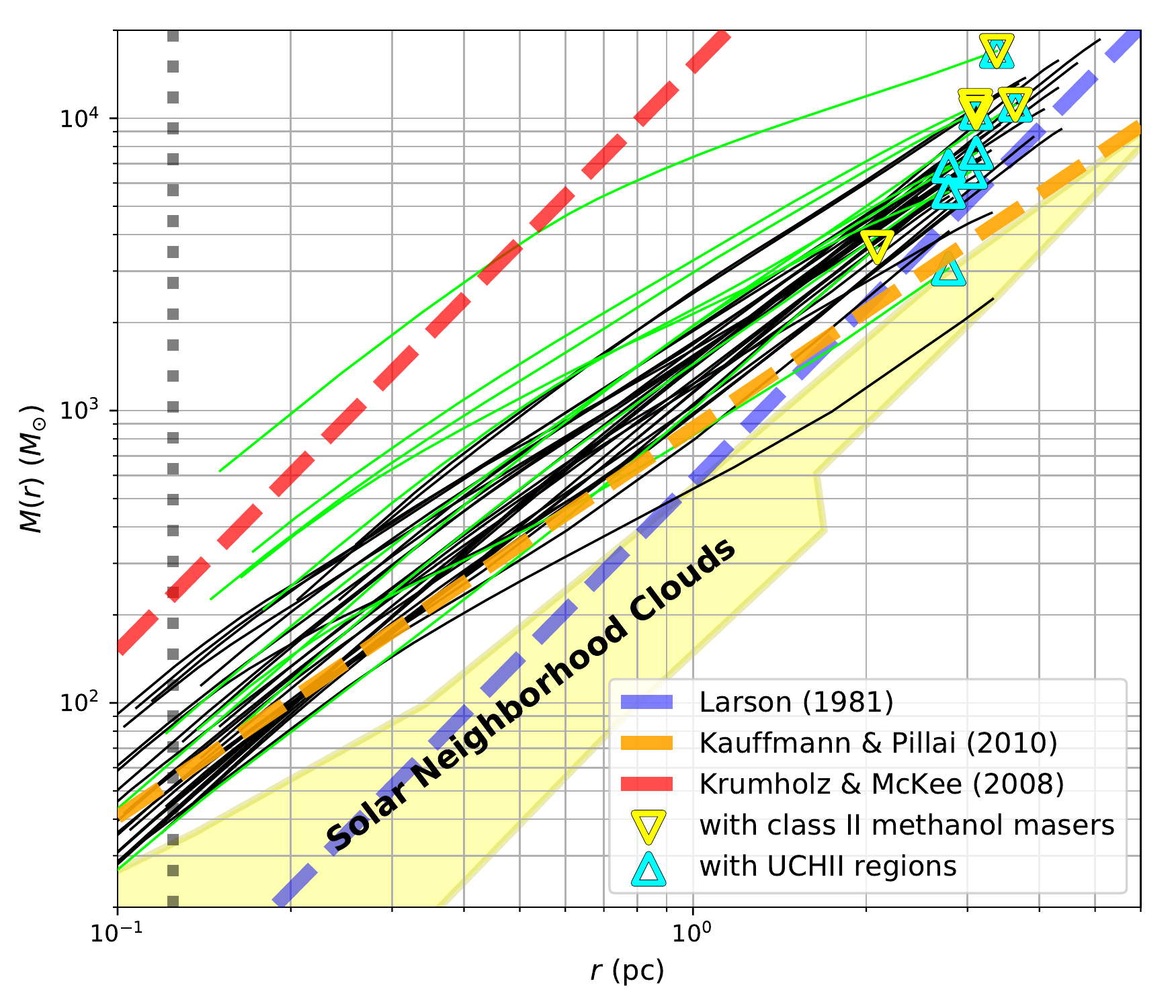} 
\caption{$M(r)$--$r$ relations of the fields as solid curves (see Sect. \ref{subsec:Field_HMSF}). Fields containing HMSF signposts have the corresponding curves colored green and attached with triangles. The vertical dotted line shows the resolution of the column density map (18\arcsec). Blue, orange, and red dashed lines represent the relations of \citet{1981MNRAS.194..809L} ($M(r)=572\ M_{\odot}(r/\rm pc)^2$), \citet{2010ApJ...723L...7K} ($M(r)=870\ M_{\odot}(r/\rm pc)^{1.33}$), and \citet{2008Natur.451.1082K} (1 $\rm g\ cm^{-2}$), respectively. The yellow range of solar neighborhood clouds were from Figure 2 of \citet{2010ApJ...723L...7K}. \label{fig:M-r}}
\end{figure}

\begin{figure*}[htb!]
\epsscale{0.8}\plotone{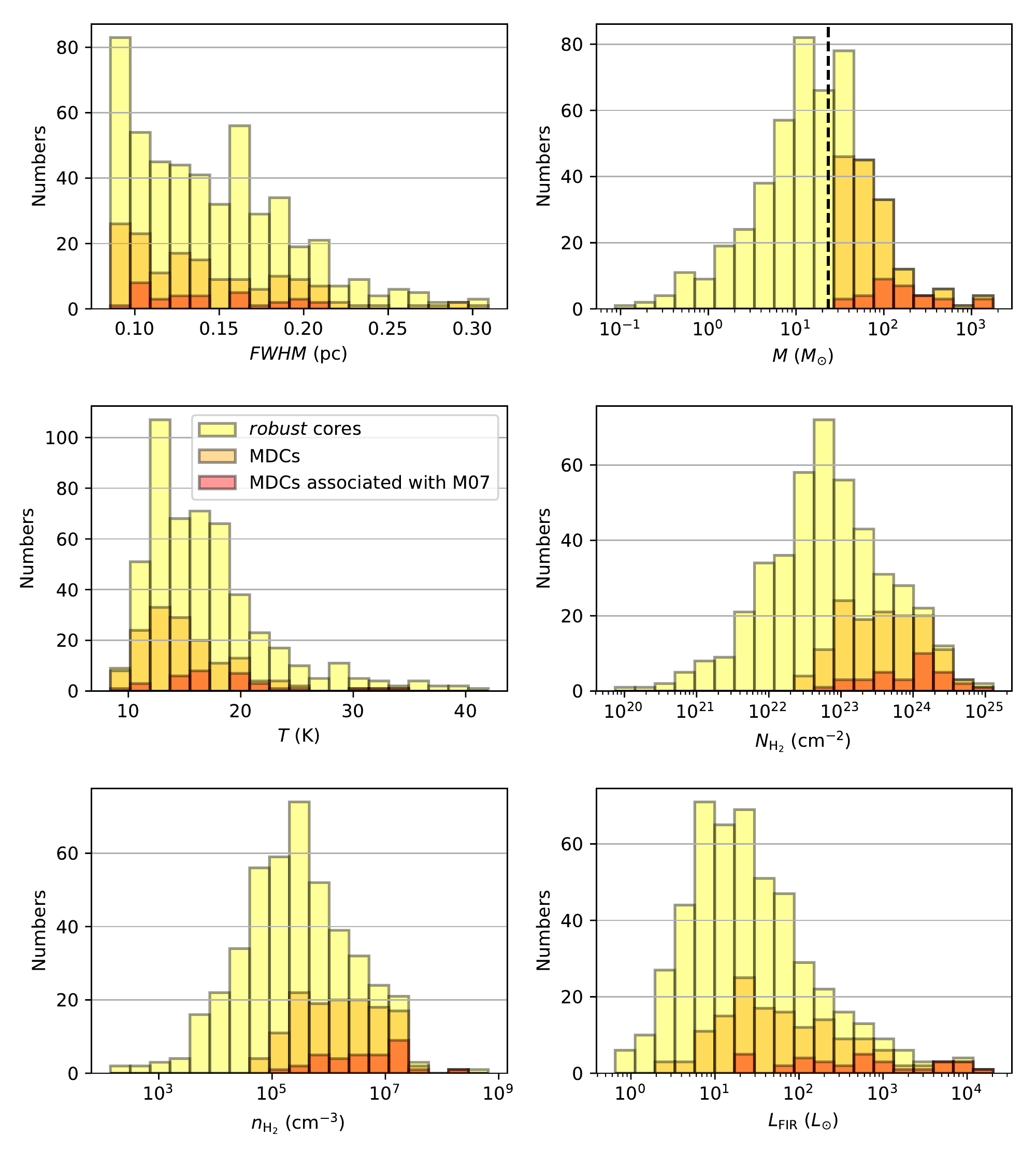}
\caption{Distributions of the \emph{convolved} FWHM sizes (at \emph{Herschel} 160 \um), masses, dust temperatures, $\rm H_2$ column and volume densities, and far-IR luminosities of the \emph{robust} core sample, the MDC sample, and the MDCs that are associated with the MDCs in \motte\ (Sect. \ref{subsec:M07_compare}). In the ``mass'' panel, cores with masses larger than the critical mass represented as the vertical dashed line (\massThresholdCompleteness) are statistically complete (see Sect. \ref{subsec:field}). \label{fig:histProp}}
\end{figure*}

As part of the \surveyAb\ project, we aimed at characterizing the physical conditions of high-mass star formation at $\sim$ 0.1 pc scales and conducted a complete survey toward the massive dense cores in Cygnus X using space telescope observations of \emph{Spitzer} and \emph{Herschel} as well as ground-based submillimeter and millimeter data. Our results and conclusions are summarized as follows: 

\begin{enumerate}

\item
We generated 5\degree$\times$6\degree\ mosaic maps of the whole Cygnus X region in the bands of \emph{Herschel} (70, 160, 250, 350, and 500 \um), JCMT (450 and 850 \um), and the IRAM 30 m telescope (1.2 mm) (see Figure \ref{fig:H70} to \ref{fig:I1200}). A high-resolution column density map (at 18\arcsec, Figure \ref{fig:panorama}) and a temperature map (at 36.7\arcsec, Figure \ref{fig:T}) of Cygnus X were also created using the \emph{Herschel} images.

\item
With a column density threshold of \ColThreshold\ (corresponding to $A_{\rm v}\ge15$) we selected \numTrueField\ high-density fields in Cygnus X (see Figure \ref{fig:panorama} and Table \ref{tab:field}). After the source extraction and SED fitting, \numCoreGood\ cores were identified with well-defined masses and temperatures. Statistical analysis (Sect. \ref{subsec:extraction}) suggests that cores with $M\ge$\massThresholdCompleteness\ (\numCoreBig\ cores) are statistically complete. 

\item 
We cross-correlated the \numCoreBig\ statistically complete cores with two kinds of HMSF signposts: class II methanol masers and UC\hii regions (see Sect. \ref{subsec:mdc} and Table \ref{tab:MDC_sign}). We determined a mass threshold of \massThresholdMDC\ for MDCs and selected \numMDC\ MDCs. These MDCs have an average size of $\sim$0.1 pc, median mass of 60.4 \msun, average temperature of 15.2 K, and a median density of $1.7\times10^6\ \rm cm^{-3}$ (see Figure \ref{fig:histProp}, Figure \ref{fig:MDC_table}, and Table \ref{tab:MDC_property}). 

\item
We matched our core sample with \motte's MDC sample (see Table \ref{tab:MDC_property} and Figure \ref{fig:venn}) and found that our MDC sample covers most of the MDCs in \motte\ and is three times larger. Compared with the results of \motte\ with distance corrected to 1.4 kpc, our MDCs have similar sizes of $\sim$0.1 pc but are on average $\sim$3.5 times more massive, showing that the masses were probably underestimated before (see Sect. \ref{subsec:M07_compare}). With more advanced telescopes such as \emph{Herschel}, we are able to find 118 new MDCs that are more quiescent than those in \motte. 

\item
We cross-correlated the MDCs with mid-IR sources of \emph{Spitzer} and \emph{MSX}, and measured their mid-IR fluxes (see Table \ref{tab:MDC_IR}). Nearly two thirds of MDCs were found to be associated with mid-IR sources. We classified the MDCs into \numMDCIRB\ IR-bright ones and \numMDCIRQ\ IR-quiet ones by comparing their mid-IR fluxes with that of a B3-type stellar embryo. We found that the average mass, temperature, and far-IR luminosity of MDCs increase along the “starless–-IR-quiet but not starless-–IR-bright” sequence, and that starless MDCs are $\sim$1.5 times larger than the other MDCs. 

\item
The large proportion of IR-quiet MDCs (90\%) is a unique feature of our sample and we have proposed two possible scenarios of HMSF in Cygnus X to interpret it. 

\begin{enumerate}
\item Assuming that all the IR-quiet MDCs are able to form high-mass stars, HMSF is probably an accelerated progress with a relatively long ($\sim200$ kyr) IR-quiet stage and a rapid IR-bright stage. 

\item Not all the IR-quiet MDCs will form high-mass stars and their quiescent appearance is partly due to their intrinsic nature of forming low- to intermediate-mass stars. 
\end{enumerate}

We can not yet determine which one is closer to the truth with our existing data (so-called ``mass--evolution degeneracy''). Future high-resolution comprehensive surveys of MDCs are necessary to provide better constraints on HMSF.

\item
To search for HMSF at the earliest stages, we identified \numMDCStarless\ starless MDCs by their lack of compact 70 \um\ emissions (see Sect. \ref{subsec:IR_class}). With the caveat of the mass--evolutionary degeneracy, it is highly unlikely that all of them will form high-mass stars. However, the 7 densest MDCs with densities exceeding 1 $\rm g\ cm^{-2}$ are among the most probable HMSF candidates for future studies. 

\item
To investigate the HMSF potential of the fields, we calculated their $M(r)$--$r$ relations (see Figure \ref{fig:M-r}). Almost all the fields have $M(r)$--$r$ relations above the HMSF thresholds $M(r)=870\ M_{\odot}(r/\rm{pc})^{1.33}$ proposed by \citet{2010ApJ...723L...7K}, indicating their high potential of HMSF. Interestingly, no obvious systematic deviations of the $M(r)$--$r$ relation have been found among the fields with and without HMSF signposts, which indicates that the fields without HMSF signposts are likely to contain HMSF sites at the earliest stages. 

\item
We found that the dense regions (covered by our fields) in Cygnus X prefer to locate on the edges of or in the space between the developed ($r\ge$10 pc) \hii\ regions (see Figure \ref{fig:panorama}). The ``cavity-like'' density structures and the increasing radial profiles of temperature of the \hii\ regions (see Figure \ref{fig:HII_profile}) further confirm their impacts on surrounding cloud materials. We here propose that large-scale structures in Cygnus X might be dominated by the expansion of developed \hii\ regions. 

\end{enumerate}

\acknowledgments 
\emph{Acknowledgments}.
Y.C., K.Q., J.L., B.H., and Y.W. are supported by National Key R\&D Program of China No. 2017YFA0402600. We acknowledge the support from National Natural Science Foundation of China (NSFC) through grants U1731237, 11473011, 11590781 and 11629302.

We are grateful to Alexander Men'shchikov for his help and discussions on \emph{getsources}. This research made use of \texttt{Astropy}, a community-developed core Python package for Astronomy \citep{2013A&A...558A..33A}. This research made use of data products from the Midcourse Space Experiment. Processing of the data was funded by the Ballistic Missile Defense Organization with additional support from NASA Office of Space Science. This research has also made use of the NASA/ IPAC Infrared Science Archive, which is operated by the Jet Propulsion Laboratory, California Institute of Technology, under contract with the National Aeronautics and Space Administration. This research used the facilities of the Canadian Astronomy Data Center operated by the the National Research Council of Canada with COMMENT the support of the Canadian Space Agency. This research has made use of the NASA/IPAC Infrared Science Archive, which is operated by the Jet Propulsion Laboratory, California Institute of Technology, under contract with the National Aeronautics and Space Administration. This research has made use of the SIMBAD database, operated at CDS, Strasbourg, France. This work uses the data of \emph{Herschel}, which is an ESA space observatory with science instruments provided by European-led Principal Investigator consortia and with important participation from NASA. This article makes use of data products from the Wide-field Infrared Survey Explorer, a joint project of the University of California, Los Angeles, and the Jet Propulsion Laboratory/California Institute of Technology, funded by the National Aeronautics and Space Administration. This work is based in part on observations made with the Spitzer Space Telescope, which is operated by the Jet Propulsion Laboratory, California Institute of Technology under a contract with NASA. The images are based on data obtained as part of the Spitzer Space Telescope Cygnus X Legacy Survey (PID 40184, P.I. Joseph L. Hora). This work uses the data of the James Clerk Maxwell Telescope, which is operated by the East Asian Observatory on behalf of The National Astronomical Observatory of Japan; Academia Sinica Institute of Astronomy and Astrophysics; the Korea Astronomy and Space Science Institute; the Operation, Maintenance and Upgrading Fund for Astronomical Telescopes and Facility Instruments, budgeted from the Ministry of Finance (MOF) of China and administrated by the Chinese Academy of Sciences (CAS), as well as the National Key R\&D Program of China (No. 2017YFA0402700). Additional funding support is provided by the Science and Technology Facilities Council of the United Kingdom and participating universities in the United Kingdom and Canada. This work is based on observations carried out with the IRAM 30 m telescope. IRAM is supported by INSU/CNRS (France), MPG (Germany) and IGN (Spain).

\facilities{ IRAM: 30m, JCMT (SCUBA2), \emph{Herschel} (PACS,SPIRE), \emph{Spitzer} (IRAC,MIPS), \emph{MSX}, \emph{WISE}}
\software{\emph{getsources}, \texttt{Astropy}}

\clearpage
\appendix

\section{Fields used for core extraction}\label{sec_A:field} 

To extract MDCs in Cygnus X, we selected \numTrueField\ high-density rectangular fields using a column density threshold of \ColThreshold. The threshold guarantees that cores with $M_{\rm core}\ge$\massThresholdCompleteness\ extracted in the fields are statistically complete (see Sect. \ref{subsec:field}). Table \ref{tab:field} lists the geometric information and the numbers of cores in the fields. 

\renewcommand{\thetable}{A}
\startlongtable
\begin{deluxetable*}{c|ccccccc}
\tabletypesize{\normalsize}
\tablecaption{Fields in Cygnus X Used for Core Extraction \label{tab:field}} 
\tablehead{
    \colhead{Name} & \colhead{$\rm RA_{J2000}$} & \colhead{$\rm DEC_{J2000}$} & \colhead{Width} & \colhead{Height} & \colhead{$N_{\rm robust}$\tablenotemark{a}} & \colhead{$N_{\rm complete}$\tablenotemark{b}} & \colhead{$N_{\rm MDC}$\tablenotemark{c}} \\
    \colhead{} & \colhead{(h m s)} & \colhead{(\degree\ \arcmin\ \arcsec)} & \colhead{(\degree)} & \colhead{(\degree)} & \colhead{} & \colhead{} & \colhead{} 
    }
\startdata
C01S&	20:29:12.9&	+40:35:00&	0.16&	0.19&	8&	2&	1\\
C03&	20:30:15.4&	+41:16:12&	0.15&	0.15&	11&	2&	1\\
C03S&	20:30:50.1&	+41:02:45&	0.15&	0.15&	4&	1&	1\\
C05&	20:32:07.9&	+41:08:15&	0.21&	0.20&	17&	6&	4\\
C08&	20:34:36.1&	+41:17:38&	0.33&	0.24&	16&	3&	3\\
C09&	20:34:44.7&	+41:33:01&	0.22&	0.22&	13&	4&	2\\
DR15&	20:31:54.8&	+40:19:08&	0.34&	0.22&	19&	10&	7\\
DR15E1&	20:34:05.3&	+39:44:11&	0.37&	0.26&	7&	3&	3\\
DR15E2&	20:37:08.3&	+39:46:55&	0.37&	0.26&	20&	6&	4\\
DR21&	20:39:04.3&	+42:22:19&	0.20&	0.30&	34&	24&	23\\
N03&	20:35:55.9&	+42:14:06&	0.21&	0.28&	12&	3&	2\\
N05&	20:36:46.3&	+41:36:24&	0.33&	0.21&	24&	8&	6\\
N07&	20:36:14.4&	+42:52:12&	0.24&	0.20&	10&	5&	2\\
N08&	20:36:27.0&	+42:33:47&	0.20&	0.20&	9&	4&	2\\
N12&	20:37:16.0&	+42:12:58&	0.21&	0.20&	17&	10&	7\\
N26&	20:38:23.2&	+42:09:31&	0.17&	0.23&	6&	1&	1\\
N58&	20:39:26.1&	+41:18:31&	0.22&	0.23&	16&	7&	4\\
N63&	20:40:18.6&	+41:33:08&	0.22&	0.25&	10&	4&	3\\
N68&	20:40:19.8&	+41:53:30&	0.22&	0.32&	20&	12&	8\\
NW01&	20:19:36.5&	+40:59:24&	0.20&	0.20&	6&	1&	1\\
NW04&	20:20:14.7&	+41:16:46&	0.32&	0.31&	14&	5&	5\\
NW04E&	20:23:48.8&	+41:18:01&	0.25&	0.20&	4&	3&	2\\
NW12&	20:24:23.7&	+42:14:05&	0.21&	0.41&	19&	2&	1\\
S01&	20:17:17.8&	+39:21:53&	0.30&	0.20&	9&	7&	6\\
S01S&	20:17:12.5&	+39:02:47&	0.27&	0.21&	15&	4&	4\\
S07&	20:20:38.2&	+39:37:43&	0.20&	0.20&	6&	2&	1\\
S11&	20:22:06.5&	+39:59:00&	0.20&	0.20&	6&	4&	3\\
S13&	20:26:38.7&	+39:59:27&	0.20&	0.20&	6&	3&	2\\
S13S&	20:26:19.5&	+39:20:56&	0.30&	0.45&	14&	4&	4\\
S29&	20:29:44.9&	+40:13:47&	0.25&	0.20&	6&	1&	1\\
S30&	20:31:12.5&	+40:03:23&	0.20&	0.20&	8&	3&	3\\
S30S&	20:30:57.4&	+39:42:47&	0.20&	0.20&	6&	2&	1\\
S32&	20:31:36.2&	+38:58:05&	0.25&	0.20&	18&	7&	4\\
S43&	20:32:41.2&	+38:45:53&	0.20&	0.20&	12&	8&	4\\
S106&	20:27:25.7&	+37:22:49&	0.20&	0.20&	10&	5&	2\\
S106N&	20:26:14.6&	+38:02:45&	0.20&	0.20&	4&	1&	1\\
S106W1&	20:23:43.0&	+37:36:39&	0.25&	0.20&	15&	8&	7\\
S106W2&	20:21:38.1&	+37:26:05&	0.20&	0.25&	16&	12&	9\\
W75N&	20:38:18.5&	+42:38:32&	0.25&	0.20&	29&	8&	6\\

\enddata
\tablenotetext{a}{Number of \emph{robust} cores (with well-defined fluxes, temperatures, and masses, see Sect. \ref{subsec:extraction}) in each field.}
\tablenotetext{b}{Number of \emph{complete} cores (\emph{robust} cores with $M_{\rm core}\ge$\massThresholdCompleteness, which means they are statistically complete in Cygnus X, see Sect. \ref{subsec:extraction}) in each field.}
\tablenotetext{c}{Number of MDCs (\emph{complete} cores with $M_{\rm core}\ge$\massThresholdMDC, see Sect. \ref{subsec:mdc}) in each field.}
\end{deluxetable*}

\clearpage
\section{Physical properties of the MDCs in Cygnus X}\label{sec_A:MDC_property} 

In this appendix, Figure \ref{fig:MDC_table} visualizes all the MDCs in Cygnus X, Figure \ref{fig:number_in_fields} shows the numbers of different MDCs in each field, and Table \ref{tab:MDC_property} lists the detailed physical properties and classifications of the MDCs. 

\begin{minipage}{\textwidth}
\figurenum{B1}\addtocounter{figure}{1}
\centering\vspace{0.5in}
\includegraphics[width=6 in]{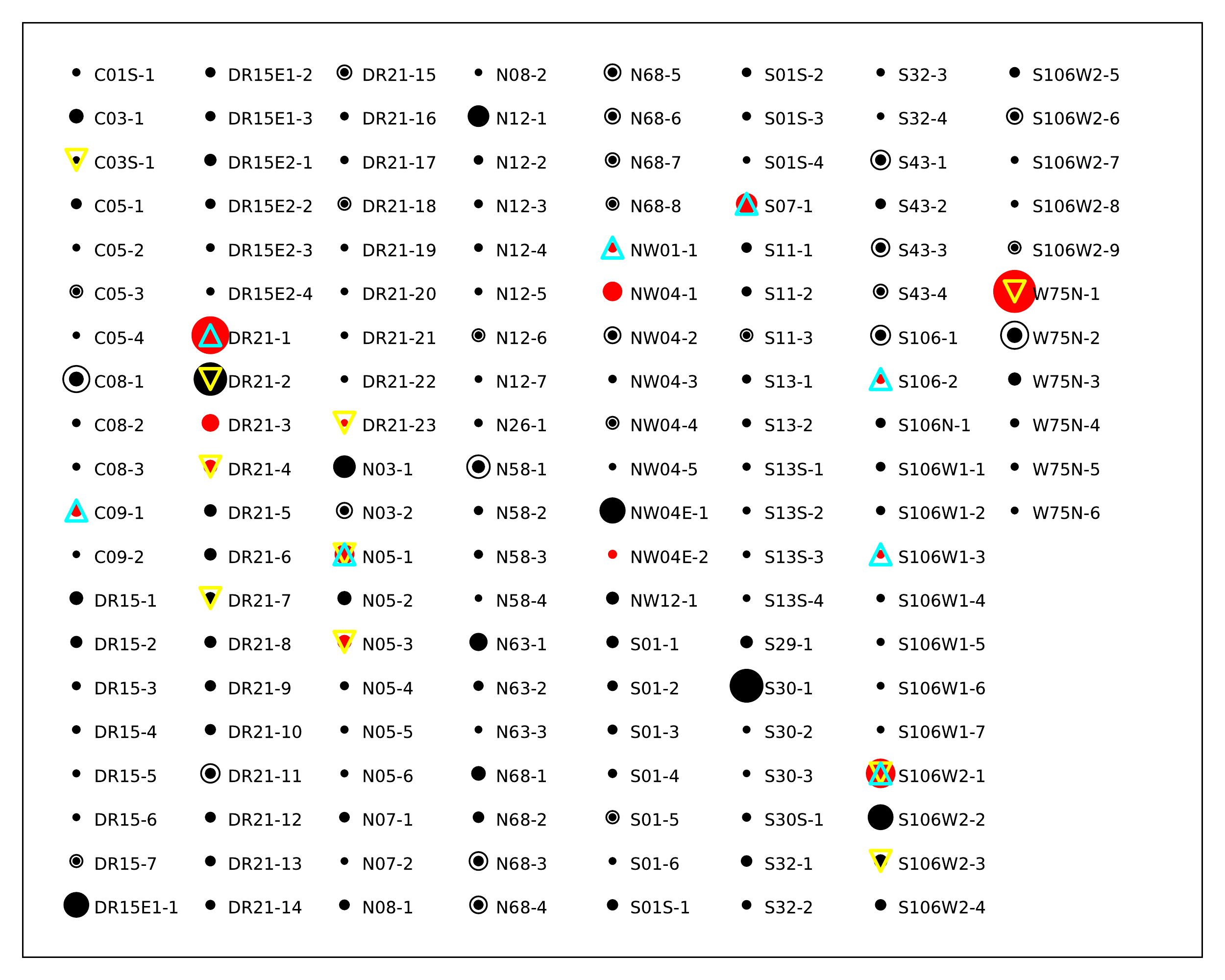}
\nfcaption{Assembly of the MDCs in Cygnus X in alphabetical order. Each (black/red) dot represents an (IR-quiet/IR-bright) MDC with its area proportional to the mass of the MDC. Dots will be attached with open circles/cyan triangles/yellow triangles if the corresponding MDCs are starless/associated with UC\hii\ regions/associated with class II methanol masers. \label{fig:MDC_table}}
\end{minipage}

\clearpage
\begin{minipage}{\textwidth}
\figurenum{B2}\addtocounter{figure}{1}
\centering\vspace{0.5in}
\includegraphics[width=6 in]{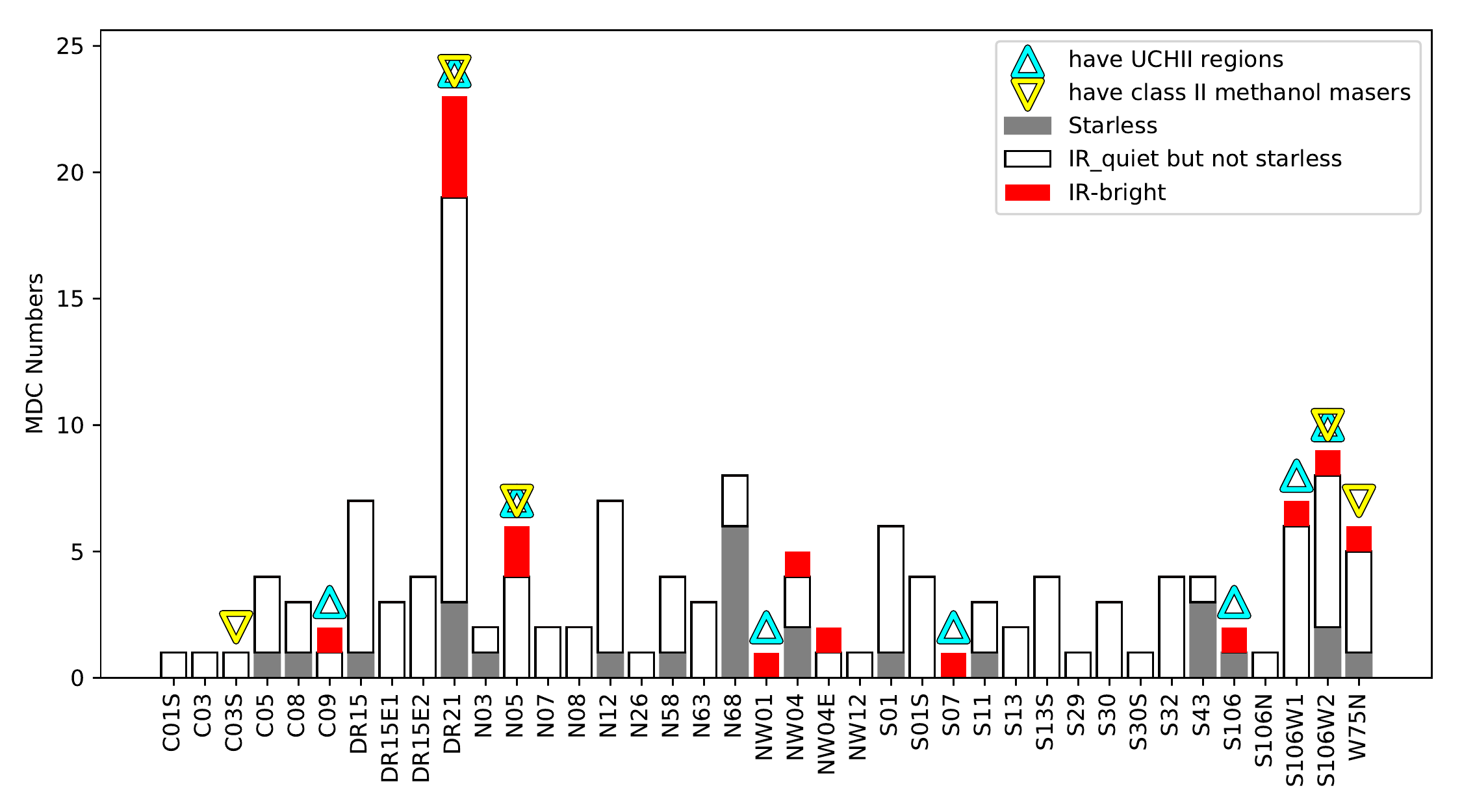}
\nfcaption{Stacked-bar plot of the numbers of MDCs in each field. MDCs in different infrared classes are described in the legend. Triangle markers will be added to the bar tops if the corresponding fields contain MDCs that are associated with HMSF signposts. \label{fig:number_in_fields}}
\end{minipage}

\renewcommand{\thetable}{B}
\startlongtable
\begin{deluxetable*}{c|c|cccccccc|ccc}
\tabletypesize{\footnotesize}
\tablecaption{Properties of MDCs in Cygnus X \label{tab:MDC_property}} 
\tablehead{
    \colhead{Name} & \colhead{Name in \motte\tablenotemark{a}} & \colhead{$\rm RA_{J2000}$} & \colhead{$\rm DEC_{J2000}$} & \colhead{$FWHM$\tablenotemark{b}} & \colhead{$T$\tablenotemark{c}} & \colhead{$M$\tablenotemark{c}} & \colhead{$N_{\rm{H_2}}$\tablenotemark{d}} & \colhead{$n_{\rm{H_2}}$\tablenotemark{e}} & \colhead{$L_{\rm FIR}$\tablenotemark{f}} & \colhead{Classification\tablenotemark{g}} \\
    \colhead{} & \colhead{} & \colhead{(h m s)} & \colhead{(\degree\ \arcmin\ \arcsec)} & \colhead{(pc)} & \colhead{(K)} & \colhead{(\msun)} & \colhead{($10^{22} \rm{cm^{-2}}$)} & \colhead{($10^5 \rm{cm^{-3}}$)} & \colhead{($L_{\odot}$)}  & \colhead{}
    }
\startdata
C01S-1&	&	20:29:25.87&	+40:36:05.2&	0.045 (0.096)&	12.1$\pm$0.4&	45.7$\pm$4.2&	118.7&	122.7&	14.1&	IR-quiet\\
\hline
C03-1&	&	20:30:28.90&	+41:15:55.4&	0.098 (0.130)&	19.7$\pm$3.6&	174$\pm$59.4&	94.1&	44.3&	977&	IR-quiet\\
\hline
C03S-1&	&	20:30:50.78&	+41:02:28.6&	0.051 (0.100)&	20.4$\pm$3.6&	36.3$\pm$11.5&	71.7&	64.6&	251&	IR-quiet\\
\hline
C05-1&	&	20:32:23.84&	+41:07:52.8&	0.221 (0.237)&	14.4$\pm$0.2&	89.0$\pm$3.7&	9.4&	2.0&	78.0&	IR-quiet\\
C05-2&	&	20:32:21.06&	+41:07:54.9&	0.037 (0.093)&	13.9$\pm$1.6&	40.6$\pm$11.8&	149.7&	184.2&	28.4&	IR-quiet\\
C05-3&	&	20:32:14.90&	+41:08:28.4&	0.172 (0.192)&	13.6$\pm$1.5&	38.6$\pm$10.7&	6.7&	1.8&	23.8&	starless\\
C05-4&	&	20:32:23.20&	+41:06:50.3&	0.066 (0.108)&	13.1$\pm$1.5&	35.8$\pm$11.0&	42.8&	30.0&	17.8&	IR-quiet\\
\hline
C08-1&	&	20:35:24.28&	+41:21:33.1&	0.260 (0.274)&	8.4$\pm$0.9&	180$\pm$64.6&	13.7&	2.4&	6.11&	starless\\
C08-2&	&	20:35:07.55&	+41:13:55.8&	0.162 (0.183)&	19.1$\pm$4.7&	49.6$\pm$23.2&	9.7&	2.8&	232&	IR-quiet\\
C08-3&	&	20:35:10.33&	+41:13:10.3&	0.195 (0.213)&	20.2$\pm$4.3&	42.9$\pm$16.6&	5.8&	1.4&	282&	IR-quiet\\
\hline
C09-1&	&	20:34:59.62&	+41:34:48.7&	0.110 (0.139)&	17.6$\pm$2.9&	125$\pm$41.5&	53.3&	22.3&	356&	IR-bright\\
C09-2&	&	20:34:50.22&	+41:33:58.8&	0.161 (0.183)&	10.8$\pm$0.7&	36.9$\pm$7.5&	7.3&	2.1&	5.79&	IR-quiet\\
\hline
DR15-1&	S34&	20:31:57.78&	+40:18:32.6&	0.060 (0.105)&	10.7$\pm$0.7&	153$\pm$28.4&	215.9&	164.5&	22.5&	IR-quiet\\
DR15-2&	S37&	20:32:21.92&	+40:20:14.5&	0.114 (0.143)&	15.6$\pm$0.2&	114$\pm$3.2&	44.7&	18.0&	159&	IR-quiet\\
DR15-3&	&	20:32:22.07&	+40:19:22.3&	0.145 (0.168)&	13.4$\pm$0.7&	55.7$\pm$7.1&	13.7&	4.3&	30.5&	IR-quiet\\
DR15-4&	S41&	20:32:33.53&	+40:16:56.6&	0.136 (0.161)&	25.3$\pm$2.1&	51.0$\pm$6.0&	14.2&	4.8&	1302&	IR-quiet\\
DR15-5&	&	20:31:37.13&	+40:19:41.0&	0.062 (0.106)&	12.3$\pm$0.9&	41.5$\pm$7.8&	55.8&	41.5&	13.8&	IR-quiet\\
DR15-6&	&	20:32:21.13&	+40:19:42.3&	0.153 (0.175)&	15.6$\pm$1.3&	40.7$\pm$7.4&	9.0&	2.7&	56.1&	IR-quiet\\
DR15-7&	&	20:32:30.34&	+40:14:07.5&	... (0.086)&	9.5$\pm$4.0&	40.3$\pm$26.0&	$\ge$31.3&	$\ge$17.8&	2.89&	starless\\
\hline
DR15E1-1&	&	20:34:42.34&	+39:44:56.5&	0.095 (0.128)&	13.8$\pm$0.7&	598$\pm$74.7&	342.5&	166.1&	401&	IR-quiet\\
DR15E1-2&	&	20:33:23.33&	+39:42:47.4&	0.098 (0.130)&	11.7$\pm$0.8&	77.3$\pm$15.6&	41.4&	19.4&	18.8&	IR-quiet\\
DR15E1-3&	&	20:33:28.25&	+39:41:14.1&	0.157 (0.179)&	11.4$\pm$0.4&	77.0$\pm$8.2&	16.1&	4.7&	16.2&	IR-quiet\\
\hline
DR15E2-1&	&	20:37:42.16&	+39:51:08.2&	0.060 (0.105)&	19.4$\pm$2.8&	117$\pm$30.8&	166.5&	127.1&	604&	IR-quiet\\
DR15E2-2&	&	20:37:49.56&	+39:49:54.2&	0.066 (0.108)&	13.0$\pm$1.6&	79.0$\pm$26.3&	94.3&	66.1&	36.0&	IR-quiet\\
DR15E2-3&	&	20:36:27.86&	+39:42:39.5&	... (0.086)&	10.5$\pm$0.9&	51.9$\pm$15.1&	$\ge$40.3&	$\ge$22.9&	6.88&	IR-quiet\\
DR15E2-4&	&	20:37:15.19&	+39:49:07.2&	0.143 (0.166)&	25.2$\pm$0.9&	46.5$\pm$2.6&	11.8&	3.8&	1156&	IR-quiet\\
\hline
DR21-1&	N46&	20:39:01.03&	+42:19:33.8&	0.195 (0.213)&	20.6$\pm$1.2&	1355$\pm$130&	183.5&	43.3&	10095&	IR-bright\\
DR21-2&	N44&	20:39:00.89&	+42:22:48.8&	0.134 (0.159)&	21.0$\pm$2.8&	1048$\pm$225&	301.7&	103.7&	8600&	IR-quiet\\
DR21-3&	N48&	20:39:01.39&	+42:22:04.2&	0.177 (0.197)&	22.6$\pm$2.8&	261$\pm$48.0&	42.8&	11.1&	3353&	IR-bright\\
DR21-4&	N51&	20:39:02.13&	+42:25:01.0&	0.078 (0.116)&	19.7$\pm$0.7&	153$\pm$9.4&	131.3&	77.8&	853&	IR-bright\\
DR21-5&	&	20:38:54.06&	+42:19:12.7&	0.226 (0.242)&	17.5$\pm$1.6&	125$\pm$21.0&	12.6&	2.6&	351&	IR-quiet\\
DR21-6&	&	20:39:01.11&	+42:18:34.7&	... (0.086)&	15.4$\pm$2.1&	121$\pm$27.3&	$\ge$94.3&	$\ge$53.6&	153&	IR-quiet\\
DR21-7&	N53&	20:39:03.03&	+42:25:50.3&	0.053 (0.100)&	17.0$\pm$0.9&	115$\pm$11.5&	213.1&	186.2&	267&	IR-quiet\\
DR21-8&	N40&	20:38:59.62&	+42:23:47.2&	0.159 (0.181)&	19.8$\pm$3.6&	114$\pm$36.1&	23.2&	6.7&	672&	IR-quiet\\
DR21-9&	N56&	20:39:16.84&	+42:16:10.0&	0.053 (0.101)&	14.6$\pm$1.4&	94.9$\pm$19.0&	175.2&	152.6&	89.7&	IR-quiet\\
DR21-10&	&	20:39:19.41&	+42:15:59.1&	0.060 (0.104)&	10.3$\pm$0.6&	94.1$\pm$16.3&	136.3&	105.1&	10.6&	IR-quiet\\
DR21-11&	&	20:39:06.25&	+42:18:16.4&	... (0.086)&	14.0$\pm$3.7&	89.9$\pm$42.1&	$\ge$69.8&	$\ge$39.7&	66.3&	starless\\
DR21-12&	&	20:38:51.65&	+42:27:14.5&	... (0.086)&	13.0$\pm$0.9&	88.5$\pm$11.7&	$\ge$68.7&	$\ge$39.1&	41.7&	IR-quiet\\
DR21-13&	N38&	20:38:58.95&	+42:22:23.7&	0.054 (0.101)&	20.7$\pm$10.3&	85.3$\pm$68.3&	152.7&	131.0&	643&	IR-quiet\\
DR21-14&	&	20:39:02.01&	+42:26:55.6&	... (0.086)&	12.6$\pm$0.3&	72.6$\pm$4.1&	$\ge$56.4&	$\ge$32.1&	27.6&	IR-quiet\\
DR21-15&	&	20:38:59.82&	+42:27:32.5&	0.180 (0.200)&	14.2$\pm$0.7&	51.1$\pm$6.3&	8.1&	2.1&	40.1&	starless\\
DR21-16&	&	20:39:00.76&	+42:18:12.7&	0.159 (0.180)&	15.9$\pm$3.2&	50.6$\pm$21.0&	10.3&	3.0&	79.9&	IR-quiet\\
DR21-17&	&	20:38:58.42&	+42:18:51.5&	0.206 (0.223)&	20.6$\pm$2.1&	47.6$\pm$8.6&	5.8&	1.3&	348&	IR-quiet\\
DR21-18&	&	20:39:25.49&	+42:15:59.6&	... (0.086)&	10.0$\pm$0.9&	40.0$\pm$2.3&	$\ge$31.1&	$\ge$17.7&	1.99&	starless\\
DR21-19&	&	20:39:02.12&	+42:16:53.0&	0.169 (0.189)&	16.2$\pm$1.3&	39.6$\pm$6.2&	7.1&	1.9&	68.6&	IR-quiet\\
DR21-20&	&	20:39:02.63&	+42:18:41.3&	0.119 (0.146)&	14.6$\pm$1.9&	37.9$\pm$11.6&	13.8&	5.4&	35.9&	IR-quiet\\
DR21-21&	N52&	20:39:02.86&	+42:26:23.3&	0.175 (0.195)&	16.2$\pm$1.7&	37.2$\pm$9.0&	6.3&	1.6&	64.8&	IR-quiet\\
DR21-22&	&	20:39:07.30&	+42:15:33.9&	0.202 (0.220)&	17.8$\pm$2.1&	36.5$\pm$8.2&	4.6&	1.0&	111&	IR-quiet\\
DR21-23&	N43&	20:39:00.35&	+42:24:35.7&	0.106 (0.136)&	34.7$\pm$7.2&	35.1$\pm$8.5&	16.1&	7.0&	5865&	IR-bright\\
\hline
N03-1&	N03&	20:35:34.23&	+42:20:10.0&	0.113 (0.142)&	14.4$\pm$1.4&	455$\pm$93.9&	184.0&	75.0&	397&	IR-quiet\\
N03-2&	&	20:35:55.55&	+42:11:19.9&	... (0.086)&	11.7$\pm$0.4&	61.2$\pm$3.7&	$\ge$47.5&	$\ge$27.0&	15.0&	starless\\
\hline
N05-1&	N10&	20:36:52.25&	+41:36:22.9&	0.088 (0.122)&	19.2$\pm$2.9&	338$\pm$97.6&	226.6&	118.9&	1619&	IR-bright\\
N05-2&	N06&	20:36:08.21&	+41:39:56.4&	0.066 (0.108)&	10.4$\pm$2.1&	162$\pm$106&	189.8&	131.6&	19.4&	IR-quiet\\
N05-3&	N14&	20:37:01.02&	+41:34:55.9&	0.069 (0.110)&	20.6$\pm$1.4&	162$\pm$19.5&	173.6&	115.3&	1186&	IR-bright\\
N05-4&	&	20:37:11.61&	+41:33:36.2&	0.166 (0.186)&	13.1$\pm$0.7&	56.1$\pm$7.5&	10.6&	2.9&	26.8&	IR-quiet\\
N05-5&	&	20:37:25.41&	+41:35:37.6&	0.117 (0.145)&	13.6$\pm$0.7&	44.7$\pm$5.6&	16.9&	6.6&	27.5&	IR-quiet\\
N05-6&	&	20:36:03.55&	+41:39:45.2&	0.097 (0.130)&	15.3$\pm$1.0&	41.8$\pm$6.3&	22.7&	10.7&	51.0&	IR-quiet\\
\hline
N07-1&	&	20:36:39.61&	+42:51:12.6&	0.195 (0.213)&	11.9$\pm$1.1&	83.9$\pm$20.1&	11.4&	2.7&	23.5&	IR-quiet\\
N07-2&	&	20:35:47.47&	+42:52:58.4&	0.073 (0.112)&	11.0$\pm$0.8&	37.6$\pm$8.3&	36.5&	23.1&	6.49&	IR-quiet\\
\hline
N08-1&	&	20:36:38.36&	+42:29:12.0&	0.102 (0.133)&	14.2$\pm$0.7&	85.7$\pm$9.7&	42.1&	18.9&	67.2&	IR-quiet\\
N08-2&	&	20:36:19.87&	+42:37:27.4&	0.058 (0.103)&	10.8$\pm$0.7&	37.0$\pm$7.3&	56.6&	44.8&	5.71&	IR-quiet\\
\hline
N12-1&	N12&	20:36:57.48&	+42:11:31.2&	0.062 (0.105)&	15.9$\pm$1.5&	415$\pm$78.0&	563.8&	421.1&	659&	IR-quiet\\
N12-2&	&	20:36:56.83&	+42:13:22.5&	0.063 (0.106)&	14.3$\pm$0.8&	64.3$\pm$8.6&	82.9&	60.3&	53.6&	IR-quiet\\
N12-3&	&	20:37:16.89&	+42:16:31.3&	0.159 (0.181)&	18.3$\pm$1.5&	53.5$\pm$7.8&	10.9&	3.2&	194&	IR-quiet\\
N12-4&	&	20:37:22.89&	+42:16:37.9&	0.088 (0.122)&	10.8$\pm$0.3&	50.9$\pm$4.6&	34.3&	18.0&	7.82&	IR-quiet\\
N12-5&	&	20:37:30.10&	+42:13:59.3&	0.076 (0.114)&	18.5$\pm$1.2&	40.7$\pm$5.0&	36.3&	22.0&	156&	IR-quiet\\
N12-6&	&	20:37:31.27&	+42:13:16.6&	0.163 (0.184)&	13.7$\pm$0.7&	39.1$\pm$4.9&	7.6&	2.1&	24.7&	starless\\
N12-7&	&	20:37:05.74&	+42:11:50.9&	0.181 (0.201)&	12.9$\pm$1.0&	37.8$\pm$7.2&	5.9&	1.5&	16.5&	IR-quiet\\
\hline
N26-1&	&	20:38:21.03&	+42:11:31.1&	0.089 (0.123)&	14.5$\pm$0.8&	49.4$\pm$5.9&	32.3&	16.8&	44.4&	IR-quiet\\
\hline
N58-1&	&	20:39:54.24&	+41:23:26.7&	... (0.086)&	12.1$\pm$1.5&	134$\pm$26.3&	$\ge$104.2&	$\ge$59.2&	40.1&	starless\\
N58-2&	&	20:39:31.02&	+41:20:04.7&	0.193 (0.211)&	17.6$\pm$0.9&	59.9$\pm$5.7&	8.3&	2.0&	171&	IR-quiet\\
N58-3&	N60&	20:39:36.15&	+41:19:37.5&	0.162 (0.183)&	17.2$\pm$0.9&	57.1$\pm$6.0&	11.2&	3.2&	141&	IR-quiet\\
N58-4&	&	20:39:03.55&	+41:17:48.2&	0.042 (0.095)&	11.5$\pm$0.4&	35.9$\pm$3.8&	103.1&	112.0&	7.96&	IR-quiet\\
\hline
N63-1&	N63&	20:40:05.43&	+41:32:13.0&	0.034 (0.092)&	14.4$\pm$0.9&	282$\pm$40.6&	1263.3&	1715.0&	247&	IR-quiet\\
N63-2&	&	20:40:03.92&	+41:27:54.3&	0.121 (0.148)&	17.1$\pm$2.7&	77.2$\pm$24.0&	27.3&	10.4&	184&	IR-quiet\\
N63-3&	&	20:40:34.26&	+41:38:46.3&	0.129 (0.154)&	13.1$\pm$1.2&	39.0$\pm$8.4&	12.2&	4.3&	18.8&	IR-quiet\\
\hline
N68-1&	N68&	20:40:33.56&	+41:59:03.0&	0.056 (0.102)&	14.1$\pm$1.0&	175$\pm$27.4&	284.2&	232.1&	131&	IR-quiet\\
N68-2&	N65&	20:40:28.43&	+41:57:11.2&	0.088 (0.122)&	17.0$\pm$1.4&	102$\pm$16.6&	68.2&	35.8&	237&	IR-quiet\\
N68-3&	&	20:40:45.49&	+41:57:50.6&	... (0.086)&	11.7$\pm$0.5&	83.4$\pm$7.3&	$\ge$64.7&	$\ge$36.8&	20.8&	starless\\
N68-4&	&	20:40:34.94&	+41:51:18.1&	... (0.086)&	12.4$\pm$2.1&	73.9$\pm$24.5&	$\ge$57.4&	$\ge$32.6&	26.4&	starless\\
N68-5&	&	20:40:27.51&	+41:56:54.3&	0.073 (0.113)&	12.1$\pm$0.7&	67.3$\pm$10.0&	64.4&	40.4&	19.9&	starless\\
N68-6&	&	20:40:29.11&	+41:49:12.9&	... (0.086)&	15.1$\pm$0.4&	58.6$\pm$2.2&	$\ge$45.5&	$\ge$25.8&	66.2&	starless\\
N68-7&	&	20:40:31.24&	+41:45:35.3&	... (0.086)&	11.4$\pm$1.8&	48.2$\pm$16.1&	$\ge$37.5&	$\ge$21.3&	10.3&	starless\\
N68-8&	&	20:40:45.36&	+42:01:41.1&	... (0.086)&	13.2$\pm$1.5&	39.4$\pm$8.1&	$\ge$30.6&	$\ge$17.4&	20.3&	starless\\
\hline
NW01-1&	NW01&	20:19:38.95&	+40:56:39.3&	0.113 (0.142)&	29.8$\pm$5.6&	72.3$\pm$18.3&	28.9&	11.7&	4887&	IR-bright\\
\hline
NW04-1&	NW05&	20:20:30.57&	+41:21:26.4&	0.099 (0.131)&	17.0$\pm$0.3&	330$\pm$12.2&	173.8&	80.8&	763&	IR-bright\\
NW04-2&	&	20:20:31.31&	+41:23:25.7&	0.154 (0.176)&	9.1$\pm$0.2&	67.5$\pm$6.4&	14.6&	4.4&	3.75&	starless\\
NW04-3&	&	20:19:39.20&	+41:11:01.8&	0.046 (0.097)&	14.7$\pm$0.5&	48.5$\pm$4.3&	118.5&	118.8&	46.3&	IR-quiet\\
NW04-4&	&	20:20:52.51&	+41:13:47.6&	... (0.086)&	9.2$\pm$0.5&	39.2$\pm$6.1&	$\ge$30.4&	$\ge$17.3&	2.30&	starless\\
NW04-5&	&	20:20:31.85&	+41:23:52.9&	0.053 (0.100)&	14.3$\pm$0.7&	36.1$\pm$4.0&	67.4&	59.1&	29.6&	IR-quiet\\
\hline
NW04E-1&	&	20:23:43.43&	+41:16:59.1&	0.126 (0.152)&	11.1$\pm$0.2&	619$\pm$35.2&	200.8&	73.3&	113&	IR-quiet\\
NW04E-2&	&	20:23:23.64&	+41:17:39.7&	0.074 (0.113)&	22.7$\pm$0.5&	56.8$\pm$2.2&	53.3&	33.0&	750&	IR-bright\\
\hline
NW12-1&	NW14&	20:24:31.63&	+42:04:21.0&	0.073 (0.113)&	19.6$\pm$0.9&	133$\pm$10.1&	127.6&	80.3&	719&	IR-quiet\\
\hline
S01-1&	&	20:16:48.57&	+39:22:16.2&	0.141 (0.165)&	10.1$\pm$0.7&	120$\pm$27.0&	31.1&	10.1&	12.2&	IR-quiet\\
S01-2&	&	20:17:55.73&	+39:20:38.4&	0.133 (0.158)&	15.1$\pm$1.3&	82.6$\pm$18.0&	24.1&	8.3&	93.0&	IR-quiet\\
S01-3&	&	20:16:58.96&	+39:21:05.7&	0.056 (0.102)&	17.6$\pm$1.0&	72.0$\pm$8.0&	119.1&	98.1&	205&	IR-quiet\\
S01-4&	&	20:17:45.79&	+39:20:40.4&	0.095 (0.128)&	14.0$\pm$0.7&	60.3$\pm$7.9&	34.4&	16.7&	43.1&	IR-quiet\\
S01-5&	&	20:17:37.19&	+39:21:38.9&	0.115 (0.143)&	12.1$\pm$0.8&	41.1$\pm$8.0&	16.2&	6.5&	12.3&	starless\\
S01-6&	&	20:16:43.64&	+39:23:19.2&	0.055 (0.102)&	14.7$\pm$1.2&	39.5$\pm$8.7&	68.0&	57.2&	38.8&	IR-quiet\\
\hline
S01S-1&	&	20:16:45.52&	+39:06:27.7&	0.059 (0.104)&	13.0$\pm$0.5&	96.7$\pm$10.9&	142.9&	111.3&	44.8&	IR-quiet\\
S01S-2&	&	20:17:36.97&	+39:03:35.0&	0.072 (0.112)&	18.7$\pm$0.4&	64.5$\pm$3.0&	64.5&	41.3&	264&	IR-quiet\\
S01S-3&	&	20:17:06.58&	+38:59:17.4&	0.277 (0.290)&	17.9$\pm$1.4&	54.9$\pm$9.7&	3.7&	0.6&	174&	IR-quiet\\
S01S-4&	&	20:16:51.91&	+39:00:13.5&	0.114 (0.142)&	11.3$\pm$0.3&	36.9$\pm$3.3&	14.7&	5.9&	7.32&	IR-quiet\\
\hline
S07-1&	S08&	20:20:39.10&	+39:37:51.1&	0.137 (0.161)&	21.9$\pm$4.1&	395$\pm$126&	108.7&	36.6&	4239&	IR-bright\\
\hline
S11-1&	&	20:22:20.15&	+39:58:18.1&	0.118 (0.146)&	14.6$\pm$1.7&	81.4$\pm$21.2&	29.9&	11.6&	75.2&	IR-quiet\\
S11-2&	&	20:21:53.56&	+39:59:33.4&	0.081 (0.117)&	11.9$\pm$0.8&	74.0$\pm$13.6&	58.8&	33.6&	20.0&	IR-quiet\\
S11-3&	&	20:21:56.83&	+39:59:38.3&	0.129 (0.155)&	11.7$\pm$2.3&	37.6$\pm$16.1&	11.6&	4.1&	9.46&	starless\\
\hline
S13-1&	&	20:26:32.31&	+39:57:20.4&	0.119 (0.146)&	16.9$\pm$0.4&	58.7$\pm$3.1&	21.5&	8.4&	130&	IR-quiet\\
S13-2&	&	20:26:34.82&	+39:57:26.5&	0.101 (0.132)&	13.9$\pm$0.4&	55.1$\pm$3.7&	28.1&	12.9&	38.1&	IR-quiet\\
\hline
S13S-1&	&	20:26:58.32&	+39:10:19.5&	0.270 (0.283)&	11.0$\pm$0.3&	44.8$\pm$4.8&	3.2&	0.5&	7.45&	IR-quiet\\
S13S-2&	&	20:25:49.02&	+39:30:14.7&	0.077 (0.115)&	16.8$\pm$1.7&	42.8$\pm$10.1&	37.7&	22.6&	92.9&	IR-quiet\\
S13S-3&	&	20:26:12.55&	+39:16:42.9&	0.042 (0.095)&	11.7$\pm$0.7&	38.7$\pm$7.8&	115.5&	127.9&	9.49&	IR-quiet\\
S13S-4&	&	20:25:47.68&	+39:29:54.2&	0.043 (0.096)&	12.3$\pm$0.2&	38.4$\pm$2.1&	106.0&	112.8&	12.8&	IR-quiet\\
\hline
S29-1&	S29&	20:29:58.20&	+40:15:58.7&	0.091 (0.125)&	10.7$\pm$0.8&	126$\pm$29.7&	77.8&	39.2&	18.2&	IR-quiet\\
\hline
S30-1&	&	20:31:11.73&	+40:03:12.8&	0.169 (0.189)&	13.7$\pm$0.3&	1068$\pm$60.4&	192.5&	52.4&	686&	IR-quiet\\
S30-2&	&	20:31:14.59&	+40:03:05.5&	0.112 (0.141)&	13.8$\pm$1.3&	40.4$\pm$9.8&	16.6&	6.8&	27.0&	IR-quiet\\
S30-3&	S30&	20:31:12.80&	+40:03:22.6&	0.057 (0.103)&	14.0$\pm$2.0&	36.3$\pm$11.6&	58.5&	47.6&	26.7&	IR-quiet\\
\hline
S30S-1&	&	20:30:58.19&	+39:42:26.2&	0.039 (0.094)&	13.5$\pm$0.7&	57.7$\pm$8.1&	191.6&	223.8&	33.6&	IR-quiet\\
\hline
S32-1&	S32&	20:31:20.94&	+38:57:17.7&	0.132 (0.157)&	15.7$\pm$1.0&	101$\pm$12.9&	29.9&	10.4&	146&	IR-quiet\\
S32-2&	&	20:32:10.71&	+38:56:35.5&	0.101 (0.132)&	11.0$\pm$0.6&	66.3$\pm$11.7&	33.7&	15.4&	11.1&	IR-quiet\\
S32-3&	&	20:31:59.21&	+38:58:38.5&	0.063 (0.106)&	15.7$\pm$0.8&	47.6$\pm$4.8&	61.7&	45.1&	67.5&	IR-quiet\\
S32-4&	&	20:31:47.25&	+39:00:48.6&	0.059 (0.104)&	13.0$\pm$1.0&	36.7$\pm$7.9&	54.4&	42.5&	17.1&	IR-quiet\\
\hline
S43-1&	&	20:32:42.44&	+38:43:54.6&	... (0.086)&	8.8$\pm$0.9&	93.6$\pm$20.3&	$\ge$72.6&	$\ge$41.3&	4.21&	starless\\
S43-2&	S43&	20:32:41.01&	+38:46:30.1&	0.132 (0.157)&	13.8$\pm$0.5&	84.5$\pm$7.2&	25.1&	8.8&	56.7&	IR-quiet\\
S43-3&	&	20:32:40.31&	+38:46:06.8&	0.086 (0.121)&	12.6$\pm$1.3&	80.7$\pm$21.2&	56.7&	30.5&	30.8&	starless\\
S43-4&	&	20:32:38.17&	+38:45:33.8&	... (0.086)&	11.0$\pm$0.4&	49.1$\pm$4.9&	$\ge$38.1&	$\ge$21.7&	8.43&	starless\\
\hline
S106-1&	&	20:27:32.23&	+37:22:02.8&	0.150 (0.173)&	13.4$\pm$1.5&	95.9$\pm$26.9&	21.9&	6.7&	54.5&	starless\\
S106-2&	S20&	20:27:27.57&	+37:22:51.6&	0.196 (0.214)&	33.1$\pm$12.1&	70.5$\pm$32.4&	9.4&	2.2&	8971&	IR-bright\\
\hline
S106N-1&	&	20:26:14.16&	+38:02:26.5&	0.045 (0.097)&	15.8$\pm$1.4&	75.4$\pm$15.2&	191.6&	195.8&	114&	IR-quiet\\
\hline
S106W1-1&	&	20:23:23.49&	+37:35:37.5&	0.077 (0.115)&	16.2$\pm$1.0&	66.8$\pm$8.6&	58.6&	35.2&	116&	IR-quiet\\
S106W1-2&	&	20:24:06.19&	+37:35:51.4&	0.297 (0.309)&	12.4$\pm$0.5&	60.4$\pm$6.8&	3.5&	0.5&	20.8&	IR-quiet\\
S106W1-3&	&	20:23:54.90&	+37:38:10.7&	0.157 (0.198)&	24.0$\pm$0.8&	51.9$\pm$2.8&	10.9&	3.2&	958&	IR-bright\\
S106W1-4&	&	20:23:58.34&	+37:37:31.7&	0.173 (0.193)&	16.0$\pm$0.4&	49.1$\pm$3.0&	8.5&	2.2&	79.1&	IR-quiet\\
S106W1-5&	&	20:23:57.46&	+37:39:50.6&	0.141 (0.165)&	18.4$\pm$1.0&	43.7$\pm$3.9&	11.3&	3.7&	162&	IR-quiet\\
S106W1-6&	&	20:23:27.80&	+37:34:25.2&	0.172 (0.192)&	12.6$\pm$1.0&	39.6$\pm$9.1&	6.9&	1.9&	14.9&	IR-quiet\\
S106W1-7&	&	20:24:02.66&	+37:36:37.8&	0.087 (0.122)&	17.0$\pm$1.2&	38.5$\pm$5.8&	26.0&	13.7&	88.0&	IR-quiet\\
\hline
S106W2-1&	&	20:21:44.31&	+37:26:41.0&	0.085 (0.121)&	18.0$\pm$0.5&	805$\pm$47.9&	567.6&	305.5&	2684&	IR-bright\\
S106W2-2&	&	20:21:39.24&	+37:25:12.0&	0.115 (0.143)&	14.8$\pm$1.1&	602$\pm$115&	235.5&	94.3&	600&	IR-quiet\\
S106W2-3&	&	20:21:41.11&	+37:25:33.8&	0.103 (0.134)&	23.4$\pm$3.6&	145$\pm$38.6&	70.6&	31.6&	2297&	IR-quiet\\
S106W2-4&	&	20:21:47.65&	+37:30:20.9&	0.103 (0.134)&	16.9$\pm$0.7&	99.1$\pm$8.8&	47.9&	21.3&	223&	IR-quiet\\
S106W2-5&	&	20:21:55.47&	+37:29:57.3&	0.128 (0.154)&	12.2$\pm$0.6&	84.1$\pm$13.2&	26.5&	9.5&	26.7&	IR-quiet\\
S106W2-6&	&	20:21:51.26&	+37:26:03.8&	0.102 (0.133)&	11.7$\pm$1.1&	63.0$\pm$18.2&	31.5&	14.3&	15.8&	starless\\
S106W2-7&	&	20:21:51.06&	+37:30:12.8&	0.195 (0.213)&	21.9$\pm$2.1&	40.2$\pm$7.2&	5.4&	1.3&	431&	IR-quiet\\
S106W2-8&	&	20:21:29.94&	+37:22:35.9&	0.097 (0.129)&	12.7$\pm$0.6&	38.5$\pm$5.8&	21.1&	10.0&	15.8&	IR-quiet\\
S106W2-9&	&	20:21:19.59&	+37:21:39.1&	0.280 (0.293)&	13.2$\pm$0.4&	38.4$\pm$3.0&	2.5&	0.4&	19.4&	starless\\
\hline
W75N-1&	N30&	20:38:36.44&	+42:37:34.1&	0.145 (0.168)&	22.2$\pm$2.0&	1762$\pm$261&	434.6&	138.3&	20413&	IR-bright\\
W75N-2&	N24&	20:38:05.93&	+42:39:54.2&	0.174 (0.194)&	9.9$\pm$0.6&	194$\pm$38.9&	32.9&	8.7&	17.9&	starless\\
W75N-3&	&	20:38:33.04&	+42:39:47.2&	0.194 (0.212)&	13.0$\pm$0.5&	138$\pm$13.6&	18.9&	4.5&	65.0&	IR-quiet\\
W75N-4&	&	20:38:10.16&	+42:38:07.1&	0.116 (0.144)&	13.7$\pm$0.9&	59.4$\pm$10.3&	22.7&	9.0&	37.4&	IR-quiet\\
W75N-5&	&	20:38:05.17&	+42:33:19.8&	0.060 (0.104)&	11.9$\pm$0.6&	44.0$\pm$7.3&	63.0&	48.3&	12.3&	IR-quiet\\
W75N-6&	&	20:37:55.06&	+42:40:54.8&	0.108 (0.138)&	14.6$\pm$1.1&	39.3$\pm$6.9&	17.4&	7.4&	36.7&	IR-quiet\\

\enddata
\tablenotetext{a}{ Names of the MDCs in \motte\ that are associated with our MDCs with a positional tolerance of 10\arcsec\ (see Sect. \ref{subsec:M07_compare}).}
\tablenotetext{b}{ Deconvolved (convolved) FWHM sizes in the \emph{Herschel} 160 \um\ band. Ellipses will show where convolved sizes are smaller than the \emph{Herschel} beam.}
\tablenotetext{c}{ Dust temperatures and masses were obtained by SED fitting (Sect. \ref{subsec:extraction}). The errors in come from the flux uncertainties and the fitting errors.}
\tablenotetext{d}{ Estimated as $N_{\rm{H_2}}=\frac{4}{\pi}\frac{M}{\mu_{\rm H_2}m_{\rm H}FWHM_{\rm dec}^2}$.}
\tablenotetext{e}{ Estimated as $n_{\rm{H_2}}=\frac{6}{\pi}\frac{M}{\mu_{\rm H_2}m_{\rm H}FWHM_{\rm dec}^3}$.}
\tablenotetext{f}{ Estimated as $L_{\rm FIR} = 4\pi D^2\int_0^{+\infty} \frac{\kappa_{\nu}B_{\nu}(T)M}{D^2}\rm{d}\nu$. }
\tablenotetext{g}{ Infrared classifications of the MDCs (see Sect. \ref{subsec:IR_class}). Note that starless MDCs also belong to IR-quiet MDCs.}
\end{deluxetable*}

\clearpage
\section{Far-infrared to millimeter fluxes of the MDCs and SED fitting}\label{sec_A:MDC_flux}

Fluxes of the MDCs in the bands of \emph{Herschel} (70, 160, 250, 350, and 500 \um), JCMT (450 and 850 \um), and the IRAM 30 m telescope (1.2 mm) were extracted by \emph{getsources} on the single-frequency maps of the fields (see Sect. \ref{subsec:extraction}).  Large-scale emissions in the maps were firstly removed before flux extraction by setting the ``maxsize'' parameter in \emph{getsources} to 35\arcsec\ (0.23 pc). The results are shown in Table \ref{tab:MDC_flux}.

We used the $\ge160$ \um\ fluxes and a modified blackbody model to do the SED fitting (see Sect. \ref{subsec:extraction}). During the fitting, we found that some fluxes derivate largely from the model curve (see Figure \ref{fig:SED_example}). These poorly-measured fluxes are due to source blending in low-resolution images (e.g., S30-2@500 \um, DR15-2@350 \um, see the corresponding panels in Figure \ref{fig:MDC}) and poor image quality (e.g., C09-2@450 \um, N05-5@450 \um). Given that the errors provided by \emph{getsources} did not reflect the poor detections, we have manually removed these points from the SED fitting after careful by-eye inspection. The removed fluxes are enclosed in brackets in Table \ref{tab:MDC_flux}.

\clearpage
\begin{minipage}{\textwidth}
\figurenum{C}\addtocounter{figure}{1}
\centering\vspace{0.5in}
\includegraphics[width=6 in]{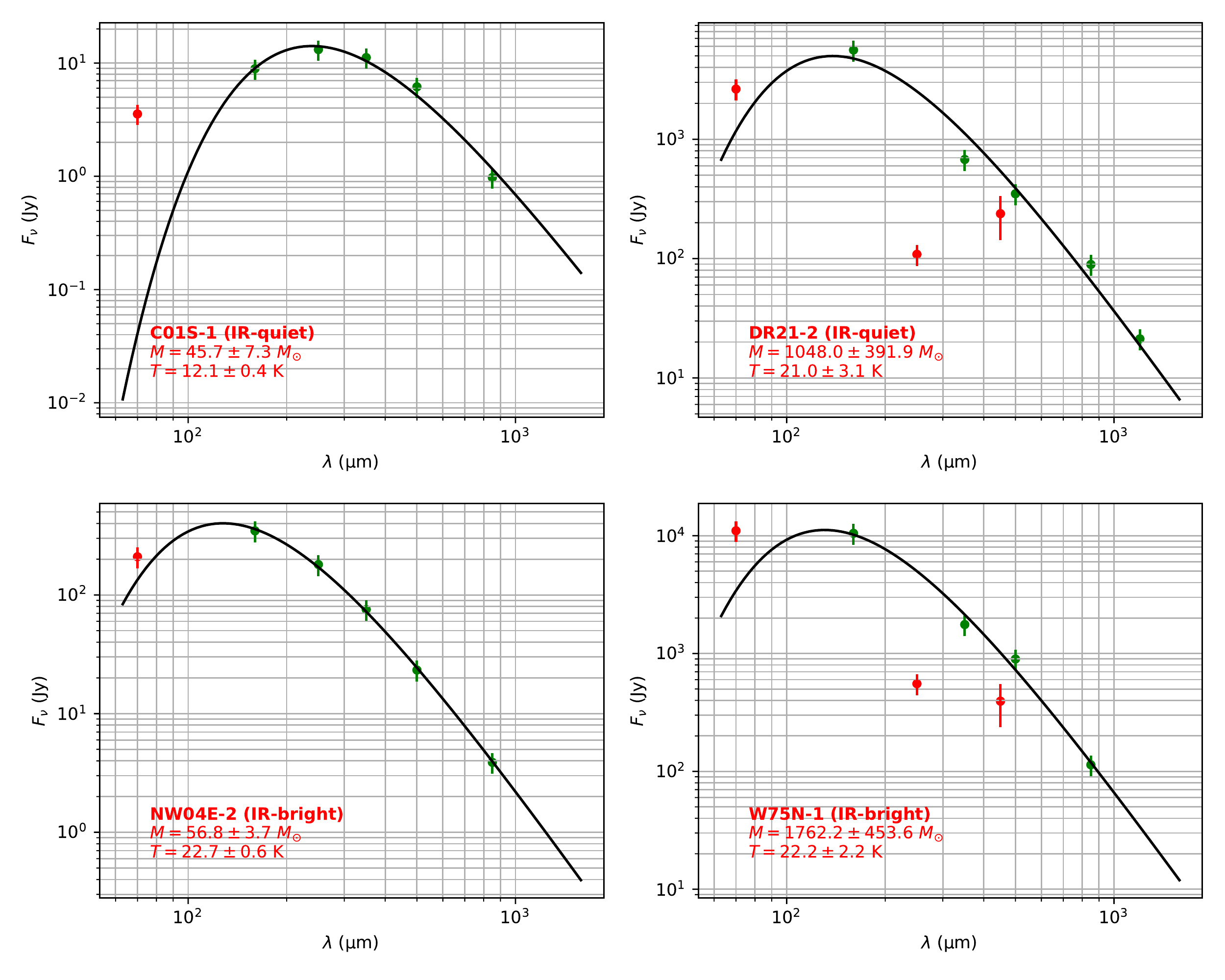}
\nfcaption{SED fittings of four example MDCs (two IR-bright and two IR-quiet). Data in green were used for the fitting while data in red were not due to their large deviations from the model curves (see Sect. \ref{subsec:extraction}). \label{fig:SED_example}}
\end{minipage}

\clearpage
\renewcommand{\thetable}{C}
\startlongtable
\begin{deluxetable*}{c|cccccccc}
\tabletypesize{\footnotesize}
\tablecaption{Fluxes of the MDCs in Cygnus X Obtained by \emph{getsources} (in Jy) \label{tab:MDC_flux}}  
\tablehead{ 
\colhead{Name}& \colhead{$F_{\nu}(70\ \rm \mu m)$}& \colhead{$F_{\nu}(160\ \rm \mu m)$}& \colhead{$F_{\nu}(250\ \rm \mu m)$}& \colhead{$F_{\nu}(350\ \rm \mu m)$}& \colhead{$F_{\nu}(450\ \rm \mu m)$}& \colhead{$F_{\nu}(500\ \rm \mu m)$}& \colhead{$F_{\nu}(850\ \rm \mu m)$}& \colhead{$F_{\nu}(1.2\ \rm mm)$} }
\startdata
C01S-1&	($3.55\pm0.03$)&	$8.91\pm0.28$&	$13.2\pm0.63$&	$11.2\pm0.64$&	...&	$6.17\pm0.78$&	$0.975\pm0.04$&	...\\
\hline
C03-1&	($267\pm1.15$)&	$420\pm0.83$&	$560\pm0.32$&	$218\pm0.23$&	...&	$63.6\pm0.49$&	$5.73\pm0.05$&	...\\
\hline
C03S-1&	($68.7\pm0.07$)&	$120\pm0.24$&	$88.6\pm0.25$&	$48.9\pm0.32$&	...&	$18.7\pm0.31$&	$1.20\pm0.04$&	...\\
\hline
C05-1&	($6.80\pm0.91$)&	$57.6\pm4.10$&	$59.8\pm2.87$&	$33.1\pm0.43$&	($5.90\pm0.13$)&	$16.7\pm0.89$&	$3.00\pm0.11$&	...\\
C05-2&	($14.2\pm0.25$)&	$19.1\pm1.86$&	$22.2\pm2.18$&	$15.1\pm0.83$&	($0.962\pm0.06$)&	$11.9\pm0.99$&	$0.763\pm0.07$&	...\\
C05-3&	...&	$14.9\pm3.84$&	$20.8\pm3.59$&	$15.2\pm2.27$&	($2.69\pm0.09$)&	$9.64\pm1.27$&	$0.684\pm0.14$&	...\\
C05-4&	($2.13\pm0.56$)&	$9.76\pm1.48$&	$18.9\pm1.17$&	$14.6\pm0.49$&	($0.910\pm0.06$)&	$7.39\pm0.45$&	$0.547\pm0.04$&	...\\
\hline
C08-1&	...&	...&	$6.06\pm0.67$&	$11.1\pm0.79$&	...&	$5.80\pm0.62$&	...&	...\\
C08-2&	($182\pm2.49$)&	$140\pm7.13$&	$65.6\pm4.71$&	$61.4\pm2.17$&	...&	$33.5\pm0.91$&	$1.23\pm0.16$&	...\\
C08-3&	($214\pm2.00$)&	$170\pm5.66$&	$63.6\pm3.90$&	$54.7\pm1.91$&	...&	$28.4\pm1.30$&	$1.43\pm0.13$&	...\\
\hline
C09-1&	($464\pm0.68$)&	$170\pm0.51$&	$290\pm0.75$&	$122\pm0.62$&	($11.7\pm0.24$)&	$33.4\pm0.51$&	$3.59\pm0.06$&	...\\
C09-2&	...&	$3.00\pm0.63$&	$5.30\pm0.84$&	$6.38\pm0.69$&	...&	$4.64\pm0.59$&	$0.523\pm0.04$&	...\\
\hline
DR15-1&	($5.87\pm0.17$)&	$10.6\pm0.20$&	$27.2\pm0.96$&	$20.0\pm1.77$&	($2.69\pm0.40$)&	$15.2\pm1.19$&	$4.84\pm0.08$&	$0.570\pm0.02$\\
DR15-2&	($53.9\pm0.19$)&	$114\pm2.24$&	$105\pm2.70$&	($334\pm0.94$)&	($10.3\pm0.35$)&	$24.8\pm1.16$&	$4.50\pm0.15$&	$1.27\pm0.03$\\
DR15-3&	($0.445\pm0.23$)&	$21.1\pm3.38$&	$24.8\pm2.50$&	($6.09\pm1.74$)&	($3.74\pm0.35$)&	$11.3\pm1.19$&	$1.68\pm0.14$&	$0.410\pm0.03$\\
DR15-4&	($1083\pm43.5$)&	$519\pm18.3$&	$160\pm8.56$&	$102\pm2.92$&	($9.34\pm0.29$)&	$27.3\pm1.78$&	$4.41\pm0.15$&	$1.03\pm0.05$\\
DR15-5&	($3.22\pm0.28$)&	$9.58\pm0.20$&	$10.7\pm0.77$&	$9.28\pm1.25$&	($3.41\pm0.30$)&	$6.54\pm0.89$&	$1.65\pm0.07$&	$0.207\pm0.02$\\
DR15-6&	($14.0\pm1.53$)&	$42.3\pm3.68$&	$29.8\pm2.39$&	...&	...&	$13.4\pm1.01$&	$1.56\pm0.12$&	$0.364\pm0.03$\\
DR15-7&	...&	...&	...&	...&	...&	$2.03\pm2.44$&	$0.921\pm0.11$&	$0.166\pm0.02$\\
\hline
DR15E1-1&	($31.8\pm0.41$)&	$248\pm1.04$&	$441\pm0.78$&	$217\pm0.67$&	($14.0\pm0.30$)&	$86.6\pm0.66$&	$17.3\pm0.03$&	...\\
DR15E1-2&	($1.42\pm0.22$)&	$10.8\pm0.81$&	$17.1\pm0.98$&	$18.6\pm0.98$&	...&	$11.4\pm0.51$&	$1.23\pm0.03$&	...\\
DR15E1-3&	...&	$10.1\pm0.86$&	$14.2\pm1.06$&	$13.5\pm1.03$&	...&	$9.10\pm0.51$&	$1.61\pm0.03$&	...\\
\hline
DR15E2-1&	($165\pm0.23$)&	$331\pm1.03$&	($124\pm1.48$)&	$161\pm0.55$&	...&	$37.8\pm0.39$&	$4.75\pm0.05$&	...\\
DR15E2-2&	($3.36\pm0.22$)&	$22.9\pm1.81$&	$27.8\pm1.80$&	$29.8\pm0.97$&	...&	$19.6\pm0.38$&	$1.17\pm0.08$&	...\\
DR15E2-3&	($0.764\pm0.04$)&	$2.67\pm0.16$&	$8.05\pm0.39$&	$14.4\pm0.44$&	...&	$3.51\pm0.45$&	$0.661\pm0.04$&	...\\
DR15E2-4&	($178\pm0.43$)&	$418\pm0.46$&	$201\pm0.52$&	$77.5\pm0.55$&	...&	$23.6\pm0.52$&	$3.53\pm0.03$&	...\\
\hline
DR21-1&	($7666\pm28.7$)&	$5533\pm27.7$&	...&	$1396\pm3.70$&	($264\pm1.64$)&	$617\pm1.83$&	$70.3\pm0.68$&	$22.6\pm0.14$\\
DR21-2&	($2635\pm7.46$)&	$5595\pm6.61$&	($109\pm5.33$)&	$677\pm3.82$&	($238\pm1.60$)&	$350\pm2.16$&	$89.3\pm0.31$&	$21.3\pm0.07$\\
DR21-3&	($364\pm13.9$)&	$1831\pm26.1$&	$553\pm18.1$&	$409\pm3.61$&	$107\pm1.80$&	$167\pm1.99$&	$16.0\pm0.67$&	$5.52\pm0.12$\\
DR21-4&	($224\pm1.27$)&	$527\pm6.61$&	$296\pm6.39$&	($371\pm3.52$)&	($31.0\pm0.58$)&	$48.2\pm2.32$&	$10.1\pm0.50$&	$2.29\pm0.09$\\
DR21-5&	($220\pm12.6$)&	$290\pm35.0$&	$144\pm22.7$&	$71.9\pm4.68$&	$42.1\pm2.32$&	$25.2\pm1.70$&	$10.7\pm0.99$&	$1.58\pm0.19$\\
DR21-6&	...&	...&	$116\pm20.9$&	$42.9\pm3.21$&	$34.6\pm0.64$&	...&	$6.37\pm0.80$&	$1.14\pm0.15$\\
DR21-7&	($54.2\pm0.90$)&	$203\pm6.96$&	$121\pm6.68$&	$72.3\pm3.73$&	($14.8\pm0.58$)&	$28.0\pm2.43$&	$6.87\pm0.41$&	$1.28\pm0.08$\\
DR21-8&	...&	$481\pm9.27$&	$151\pm12.2$&	($498\pm3.51$)&	($18.5\pm1.68$)&	$44.3\pm2.35$&	$12.2\pm0.59$&	$1.13\pm0.10$\\
DR21-9&	($38.7\pm0.23$)&	$79.9\pm0.77$&	$59.9\pm1.30$&	$34.0\pm1.54$&	$18.8\pm0.10$&	$10.5\pm1.60$&	$6.94\pm0.05$&	$1.01\pm0.02$\\
DR21-10&	...&	$4.91\pm0.38$&	$10.8\pm1.13$&	$12.3\pm1.12$&	($3.73\pm0.13$)&	$5.82\pm1.21$&	$3.04\pm0.05$&	$0.377\pm0.02$\\
DR21-11&	...&	...&	$41.4\pm6.71$&	$41.7\pm2.36$&	$21.4\pm0.37$&	($4.15\pm1.50$)&	$5.87\pm0.19$&	$0.416\pm0.11$\\
DR21-12&	...&	...&	$46.3\pm5.13$&	$20.6\pm3.27$&	$13.7\pm0.56$&	$10.7\pm2.28$&	$2.77\pm0.27$&	$0.945\pm0.09$\\
DR21-13&	($42.1\pm6.39$)&	$112\pm11.5$&	$473\pm15.3$&	$278\pm3.80$&	$103\pm1.64$&	$109\pm2.25$&	$1.48\pm0.30$&	$0.264\pm0.07$\\
DR21-14&	...&	...&	$26.8\pm5.04$&	...&	$10.9\pm0.59$&	...&	$2.03\pm0.23$&	...\\
DR21-15&	...&	$30.8\pm2.44$&	$30.0\pm3.98$&	...&	$9.12\pm0.30$&	...&	$2.37\pm0.18$&	$0.463\pm0.05$\\
DR21-16&	...&	$70.1\pm10.4$&	$25.2\pm11.8$&	$56.1\pm2.24$&	$9.07\pm0.51$&	...&	$5.00\pm0.56$&	$0.320\pm0.06$\\
DR21-17&	($72.0\pm3.07$)&	$240\pm23.3$&	$76.7\pm24.1$&	$45.3\pm3.56$&	$24.3\pm0.90$&	...&	$3.49\pm0.64$&	...\\
DR21-18&	...&	...&	...&	...&	$5.51\pm0.22$&	$2.58\pm1.23$&	$2.27\pm0.16$&	...\\
DR21-19&	...&	$56.0\pm11.5$&	$42.6\pm7.31$&	$17.3\pm2.52$&	$8.10\pm0.79$&	...&	$2.35\pm0.30$&	$0.550\pm0.08$\\
DR21-20&	...&	$28.2\pm11.5$&	...&	$9.64\pm2.81$&	$14.2\pm1.06$&	...&	$1.29\pm0.51$&	($0.054\pm0.09$)\\
DR21-21&	...&	$47.9\pm7.14$&	$42.4\pm7.47$&	($0.929\pm3.64$)&	$7.46\pm0.89$&	($1.39\pm2.50$)&	$2.06\pm0.45$&	...\\
DR21-22&	($116\pm1.09$)&	$91.3\pm4.33$&	$56.6\pm3.83$&	$14.6\pm2.69$&	$10.5\pm0.24$&	...&	$1.90\pm0.15$&	$0.838\pm0.05$\\
DR21-23&	($487\pm2.71$)&	$758\pm6.78$&	$382\pm8.38$&	($24.0\pm3.30$)&	($8.60\pm0.32$)&	$40.0\pm2.24$&	$3.71\pm0.40$&	$0.921\pm0.06$\\
\hline
N03-1&	($70.8\pm1.56$)&	$329\pm3.11$&	$366\pm2.23$&	$132\pm1.88$&	$65.2\pm0.45$&	$57.4\pm1.53$&	$30.4\pm0.07$&	$5.43\pm0.03$\\
N03-2&	...&	...&	...&	$12.2\pm1.07$&	...&	...&	$1.38\pm0.15$&	$0.490\pm0.04$\\
\hline
N05-1&	($1088\pm3.45$)&	$1017\pm4.60$&	$456\pm3.51$&	$430\pm2.03$&	($28.7\pm0.24$)&	$150\pm1.12$&	$11.9\pm0.09$&	...\\
N05-2&	($40.3\pm0.87$)&	$4.38\pm0.91$&	$40.9\pm0.96$&	$70.4\pm0.61$&	($3.35\pm0.13$)&	$12.5\pm0.36$&	$0.906\pm0.08$&	...\\
N05-3&	($591\pm2.19$)&	$594\pm4.03$&	$452\pm2.86$&	$179\pm2.20$&	($14.2\pm0.19$)&	$58.1\pm1.19$&	$8.34\pm0.07$&	...\\
N05-4&	...&	$21.7\pm3.62$&	$18.1\pm2.92$&	$15.0\pm2.02$&	($3.55\pm0.17$)&	$9.05\pm1.12$&	$1.71\pm0.07$&	...\\
N05-5&	($13.5\pm0.23$)&	$23.0\pm1.46$&	$17.6\pm1.03$&	$13.5\pm0.98$&	($2.04\pm0.17$)&	$7.41\pm1.04$&	$1.56\pm0.03$&	...\\
N05-6&	($3.18\pm1.81$)&	$38.3\pm1.08$&	$32.7\pm0.64$&	$23.9\pm0.57$&	($3.18\pm0.16$)&	$5.93\pm0.38$&	$1.88\pm0.06$&	...\\
\hline
N07-1&	...&	$13.4\pm1.71$&	$20.1\pm2.23$&	$21.0\pm1.60$&	...&	$19.2\pm1.23$&	$1.97\pm0.16$&	$0.368\pm0.03$\\
N07-2&	($0.827\pm0.05$)&	$3.33\pm0.51$&	$5.48\pm0.73$&	$7.48\pm0.76$&	...&	$6.15\pm0.66$&	$0.822\pm0.07$&	$0.147\pm0.02$\\
\hline
N08-1&	($17.3\pm0.07$)&	$52.7\pm0.48$&	$44.7\pm1.17$&	$30.6\pm1.00$&	($6.78\pm0.35$)&	$18.1\pm0.78$&	$2.63\pm0.06$&	...\\
N08-2&	($0.487\pm0.11$)&	$2.57\pm0.49$&	$6.86\pm0.66$&	$6.40\pm0.53$&	...&	$4.25\pm0.37$&	$0.489\pm0.05$&	...\\
\hline
N12-1&	($146\pm0.18$)&	$358\pm1.26$&	$586\pm0.88$&	$272\pm0.73$&	($57.7\pm0.21$)&	$119\pm1.10$&	$11.9\pm0.17$&	$4.15\pm0.02$\\
N12-2&	($5.99\pm0.12$)&	$34.2\pm1.37$&	$44.5\pm0.47$&	$29.1\pm0.48$&	($6.01\pm0.22$)&	$14.4\pm0.39$&	$2.24\pm0.12$&	$0.458\pm0.03$\\
N12-3&	($145\pm0.42$)&	$150\pm1.92$&	$73.6\pm2.02$&	$31.9\pm1.62$&	($8.86\pm0.16$)&	$14.2\pm1.13$&	$4.04\pm0.06$&	$0.732\pm0.02$\\
N12-4&	...&	$4.17\pm1.58$&	$7.95\pm2.46$&	$7.99\pm1.91$&	$4.67\pm0.24$&	...&	$1.38\pm0.08$&	$0.284\pm0.02$\\
N12-5&	($42.4\pm0.21$)&	$112\pm0.87$&	$71.5\pm1.41$&	$21.9\pm1.26$&	($6.16\pm0.15$)&	$11.5\pm1.17$&	$2.33\pm0.10$&	$0.713\pm0.02$\\
N12-6&	...&	$19.9\pm1.06$&	$17.9\pm1.20$&	$10.1\pm1.25$&	($4.27\pm0.21$)&	$7.55\pm0.94$&	$1.57\pm0.09$&	$0.320\pm0.02$\\
N12-7&	...&	$10.2\pm1.96$&	($0.677\pm0.63$)&	$11.6\pm0.46$&	($3.05\pm0.24$)&	$7.13\pm0.74$&	$1.26\pm0.21$&	$0.201\pm0.04$\\
\hline
N26-1&	($12.8\pm0.50$)&	$34.0\pm2.77$&	$30.1\pm2.65$&	$18.2\pm2.13$&	($7.12\pm0.17$)&	$10.5\pm1.50$&	$2.24\pm0.19$&	$0.392\pm0.02$\\
\hline
N58-1&	...&	...&	...&	$29.3\pm2.21$&	...&	...&	$3.94\pm0.14$&	$0.951\pm0.05$\\
N58-2&	...&	$124\pm12.2$&	$74.6\pm7.54$&	$39.8\pm3.42$&	...&	$18.9\pm1.75$&	$2.26\pm0.21$&	$0.981\pm0.04$\\
N58-3&	($95.0\pm5.72$)&	$118\pm9.12$&	$59.2\pm6.80$&	$30.6\pm4.59$&	...&	$13.7\pm1.88$&	$3.08\pm0.13$&	$0.871\pm0.05$\\
N58-4&	($0.635\pm0.43$)&	$3.96\pm1.22$&	$10.1\pm1.51$&	$7.22\pm1.26$&	...&	$4.31\pm0.61$&	$0.812\pm0.13$&	$0.200\pm0.03$\\
\hline
N63-1&	($67.4\pm0.09$)&	$211\pm0.85$&	$170\pm1.45$&	$93.2\pm1.44$&	($29.5\pm0.26$)&	$36.1\pm1.21$&	$14.6\pm0.05$&	$3.10\pm0.01$\\
N63-2&	...&	$86.5\pm3.35$&	$121\pm1.59$&	$104\pm0.85$&	($9.70\pm0.66$)&	$24.9\pm1.48$&	$3.05\pm0.12$&	$0.519\pm0.03$\\
N63-3&	($1.47\pm0.14$)&	$9.51\pm0.53$&	$26.6\pm0.56$&	$11.1\pm0.64$&	...&	$8.43\pm0.51$&	$1.04\pm0.10$&	$0.220\pm0.02$\\
\hline
N68-1&	($23.5\pm0.50$)&	$101\pm1.58$&	$121\pm1.10$&	$45.0\pm1.00$&	($15.8\pm0.42$)&	$23.2\pm0.93$&	$8.90\pm0.06$&	$1.68\pm0.01$\\
N68-2&	($51.1\pm0.34$)&	$158\pm1.29$&	$148\pm1.18$&	$58.8\pm1.09$&	($17.0\pm0.51$)&	$21.6\pm0.87$&	$7.22\pm0.13$&	$0.991\pm0.03$\\
N68-3&	...&	...&	$20.2\pm2.61$&	$18.1\pm1.24$&	($6.41\pm0.46$)&	$8.83\pm1.67$&	$2.24\pm0.15$&	$0.548\pm0.03$\\
N68-4&	...&	...&	$18.0\pm4.13$&	$26.5\pm3.96$&	($0.498\pm0.58$)&	$12.6\pm2.50$&	$2.33\pm0.23$&	$0.365\pm0.06$\\
N68-5&	...&	$11.5\pm1.45$&	$17.9\pm1.27$&	$19.5\pm1.14$&	($6.38\pm0.34$)&	$10.4\pm0.91$&	$1.72\pm0.11$&	$0.356\pm0.02$\\
N68-6&	...&	...&	...&	$26.4\pm5.98$&	...&	$11.4\pm2.62$&	$2.10\pm0.28$&	$0.674\pm0.09$\\
N68-7&	...&	...&	$8.56\pm1.56$&	$9.30\pm1.49$&	($0.379\pm0.37$)&	$9.01\pm1.58$&	$1.12\pm0.08$&	$0.218\pm0.02$\\
N68-8&	...&	...&	$16.1\pm2.30$&	$11.8\pm1.37$&	($4.54\pm0.45$)&	$7.63\pm1.68$&	$1.35\pm0.08$&	$0.261\pm0.02$\\
\hline
NW01-1&	($1039\pm5.02$)&	$1386\pm2.54$&	$379\pm1.79$&	$108\pm1.38$&	($8.03\pm0.38$)&	$65.1\pm0.91$&	$7.92\pm0.05$&	...\\
\hline
NW04-1&	($198\pm2.27$)&	$552\pm0.49$&	$377\pm0.49$&	$223\pm0.48$&	($5.16\pm0.38$)&	$78.8\pm0.40$&	$15.1\pm0.03$&	...\\
NW04-2&	...&	$1.12\pm1.13$&	...&	$5.38\pm0.73$&	...&	$3.04\pm0.55$&	...&	...\\
NW04-3&	($8.85\pm0.06$)&	$33.0\pm0.25$&	$35.3\pm0.29$&	$22.5\pm0.35$&	...&	$7.79\pm0.27$&	($0.551\pm0.04$)&	...\\
NW04-4&	...&	...&	$2.54\pm0.76$&	$3.26\pm0.89$&	...&	$1.86\pm0.55$&	...&	...\\
NW04-5&	($7.26\pm0.11$)&	$24.1\pm0.65$&	$18.3\pm0.72$&	$13.3\pm0.75$&	($3.37\pm0.35$)&	$7.31\pm0.56$&	$1.20\pm0.04$&	...\\
\hline
NW04E-1&	($10.3\pm0.76$)&	$61.5\pm2.25$&	($45.2\pm1.60$)&	$109\pm1.46$&	...&	$52.7\pm1.12$&	($1.58\pm0.07$)&	...\\
NW04E-2&	($210\pm0.41$)&	$347\pm1.38$&	$180\pm1.14$&	$75.4\pm0.90$&	...&	$23.3\pm0.79$&	$3.88\pm0.04$&	...\\
\hline
NW12-1&	($290\pm0.12$)&	$442\pm0.34$&	$268\pm0.60$&	$114\pm0.96$&	...&	$39.5\pm0.90$&	$9.25\pm0.03$&	$1.92\pm0.02$\\
\hline
S01-1&	...&	$4.71\pm0.89$&	$12.9\pm1.70$&	$16.2\pm1.48$&	...&	$12.9\pm0.64$&	...&	$0.450\pm0.03$\\
S01-2&	($238\pm1.36$)&	$65.2\pm2.62$&	$81.5\pm1.28$&	$28.4\pm0.85$&	...&	$17.3\pm0.87$&	...&	...\\
S01-3&	($96.4\pm0.16$)&	$139\pm0.62$&	$97.3\pm1.34$&	$48.3\pm1.34$&	($3.30\pm0.31$)&	$24.4\pm0.62$&	...&	$0.854\pm0.02$\\
S01-4&	($5.99\pm1.31$)&	$31.1\pm3.97$&	$30.9\pm2.79$&	$23.2\pm2.51$&	...&	$12.8\pm1.05$&	...&	$0.488\pm0.04$\\
S01-5&	...&	$7.23\pm0.96$&	$11.2\pm1.69$&	$11.0\pm1.82$&	($1.90\pm0.29$)&	$6.75\pm1.30$&	...&	$0.224\pm0.02$\\
S01-6&	($23.6\pm0.04$)&	$32.0\pm0.47$&	$23.9\pm1.46$&	$13.7\pm0.90$&	($1.38\pm0.24$)&	$9.22\pm0.74$&	...&	...\\
\hline
S01S-1&	($4.60\pm0.12$)&	$33.2\pm0.17$&	$34.2\pm0.12$&	$29.4\pm0.09$&	($0.297\pm0.03$)&	$13.6\pm0.06$&	...&	...\\
S01S-2&	($27.9\pm0.10$)&	$169\pm0.11$&	$110\pm0.07$&	$55.2\pm0.05$&	($0.500\pm0.02$)&	$18.3\pm0.04$&	...&	...\\
S01S-3&	($41.8\pm0.54$)&	$111\pm0.82$&	$93.1\pm0.71$&	($91.8\pm0.27$)&	...&	$13.9\pm0.49$&	...&	...\\
S01S-4&	($0.454\pm0.11$)&	$4.38\pm0.82$&	$6.98\pm0.59$&	$6.28\pm0.48$&	...&	$3.72\pm0.34$&	...&	...\\
\hline
S07-1&	($1466\pm0.34$)&	$1882\pm1.59$&	$1084\pm2.59$&	$629\pm1.70$&	...&	$240\pm1.46$&	$14.8\pm0.09$&	...\\
\hline
S11-1&	...&	$60.1\pm3.42$&	$38.9\pm1.69$&	$49.4\pm1.42$&	...&	$11.4\pm1.12$&	$5.49\pm0.05$&	$0.519\pm0.02$\\
S11-2&	($3.56\pm0.40$)&	$11.7\pm1.17$&	$15.2\pm1.30$&	$25.6\pm0.98$&	...&	$10.8\pm0.79$&	$1.65\pm0.06$&	$0.403\pm0.02$\\
S11-3&	...&	($0.793\pm0.92$)&	$7.74\pm1.11$&	$9.39\pm0.91$&	...&	$6.04\pm0.78$&	$0.588\pm0.06$&	...\\
\hline
S13-1&	($46.7\pm0.78$)&	$98.4\pm2.67$&	$60.8\pm1.48$&	$36.0\pm1.10$&	...&	$15.3\pm0.84$&	$2.82\pm0.11$&	$0.734\pm0.02$\\
S13-2&	($1.48\pm0.40$)&	$30.8\pm2.48$&	$26.0\pm1.53$&	$17.6\pm1.04$&	...&	$9.76\pm0.86$&	$1.88\pm0.09$&	$0.558\pm0.02$\\
\hline
S13S-1&	...&	$3.84\pm0.45$&	$8.87\pm0.76$&	$6.62\pm0.70$&	...&	...&	...&	...\\
S13S-2&	($40.1\pm0.04$)&	$59.5\pm0.17$&	$69.8\pm0.27$&	$21.7\pm0.37$&	...&	$10.1\pm0.36$&	...&	...\\
S13S-3&	($3.95\pm0.03$)&	$6.48\pm0.10$&	$7.83\pm0.21$&	$6.82\pm0.29$&	...&	$4.92\pm0.18$&	...&	...\\
S13S-4&	($4.03\pm0.04$)&	$8.37\pm0.09$&	($0.307\pm0.24$)&	$8.67\pm0.36$&	...&	$4.67\pm0.37$&	...&	...\\
\hline
S29-1&	($0.527\pm0.13$)&	$8.94\pm1.72$&	$16.5\pm2.51$&	$22.9\pm2.53$&	($2.95\pm0.26$)&	$15.6\pm1.48$&	$1.60\pm0.12$&	...\\
\hline
S30-1&	($494\pm0.72$)&	$512\pm0.60$&	$534\pm0.50$&	$355\pm0.39$&	($15.2\pm0.48$)&	$174\pm0.71$&	($6.55\pm0.08$)&	($1.59\pm0.02$)\\
S30-2&	($1.78\pm0.43$)&	$19.2\pm1.57$&	...&	...&	...&	...&	$1.76\pm0.12$&	$0.299\pm0.02$\\
S30-3&	...&	$22.2\pm1.90$&	$26.8\pm0.57$&	($2.89\pm0.41$)&	$3.45\pm0.36$&	$3.07\pm0.52$&	$2.09\pm0.12$&	$0.625\pm0.02$\\
\hline
S30S-1&	($5.21\pm0.06$)&	$22.5\pm0.25$&	$26.9\pm0.33$&	$25.7\pm0.35$&	($1.36\pm0.11$)&	$8.81\pm0.29$&	$1.42\pm0.02$&	...\\
\hline
S32-1&	($32.2\pm0.78$)&	$115\pm1.90$&	$80.0\pm2.46$&	$47.2\pm1.53$&	($12.0\pm0.15$)&	$24.4\pm1.28$&	$5.50\pm0.06$&	$0.904\pm0.03$\\
S32-2&	($2.85\pm0.35$)&	$6.08\pm0.84$&	($3.71\pm0.55$)&	$8.71\pm1.33$&	...&	$6.52\pm1.60$&	...&	...\\
S32-3&	($12.8\pm0.32$)&	$48.1\pm1.66$&	$42.2\pm1.34$&	$24.4\pm0.90$&	($3.62\pm0.13$)&	$11.9\pm0.57$&	$2.21\pm0.06$&	$0.421\pm0.02$\\
S32-4&	($4.28\pm0.54$)&	$13.9\pm2.31$&	$10.7\pm0.93$&	$11.9\pm0.78$&	...&	...&	$1.16\pm0.05$&	...\\
\hline
S43-1&	...&	...&	...&	$6.61\pm1.97$&	...&	$3.29\pm0.98$&	$1.06\pm0.15$&	$0.554\pm0.04$\\
S43-2&	...&	$41.5\pm0.75$&	$46.1\pm1.50$&	$28.3\pm1.49$&	($6.15\pm0.22$)&	$12.8\pm1.12$&	$3.55\pm0.08$&	$0.686\pm0.02$\\
S43-3&	...&	$15.6\pm0.87$&	$31.7\pm1.46$&	$30.6\pm1.49$&	($2.69\pm0.22$)&	$19.1\pm1.01$&	$1.90\pm0.07$&	$0.348\pm0.02$\\
S43-4&	...&	...&	$8.82\pm1.56$&	$8.50\pm1.37$&	...&	($0.714\pm1.02$)&	$0.999\pm0.09$&	...\\
\hline
S106-1&	...&	$37.3\pm10.3$&	$40.5\pm7.51$&	$34.6\pm4.44$&	($6.09\pm0.42$)&	$25.0\pm3.33$&	$1.74\pm0.19$&	...\\
S106-2&	($3180\pm8.97$)&	$2124\pm13.9$&	$377\pm11.3$&	$106\pm8.32$&	($10.5\pm0.64$)&	$93.5\pm3.15$&	$9.22\pm0.20$&	...\\
\hline
S106N-1&	($21.9\pm0.02$)&	$79.8\pm0.09$&	$66.0\pm0.19$&	$44.3\pm0.19$&	...&	$23.4\pm0.18$&	$2.08\pm0.04$&	...\\
\hline
S106W1-1&	($52.3\pm0.09$)&	$79.5\pm0.74$&	$78.2\pm0.99$&	$32.3\pm1.12$&	...&	$18.8\pm0.55$&	$2.37\pm0.06$&	...\\
S106W1-2&	...&	$15.2\pm2.00$&	$16.5\pm3.17$&	($0.529\pm2.34$)&	...&	$7.17\pm0.90$&	$1.75\pm0.10$&	...\\
S106W1-3&	($585\pm1.81$)&	$420\pm4.62$&	$165\pm3.67$&	$78.1\pm2.45$&	...&	$25.8\pm0.84$&	$3.79\pm0.08$&	...\\
S106W1-4&	($1.83\pm0.74$)&	$56.0\pm5.59$&	$47.9\pm3.63$&	$30.5\pm2.30$&	...&	$10.2\pm0.44$&	$1.90\pm0.11$&	...\\
S106W1-5&	...&	...&	$67.4\pm8.23$&	$36.2\pm4.35$&	...&	($2.33\pm0.65$)&	$2.17\pm0.20$&	...\\
S106W1-6&	...&	$11.4\pm0.91$&	$11.4\pm0.99$&	$8.48\pm0.91$&	...&	$6.37\pm1.29$&	($0.108\pm0.08$)&	...\\
S106W1-7&	($39.9\pm0.92$)&	$67.5\pm1.96$&	$41.1\pm2.26$&	$20.1\pm2.41$&	...&	$12.9\pm0.82$&	$1.60\pm0.05$&	...\\
\hline
S106W2-1&	($3474\pm6.14$)&	$1872\pm14.3$&	$1116\pm4.01$&	$601\pm3.19$&	...&	$234\pm2.11$&	...&	...\\
S106W2-2&	($206\pm8.40$)&	$458\pm7.40$&	$448\pm4.38$&	$198\pm2.71$&	...&	$133\pm1.77$&	...&	...\\
S106W2-3&	($1513\pm7.58$)&	$1115\pm12.1$&	$358\pm4.16$&	$256\pm2.59$&	...&	$61.9\pm1.74$&	...&	...\\
S106W2-4&	($102\pm4.03$)&	$162\pm9.61$&	$119\pm4.61$&	$55.4\pm5.11$&	...&	$26.5\pm2.42$&	...&	...\\
S106W2-5&	($1.81\pm0.20$)&	$19.0\pm1.76$&	$21.6\pm2.07$&	$18.4\pm2.43$&	...&	$11.2\pm2.33$&	...&	...\\
S106W2-6&	...&	$10.6\pm5.19$&	$11.3\pm4.60$&	$17.9\pm0.98$&	...&	$6.00\pm0.95$&	...&	...\\
S106W2-7&	($289\pm4.25$)&	$230\pm6.51$&	$108\pm4.81$&	$38.7\pm5.21$&	...&	$19.3\pm2.54$&	...&	...\\
S106W2-8&	($3.51\pm0.13$)&	$11.8\pm1.14$&	$12.0\pm1.44$&	$10.1\pm1.69$&	...&	$5.59\pm1.87$&	...&	...\\
S106W2-9&	...&	$13.0\pm0.94$&	$18.3\pm1.46$&	$12.4\pm1.76$&	...&	$4.86\pm1.24$&	...&	...\\
\hline
W75N-1&	($11010\pm0.34$)&	$10470\pm2.19$&	($554\pm7.71$)&	$1762\pm1.63$&	($394\pm0.14$)&	$898\pm1.09$&	$114\pm0.08$&	...\\
W75N-2&	...&	$9.38\pm1.65$&	$12.3\pm2.13$&	$20.4\pm1.68$&	($4.67\pm0.25$)&	$14.9\pm1.03$&	$3.59\pm0.07$&	...\\
W75N-3&	($0.966\pm0.24$)&	$49.8\pm2.52$&	$52.7\pm7.01$&	$32.8\pm1.36$&	($12.6\pm0.35$)&	$20.9\pm1.06$&	$4.36\pm0.16$&	...\\
W75N-4&	($10.2\pm0.79$)&	$27.9\pm2.04$&	$25.9\pm2.68$&	$20.6\pm2.22$&	($4.11\pm0.22$)&	$13.2\pm0.98$&	$1.45\pm0.08$&	...\\
W75N-5&	($0.611\pm0.16$)&	$7.15\pm1.00$&	$12.3\pm1.18$&	$12.5\pm0.60$&	($3.43\pm0.13$)&	($1.32\pm0.75$)&	$0.902\pm0.05$&	...\\
W75N-6&	($16.7\pm0.08$)&	$29.6\pm0.63$&	$21.8\pm1.73$&	$14.9\pm1.33$&	...&	$10.2\pm0.81$&	$1.17\pm0.09$&	...\\

\enddata
\tablecomments{Fluxes in brackets were not used for SED fitting since they deviate largely from the model or the wavelengths are at 70 \um\ (see Sect. \ref{subsec:extraction}). Ellipses mean the data are not available in the continuum images used for source extraction. }
\end{deluxetable*}

\section{Sources associated with the MDCs in Cygnus X}\label{sec_A:MDC_source}

This appendix presents the mid-IR sources and signposts of HMSF that are associated with the MDCs. The positional tolerances were all set to 10\arcsec. 

\renewcommand{\thetable}{D1}
\startlongtable
\begin{deluxetable*}{c|c|c|c|c|c}
\tabletypesize{\small}
\tablecaption{Mid-infrared Sources Associated with the MDCs in Cygnus X \label{tab:MDC_IR}}  
\tablehead{ 
    \colhead{Name}& \colhead{Method\tablenotemark{a}}& \colhead{$F_{\nu}(Spitzer\ 24\mu m)$}& \colhead{\emph{Spitzer} ID}& \colhead{$F_{\nu}(MSX\ 21\mu m)$}& \colhead{\emph{MSX} ID}\\
    \colhead{}& \colhead{}& \colhead{(Jy)}& \colhead{}& \colhead{(Jy)}& \colhead{}
    }
\startdata
\multirow{1}{*}{C01S-1}&	\multirow{1}{*}{catalog}&	0.046&	J202925.88+403605.1&	\multirow{1}{*}{...}&	\multirow{1}{*}{...}\\\hline
\multirow{1}{*}{C03-1}&	\multirow{1}{*}{catalog}&	2.446&	J203029.40+411557.9&	\multirow{1}{*}{...}&	\multirow{1}{*}{...}\\\hline
\multirow{1}{*}{C03S-1}&	\multirow{1}{*}{catalog}&	3.986&	J203050.66+410227.6&	\multirow{1}{*}{2.295}&	\multirow{1}{*}{G079.7358+00.9905}\\\hline
\multirow{2}{*}{C05-1}&	\multirow{2}{*}{catalog}&	0.084&	J203223.76+410757.5&	\multirow{2}{*}{...}&	\multirow{2}{*}{...}\\	&	&	0.088&	J203224.72+410803.4&	&	\\\hline
\multirow{1}{*}{C05-2}&	\multirow{1}{*}{catalog}&	3.213&	J203221.06+410754.5&	\multirow{1}{*}{2.691}&	\multirow{1}{*}{G079.9766+00.8153}\\\hline
\multirow{1}{*}{C05-4}&	\multirow{1}{*}{catalog}&	0.033&	J203223.21+410651.7&	\multirow{1}{*}{...}&	\multirow{1}{*}{...}\\\hline
\multirow{1}{*}{C08-2}&	\multirow{1}{*}{catalog}&	...&	...&	\multirow{1}{*}{6.649}&	\multirow{1}{*}{G080.3656+00.4573}\\\hline
\multirow{1}{*}{C08-3}&	\multirow{1}{*}{catalog}&	...&	...&	\multirow{1}{*}{11.089}&	\multirow{1}{*}{G080.3597+00.4419}\\\hline
\multirow{1}{*}{C09-1}&	\multirow{1}{*}{catalog}&	...&	...&	\multirow{1}{*}{18.475}&	\multirow{1}{*}{G080.6344+00.6822}\\\hline
\multirow{1}{*}{C09-2}&	\multirow{1}{*}{catalog}&	0.021&	J203449.97+413359.0&	\multirow{1}{*}{...}&	\multirow{1}{*}{...}\\\hline
\multirow{1}{*}{DR15-1}&	\multirow{1}{*}{catalog}&	0.241&	J203158.15+401836.1&	\multirow{1}{*}{...}&	\multirow{1}{*}{...}\\\hline
\multirow{3}{*}{DR15-2}&	\multirow{3}{*}{catalog}&	4.962&	J203222.10+402017.1&	\multirow{3}{*}{4.441}&	\multirow{3}{*}{G079.3398+00.3417}\\	&	&	1.545&	J203222.99+402021.4&	&	\\	&	&	0.846&	J203221.13+402025.6&	&	\\\hline
\multirow{1}{*}{DR15-3}&	\multirow{1}{*}{catalog}&	0.476&	J203223.04+401922.7&	\multirow{1}{*}{...}&	\multirow{1}{*}{...}\\\hline
\multirow{1}{*}{DR15-5}&	\multirow{1}{*}{catalog}&	0.159&	J203137.00+401939.3&	\multirow{1}{*}{...}&	\multirow{1}{*}{...}\\\hline
\multirow{2}{*}{DR15-6}&	\multirow{2}{*}{catalog}&	0.612&	J203221.96+401937.7&	\multirow{2}{*}{...}&	\multirow{2}{*}{...}\\	&	&	0.990&	J203220.60+401950.1&	&	\\\hline
\multirow{1}{*}{DR15E1-1}&	\multirow{1}{*}{catalog}&	0.142&	J203442.23+394505.8&	\multirow{1}{*}{...}&	\multirow{1}{*}{...}\\\hline
\multirow{2}{*}{DR15E1-2}&	\multirow{2}{*}{catalog}&	0.262&	J203323.60+394245.3&	\multirow{2}{*}{...}&	\multirow{2}{*}{...}\\	&	&	0.042&	J203323.95+394256.7&	&	\\\hline
\multirow{1}{*}{DR15E1-3}&	\multirow{1}{*}{catalog}&	0.009&	J203328.73+394119.1&	\multirow{1}{*}{...}&	\multirow{1}{*}{...}\\\hline
\multirow{1}{*}{DR15E2-1}&	\multirow{1}{*}{photometry}&	0.305&	...&	...&	...\\\hline
\multirow{1}{*}{DR15E2-2}&	\multirow{1}{*}{catalog}&	0.028&	J203750.17+394952.0&	\multirow{1}{*}{...}&	\multirow{1}{*}{...}\\\hline
\multirow{1}{*}{DR15E2-3}&	\multirow{1}{*}{catalog}&	0.124&	J203627.76+394239.3&	\multirow{1}{*}{...}&	\multirow{1}{*}{...}\\\hline
\multirow{1}{*}{DR15E2-4}&	\multirow{1}{*}{catalog}&	4.894&	J203714.82+394855.5&	\multirow{1}{*}{9.669}&	\multirow{1}{*}{G079.4822-00.7174}\\\hline
\multirow{1}{*}{DR21-1}&	\multirow{1}{*}{catalog}&	...&	...&	\multirow{1}{*}{543.910}&	\multirow{1}{*}{G081.6802+00.5405}\\\hline
\multirow{1}{*}{DR21-2}&	\multirow{1}{*}{catalog}&	...&	...&	\multirow{1}{*}{6.777}&	\multirow{1}{*}{G081.7220+00.5699}\\\hline
\multirow{1}{*}{DR21-3}&	\multirow{1}{*}{catalog}&	...&	...&	\multirow{1}{*}{28.401}&	\multirow{1}{*}{G081.7133+00.5589}\\\hline
\multirow{1}{*}{DR21-4}&	\multirow{1}{*}{catalog}&	...&	...&	\multirow{1}{*}{26.502}&	\multirow{1}{*}{G081.7522+00.5906}\\\hline
\multirow{1}{*}{DR21-6}&	\multirow{1}{*}{catalog}&	1.083&	J203901.91+421837.4&	\multirow{1}{*}{...}&	\multirow{1}{*}{...}\\\hline
\multirow{1}{*}{DR21-7}&	\multirow{1}{*}{photometry}&	0.363&	...&	...&	...\\\hline
\multirow{1}{*}{DR21-8}&	\multirow{1}{*}{catalog}&	0.073&	J203859.53+422343.6&	\multirow{1}{*}{...}&	\multirow{1}{*}{...}\\\hline
\multirow{1}{*}{DR21-9}&	\multirow{1}{*}{catalog}&	5.645&	J203916.72+421609.3&	\multirow{1}{*}{3.677}&	\multirow{1}{*}{G081.6632+00.4651}\\\hline
\multirow{1}{*}{DR21-12}&	\multirow{1}{*}{catalog}&	0.024&	J203850.80+422716.4&	\multirow{1}{*}{...}&	\multirow{1}{*}{...}\\\hline
\multirow{2}{*}{DR21-14}&	\multirow{2}{*}{catalog}&	0.007&	J203901.80+422656.6&	\multirow{2}{*}{...}&	\multirow{2}{*}{...}\\	&	&	0.021&	J203902.18+422644.4&	&	\\\hline
\multirow{2}{*}{DR21-16}&	\multirow{2}{*}{catalog}&	0.226&	J203901.44+421809.6&	\multirow{2}{*}{...}&	\multirow{2}{*}{...}\\	&	&	0.102&	J203900.55+421821.0&	&	\\\hline
\multirow{1}{*}{DR21-17}&	\multirow{1}{*}{photometry}&	0.539&	...&	...&	...\\\hline
\multirow{1}{*}{DR21-20}&	\multirow{1}{*}{catalog}&	1.083&	J203901.91+421837.4&	\multirow{1}{*}{...}&	\multirow{1}{*}{...}\\\hline
\multirow{1}{*}{DR21-21}&	\multirow{1}{*}{catalog}&	0.135&	J203903.13+422627.7&	\multirow{1}{*}{...}&	\multirow{1}{*}{...}\\\hline
\multirow{1}{*}{DR21-22}&	\multirow{1}{*}{catalog}&	1.359&	J203907.31+421534.9&	\multirow{1}{*}{7.252}&	\multirow{1}{*}{G081.6411+00.4812}\\\hline
\multirow{1}{*}{DR21-23}&	\multirow{1}{*}{photometry}&	...&	...&	20.148&	...\\\hline
\multirow{3}{*}{N03-1}&	\multirow{3}{*}{catalog}&	1.250&	J203533.56+422015.1&	\multirow{3}{*}{3.737}&	\multirow{3}{*}{G081.3039+01.0520}\\	&	&	2.448&	J203534.75+422016.7&	&	\\	&	&	1.714&	J203534.44+422006.8&	&	\\\hline
\multirow{1}{*}{N05-1}&	\multirow{1}{*}{catalog}&	...&	...&	\multirow{1}{*}{99.331}&	\multirow{1}{*}{G080.8645+00.4197}\\\hline
\multirow{1}{*}{N05-3}&	\multirow{1}{*}{catalog}&	...&	...&	\multirow{1}{*}{19.714}&	\multirow{1}{*}{G080.8624+00.3827}\\\hline
\multirow{1}{*}{N05-4}&	\multirow{1}{*}{catalog}&	0.091&	J203712.23+413334.6&	\multirow{1}{*}{...}&	\multirow{1}{*}{...}\\\hline
\multirow{2}{*}{N05-5}&	\multirow{2}{*}{catalog}&	0.668&	J203724.92+413535.6&	\multirow{2}{*}{2.334}&	\multirow{2}{*}{G080.9171+00.3289}\\	&	&	1.442&	J203726.02+413540.9&	&	\\\hline
\multirow{1}{*}{N05-6}&	\multirow{1}{*}{catalog}&	0.774&	J203603.28+413939.6&	\multirow{1}{*}{...}&	\multirow{1}{*}{...}\\\hline
\multirow{1}{*}{N07-1}&	\multirow{1}{*}{catalog}&	0.016&	J203639.89+425110.6&	\multirow{1}{*}{...}&	\multirow{1}{*}{...}\\\hline
\multirow{1}{*}{N07-2}&	\multirow{1}{*}{catalog}&	0.033&	J203547.51+425254.6&	\multirow{1}{*}{...}&	\multirow{1}{*}{...}\\\hline
\multirow{2}{*}{N08-1}&	\multirow{2}{*}{catalog}&	0.336&	J203638.56+422907.3&	\multirow{2}{*}{...}&	\multirow{2}{*}{...}\\	&	&	0.143&	J203638.34+422915.7&	&	\\\hline
\multirow{1}{*}{N08-2}&	\multirow{1}{*}{catalog}&	0.188&	J203619.91+423728.5&	\multirow{1}{*}{...}&	\multirow{1}{*}{...}\\\hline
\multirow{1}{*}{N12-1}&	\multirow{1}{*}{photometry}&	1.096&	...&	...&	...\\\hline
\multirow{1}{*}{N12-2}&	\multirow{1}{*}{photometry}&	0.019&	...&	...&	...\\\hline
\multirow{1}{*}{N12-3}&	\multirow{1}{*}{catalog}&	3.225&	J203717.67+421637.5&	\multirow{1}{*}{3.872}&	\multirow{1}{*}{G081.4452+00.7635}\\\hline
\multirow{1}{*}{N12-5}&	\multirow{1}{*}{catalog}&	2.283&	J203730.51+421359.1&	\multirow{1}{*}{...}&	\multirow{1}{*}{...}\\\hline
\multirow{2}{*}{N12-7}&	\multirow{2}{*}{catalog}&	0.022&	J203704.82+421150.7&	\multirow{2}{*}{...}&	\multirow{2}{*}{...}\\	&	&	0.075&	J203705.70+421203.0&	&	\\\hline
\multirow{3}{*}{N26-1}&	\multirow{3}{*}{catalog}&	0.837&	J203821.33+421120.7&	\multirow{3}{*}{...}&	\multirow{3}{*}{...}\\	&	&	0.909&	J203820.96+421125.9&	&	\\	&	&	0.421&	J203822.29+421135.4&	&	\\\hline
\multirow{2}{*}{N58-2}&	\multirow{2}{*}{catalog}&	1.011&	J203931.50+412003.0&	\multirow{2}{*}{2.809}&	\multirow{2}{*}{G080.9473-00.1419}\\	&	&	0.908&	J203930.96+412002.8&	&	\\\hline
\multirow{1}{*}{N58-3}&	\multirow{1}{*}{catalog}&	...&	...&	\multirow{1}{*}{3.332}&	\multirow{1}{*}{G080.9500-00.1556}\\\hline
\multirow{1}{*}{N58-4}&	\multirow{1}{*}{photometry}&	0.031&	...&	...&	...\\\hline
\multirow{2}{*}{N63-1}&	\multirow{2}{*}{catalog}&	0.442&	J204005.54+413212.8&	\multirow{2}{*}{...}&	\multirow{2}{*}{...}\\	&	&	0.162&	J204005.58+413221.3&	&	\\\hline
\multirow{2}{*}{N63-3}&	\multirow{2}{*}{catalog}&	0.068&	J204033.51+413839.5&	\multirow{2}{*}{...}&	\multirow{2}{*}{...}\\	&	&	0.090&	J204034.91+413845.1&	&	\\\hline
\multirow{1}{*}{N68-1}&	\multirow{1}{*}{catalog}&	0.195&	J204033.49+415900.6&	\multirow{1}{*}{...}&	\multirow{1}{*}{...}\\\hline
\multirow{2}{*}{N68-2}&	\multirow{2}{*}{catalog}&	2.750&	J204028.49+415712.0&	\multirow{2}{*}{-1.415}&	\multirow{2}{*}{G081.5486+00.0945}\\	&	&	2.395&	J204028.98+415707.9&	&	\\\hline
\multirow{1}{*}{NW01-1}&	\multirow{1}{*}{catalog}&	...&	...&	\multirow{1}{*}{400.660}&	\multirow{1}{*}{G078.4373+02.6584}\\\hline
\multirow{1}{*}{NW04-1}&	\multirow{1}{*}{catalog}&	...&	...&	\multirow{1}{*}{85.851}&	\multirow{1}{*}{G078.8699+02.7602}\\\hline
\multirow{1}{*}{NW04-3}&	\multirow{1}{*}{catalog}&	0.274&	J201939.16+411102.5&	\multirow{1}{*}{...}&	\multirow{1}{*}{...}\\\hline
\multirow{1}{*}{NW04-5}&	\multirow{1}{*}{photometry}&	0.159&	...&	...&	...\\\hline
\multirow{1}{*}{NW04E-1}&	\multirow{1}{*}{catalog}&	...&	...&	\multirow{1}{*}{1.609}&	\multirow{1}{*}{G079.1557+02.2197}\\\hline
\multirow{1}{*}{NW04E-2}&	\multirow{1}{*}{catalog}&	...&	...&	\multirow{1}{*}{27.468}&	\multirow{1}{*}{G079.1272+02.2782}\\\hline
\multirow{1}{*}{NW12-1}&	\multirow{1}{*}{catalog}&	...&	...&	\multirow{1}{*}{15.582}&	\multirow{1}{*}{G079.8855+02.5517}\\\hline
\multirow{1}{*}{S01-1}&	\multirow{1}{*}{catalog}&	0.067&	J201648.81+392208.7&	\multirow{1}{*}{...}&	\multirow{1}{*}{...}\\\hline
\multirow{1}{*}{S01-2}&	\multirow{1}{*}{catalog}&	...&	...&	\multirow{1}{*}{12.621}&	\multirow{1}{*}{G076.9283+02.0329}\\\hline
\multirow{1}{*}{S01-3}&	\multirow{1}{*}{catalog}&	6.554&	J201659.13+392103.9&	\multirow{1}{*}{5.271}&	\multirow{1}{*}{G076.8322+02.1876}\\\hline
\multirow{1}{*}{S01-4}&	\multirow{1}{*}{catalog}&	0.511&	J201745.91+392040.1&	\multirow{1}{*}{...}&	\multirow{1}{*}{...}\\\hline
\multirow{1}{*}{S01-6}&	\multirow{1}{*}{catalog}&	3.873&	J201643.74+392320.5&	\multirow{1}{*}{3.085}&	\multirow{1}{*}{G076.8356+02.2494}\\\hline
\multirow{1}{*}{S01S-2}&	\multirow{1}{*}{catalog}&	0.320&	J201737.09+390334.4&	\multirow{1}{*}{...}&	\multirow{1}{*}{...}\\\hline
\multirow{1}{*}{S01S-4}&	\multirow{1}{*}{catalog}&	0.079&	J201651.59+390011.7&	\multirow{1}{*}{...}&	\multirow{1}{*}{...}\\\hline
\multirow{1}{*}{S07-1}&	\multirow{1}{*}{catalog}&	...&	...&	\multirow{1}{*}{145.480}&	\multirow{1}{*}{G077.4622+01.7600}\\\hline
\multirow{1}{*}{S11-1}&	\multirow{1}{*}{catalog}&	0.691&	J202220.04+395823.0&	\multirow{1}{*}{...}&	\multirow{1}{*}{...}\\\hline
\multirow{1}{*}{S11-2}&	\multirow{1}{*}{catalog}&	0.955&	J202153.50+395936.7&	\multirow{1}{*}{...}&	\multirow{1}{*}{...}\\\hline
\multirow{1}{*}{S13-1}&	\multirow{1}{*}{catalog}&	5.055&	J202632.37+395720.8&	\multirow{1}{*}{4.843}&	\multirow{1}{*}{G078.3762+01.0191}\\\hline
\multirow{1}{*}{S13S-1}&	\multirow{1}{*}{catalog}&	0.220&	J202658.91+391023.6&	\multirow{1}{*}{...}&	\multirow{1}{*}{...}\\\hline
\multirow{1}{*}{S13S-2}&	\multirow{1}{*}{catalog}&	...&	...&	\multirow{1}{*}{8.738}&	\multirow{1}{*}{G077.9280+00.8711}\\\hline
\multirow{1}{*}{S13S-3}&	\multirow{1}{*}{catalog}&	0.275&	J202612.63+391641.9&	\multirow{1}{*}{...}&	\multirow{1}{*}{...}\\\hline
\multirow{1}{*}{S13S-4}&	\multirow{1}{*}{catalog}&	0.112&	J202547.75+392954.2&	\multirow{1}{*}{...}&	\multirow{1}{*}{...}\\\hline
\multirow{1}{*}{S29-1}&	\multirow{1}{*}{catalog}&	0.316&	J202957.68+401604.2&	\multirow{1}{*}{...}&	\multirow{1}{*}{...}\\\hline
\multirow{1}{*}{S30-2}&	\multirow{1}{*}{catalog}&	0.219&	J203114.63+400318.4&	\multirow{1}{*}{...}&	\multirow{1}{*}{...}\\\hline
\multirow{1}{*}{S30-3}&	\multirow{1}{*}{catalog}&	2.073&	J203112.91+400323.1&	\multirow{1}{*}{...}&	\multirow{1}{*}{...}\\\hline
\multirow{3}{*}{S30S-1}&	\multirow{3}{*}{catalog}&	0.113&	J203058.65+394229.0&	\multirow{3}{*}{...}&	\multirow{3}{*}{...}\\	&	&	0.250&	J203057.88+394220.8&	&	\\	&	&	0.245&	J203058.04+394226.7&	&	\\\hline
\multirow{1}{*}{S32-1}&	\multirow{1}{*}{catalog}&	0.120&	J203121.71+385716.6&	\multirow{1}{*}{...}&	\multirow{1}{*}{...}\\\hline
\multirow{1}{*}{S32-4}&	\multirow{1}{*}{catalog}&	0.239&	J203147.25+390048.2&	\multirow{1}{*}{2.641}&	\multirow{1}{*}{G078.2049-00.3555}\\\hline
\multirow{1}{*}{S43-2}&	\multirow{1}{*}{catalog}&	0.105&	J203240.77+384639.0&	\multirow{1}{*}{...}&	\multirow{1}{*}{...}\\\hline
\multirow{1}{*}{S106-2}&	\multirow{1}{*}{catalog}&	...&	...&	\multirow{1}{*}{1240.900}&	\multirow{1}{*}{G076.3829-00.6210}\\\hline
\multirow{2}{*}{S106N-1}&	\multirow{2}{*}{catalog}&	0.169&	J202614.94+380224.5&	\multirow{2}{*}{...}&	\multirow{2}{*}{...}\\	&	&	0.410&	J202614.09+380224.2&	&	\\\hline
\multirow{2}{*}{S106W1-1}&	\multirow{2}{*}{catalog}&	0.124&	J202322.97+373548.8&	\multirow{2}{*}{...}&	\multirow{2}{*}{...}\\	&	&	2.184&	J202323.73+373535.2&	&	\\\hline
\multirow{1}{*}{S106W1-2}&	\multirow{1}{*}{catalog}&	0.036&	J202406.91+373543.6&	\multirow{1}{*}{...}&	\multirow{1}{*}{...}\\\hline
\multirow{1}{*}{S106W1-3}&	\multirow{1}{*}{catalog}&	...&	...&	\multirow{1}{*}{54.009}&	\multirow{1}{*}{G076.1877+00.0974}\\\hline
\multirow{1}{*}{S106W1-4}&	\multirow{1}{*}{catalog}&	0.197&	J202358.42+373729.0&	\multirow{1}{*}{...}&	\multirow{1}{*}{...}\\\hline
\multirow{1}{*}{S106W1-6}&	\multirow{1}{*}{catalog}&	0.150&	J202328.46+373421.7&	\multirow{1}{*}{...}&	\multirow{1}{*}{...}\\\hline
\multirow{1}{*}{S106W1-7}&	\multirow{1}{*}{catalog}&	3.288&	J202402.83+373637.3&	\multirow{1}{*}{3.494}&	\multirow{1}{*}{G076.1807+00.0619}\\\hline
\multirow{1}{*}{S106W2-1}&	\multirow{1}{*}{catalog}&	...&	...&	\multirow{1}{*}{46.419}&	\multirow{1}{*}{G075.7822+00.3421}\\\hline
\multirow{1}{*}{S106W2-4}&	\multirow{1}{*}{catalog}&	4.170&	J202147.91+373017.3&	\multirow{1}{*}{4.043}&	\multirow{1}{*}{G075.8404+00.3682}\\\hline
\multirow{1}{*}{S106W2-5}&	\multirow{1}{*}{catalog}&	0.291&	J202155.09+372953.9&	\multirow{1}{*}{...}&	\multirow{1}{*}{...}\\\hline
\multirow{1}{*}{S106W2-7}&	\multirow{1}{*}{catalog}&	...&	...&	\multirow{1}{*}{6.830}&	\multirow{1}{*}{G075.8428+00.3595}\\\hline
\multirow{1}{*}{W75N-1}&	\multirow{1}{*}{photometry}&	...&	...&	620.539&	...\\\hline
\multirow{3}{*}{W75N-3}&	\multirow{3}{*}{catalog}&	0.342&	J203832.33+423945.1&	\multirow{3}{*}{...}&	\multirow{3}{*}{...}\\	&	&	0.298&	J203833.31+423938.7&	&	\\	&	&	0.004&	J203834.09+423955.9&	&	\\\hline
\multirow{1}{*}{W75N-4}&	\multirow{1}{*}{catalog}&	0.698&	J203810.05+423809.4&	\multirow{1}{*}{...}&	\multirow{1}{*}{...}\\\hline
\multirow{1}{*}{W75N-6}&	\multirow{1}{*}{catalog}&	4.761&	J203755.41+424046.7&	\multirow{1}{*}{3.702}&	\multirow{1}{*}{G081.8375+00.9134}\\\hline

\enddata
\tablenotetext{a}{Methods used to extract fluxes. ``Catalog'' means that the fluxes were quoted from archival catalogs, and ``photometry'' means that the fluxes were obtained by manual photometry (see Sect. \ref{subsec:IR_class}).} 
\end{deluxetable*}

\renewcommand{\thetable}{D2}
\startlongtable
\begin{deluxetable*}{c|c|c||c|c|c}
\tabletypesize{\small}
\tablecaption{Signposts of High-mass Star Formation of the MDCs in Cygnus X \label{tab:MDC_sign}}  
\tablehead{ 
    \colhead{MDC}& \colhead{VLASS UC\hii\ region\tablenotemark{a}}& \colhead{$F_{\nu}(\rm 3\ GHz)$}& \colhead{Class II methanol maser}& \colhead{Refs.\tablenotemark{b}}& \colhead{$F_{\nu,\rm peak}(\rm 6.7\ GHz\ maser)$} \\
    \colhead{}& \colhead{} & \colhead{(Jy)}& \colhead{}& \colhead{}& \colhead{(Jy)}
    }
\startdata
C03S-1&	...&	...&	G079.735+0.990&	1&	101.3\\\hline
C09-1&	J203500.21+413453.2&	0.129&	...&	...&	...\\\hline
DR21-1&	J203901.39+421937.7&	4.917&	...&	...&	...\\\hline
DR21-2&	...&	...&	G081.721+0.571&	1&	16.7\\\hline
DR21-4&	...&	...&	G081.752+0.590&	1&	63.7\\\hline
DR21-7&	...&	...&	G081.765+0.597&	1&	7.6\\\hline
DR21-23&	...&	...&	G081.744+0.590&	1&	32.4\\\hline
N05-1&	J203652.15+413623.7&	0.034&	J203652.60+413633.0&	2&	10.0\\\hline
N05-3&	...&	...&	G080.861+0.383&	1&	35.3\\\hline
NW01-1&	J201939.39+405634.7&	0.071&	...&	...&	...\\\hline
S07-1&	J202039.29+393750.9&	0.026&	...&	...&	...\\\hline
S106-2&	J202726.86+372246.9&	3.090&	...&	...&	...\\\hline
S106W1-3&	J202355.08+373809.5&	0.398&	...&	...&	...\\\hline
S106W2-1&	J202144.11+372639.5&	0.030&	G075.782+0.342&	1&	263.4\\\hline
S106W2-3&	...&	...&	75.76+0.34&	3&	39.0\\\hline
W75N-1&	...&	...&	G081.871+0.780&	1&	1865.5\\\hline

\enddata
\tablenotetext{a}{VLASS catalog was unavailable by the time of publication and the sources here were identified on the VLASS images and named by their positions. The fluxes were obtained by manual radiometry. }
\tablenotemark{b}{References of the maser catalogs. 1. \hyperlink{H16}{Hu et al.(2016)} 2. \hyperlink{Malyshev03}{Malyshev \& Sobolev(2003)} 3. \hyperlink{Pestalozzi05}{Pestalozzi et al.(2005)}. }
\end{deluxetable*}

\clearpage
\section{Maps of the fields and MDCs in Cygnus X}\label{sec_A:map_field}

This appendix shows the multi-wavelength maps of each field (Figure \ref{fig:field}) and each MDC (Figure \ref{fig:MDC}), as well as the column density and temperature maps derived by \emph{getsources} \emph{hirescoldens} command.

\clearpage
\begin{minipage}{\textwidth}
\figurenum{E1}\addtocounter{figure}{1}
\centering\vspace{0.5in}
\includegraphics[width=4.3 in]{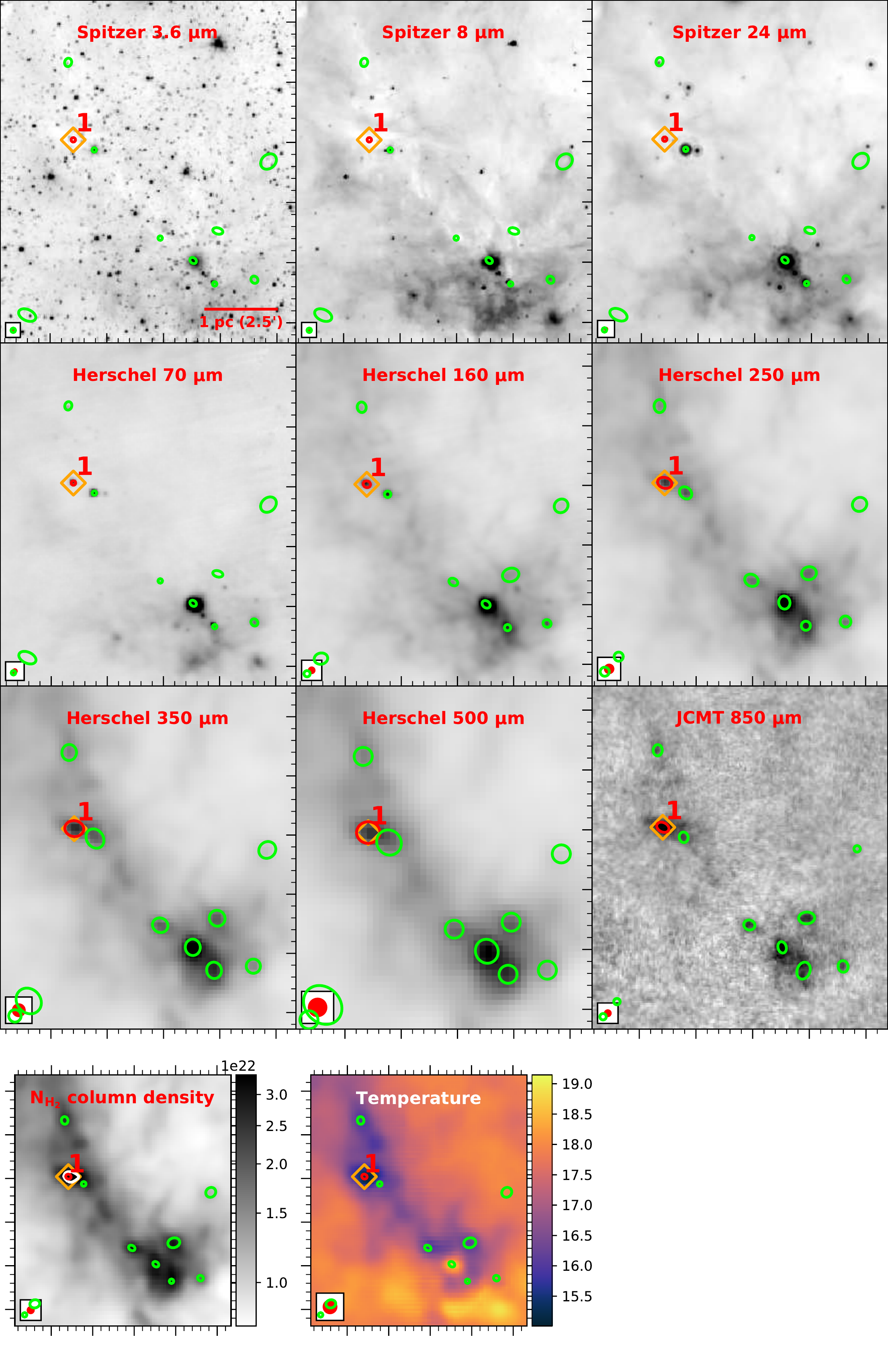}
\nfcaption{Continuum maps, column density maps, and temperature maps of the fields (here C01S) in Cygnus X. Continuum maps are in wavelengths of \emph{Spitzer} (3.6, 8, and 24 \um), \emph{Herschel} (70, 160, 250, 350, and 500 \um), JCMT (450 and 850 \um), and the IRAM 30 m telescope (1.2 mm). Panels will be omitted if the corresponding data are not available. Grayscales are logarithmically stretched for the \emph{Spitzer} bands, and are linearly stretched for the other continuum bands. A single-level contour of \ColThreshold\ is drawn as white solid lines in the column density panels. Green and red ellipses represent the FWHM sizes of cores and MDCs, respectively, obtained by \emph{getsources} for the $\ge70$ \um\ bands, or the reference sizes at 70 \um\ for the $<70$ \um\ bands as well as the column density and temperature maps. Names of MDCs are shown. \emph{Spitzer} 24 \um\ sources and \emph{MSX} 21 \um\ sources are marked as orange and yellow diamonds, respectively. UC\hii\ regions and class II methanol masers are marked as cyan and yellow triangles, respectively. {\bf The complete figure set (39 images) is available in the online journal.} \label{fig:field}}
\end{minipage}

\clearpage

\begin{minipage}{\textwidth}
\figurenum{E2}\addtocounter{figure}{1}
\centering\vspace{0.2in}
\includegraphics[width=6 in]{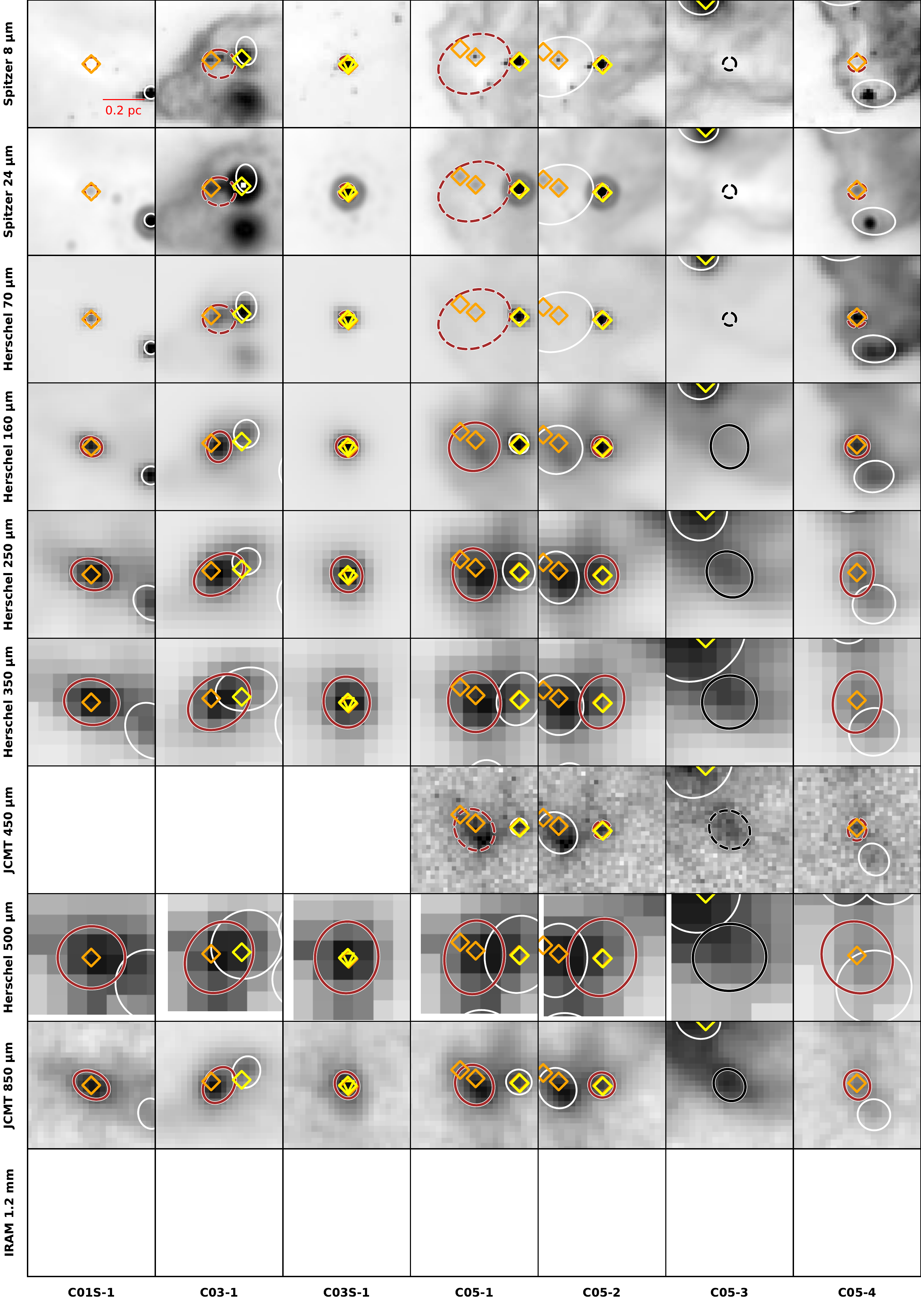}
\nfcaption{Images of the \numMDC\ MDCs in each band. Ellipses in the center of each panel are colored black, brown or red if the corresponding MDCs are starless, IR-quiet but not starless or IR-bright, respectively. Dashed ellipses mean that the corresponding fluxes were not used for SED fitting (see Sect. \ref{subsec:extraction}). Neighboring MDCs are represented as white ellipses. Other legends and figure settings are the same as in Figure \ref{fig:field}. {\bf The complete figure set (22 images) is available in the online journal.} \label{fig:MDC}}
\end{minipage}

\clearpage
\section{Maps of Cygnus X}\label{sec_A:map_CygnusX}

This appendix shows the multi-wavelength maps of the whole Cygnus X region, as well as the column density and temperature maps derived by \emph{getsources} \emph{hirescoldens} command.

\newcommand{\scaleG}{7}

\clearpage
\begin{minipage}{\textwidth}
\figurenum{F1}\addtocounter{figure}{1}
\centering\vspace{0.2in}
\includegraphics[width=\scaleG in]{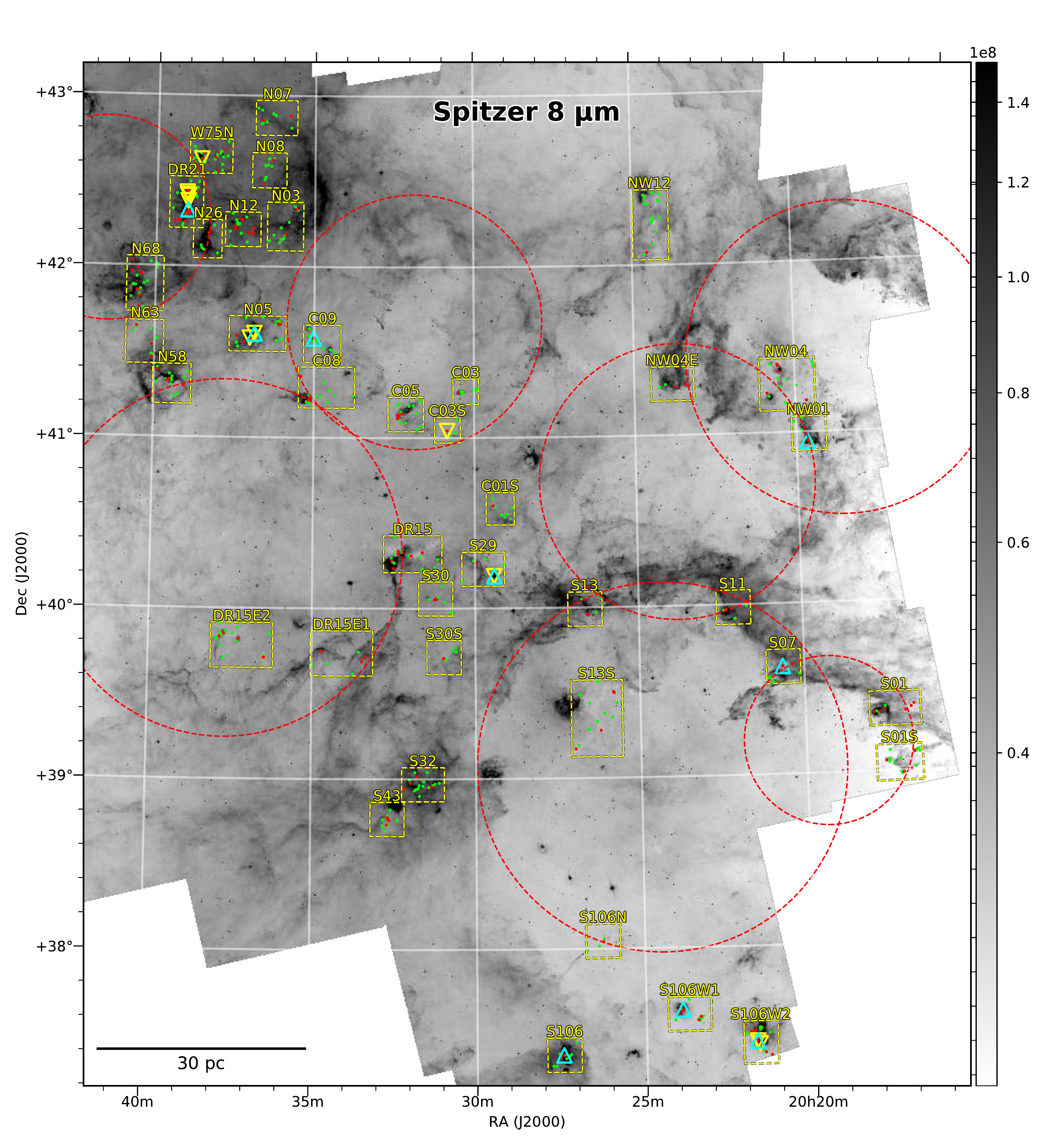}
\nfcaption{\emph{Spitzer} 8 \um\ map ($\rm Jy\ sr^{-1}$) with the same legends as in Figure \ref{fig:panorama}. The continuum data are from The \emph{Spitzer} Legacy Survey of the Cygnus-X Region. \label{fig:S8}}
\end{minipage}

\clearpage
\begin{minipage}{\textwidth}
\figurenum{F2}\addtocounter{figure}{1}
\centering\vspace{0.2in}
\includegraphics[width=\scaleG in]{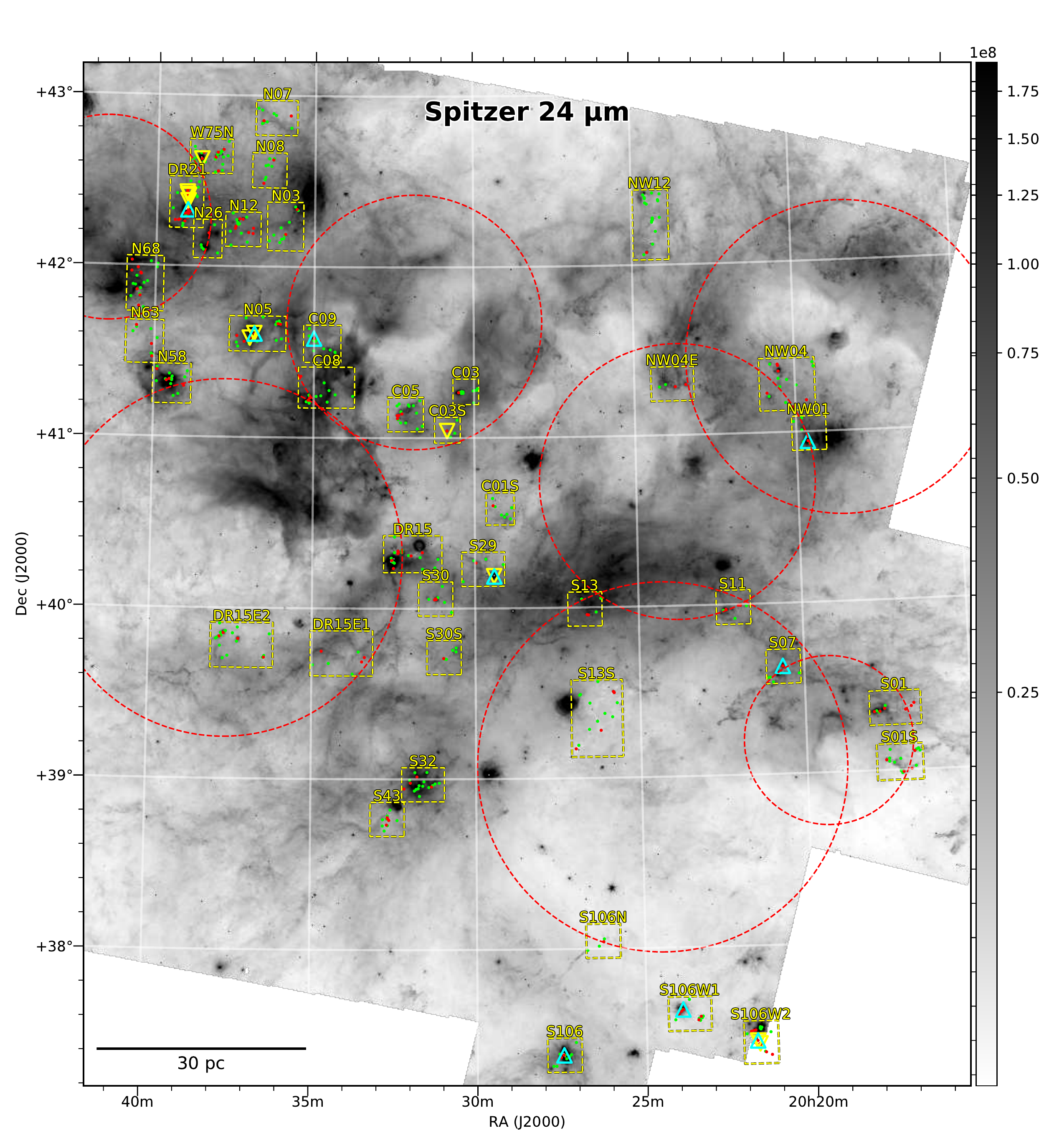}
\nfcaption{\emph{Spitzer} 24 \um\ map ($\rm Jy\ sr^{-1}$) with the same legends as in Figure \ref{fig:panorama}. The continuum data are from The \emph{Spitzer} Legacy Survey of the Cygnus-X Region.\label{fig:S24}}
\end{minipage}

\clearpage
\begin{minipage}{\textwidth}
\figurenum{F3}\addtocounter{figure}{1}
\centering\vspace{0.2in}
\includegraphics[width=\scaleG in]{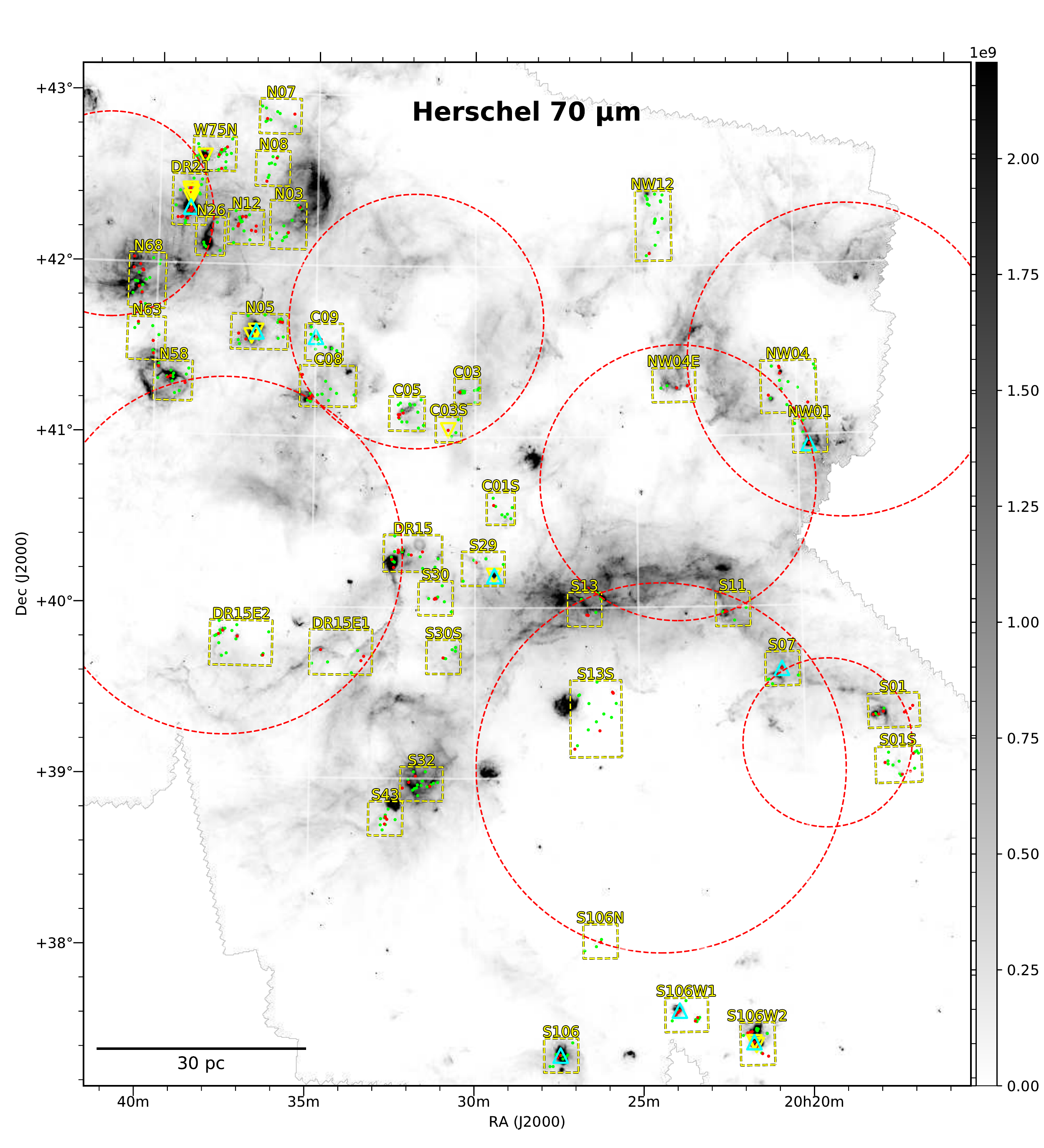}
\nfcaption{\emph{Herschel} 70 \um\ map ($\rm Jy\ sr^{-1}$) with the same legends as in Figure \ref{fig:panorama}. This map was created using the images from the Herschel Science Archive. \label{fig:H70}}
\end{minipage}

\clearpage
\begin{minipage}{\textwidth}
\figurenum{F4}\addtocounter{figure}{1}
\centering\vspace{0.2in}
\includegraphics[width=\scaleG in]{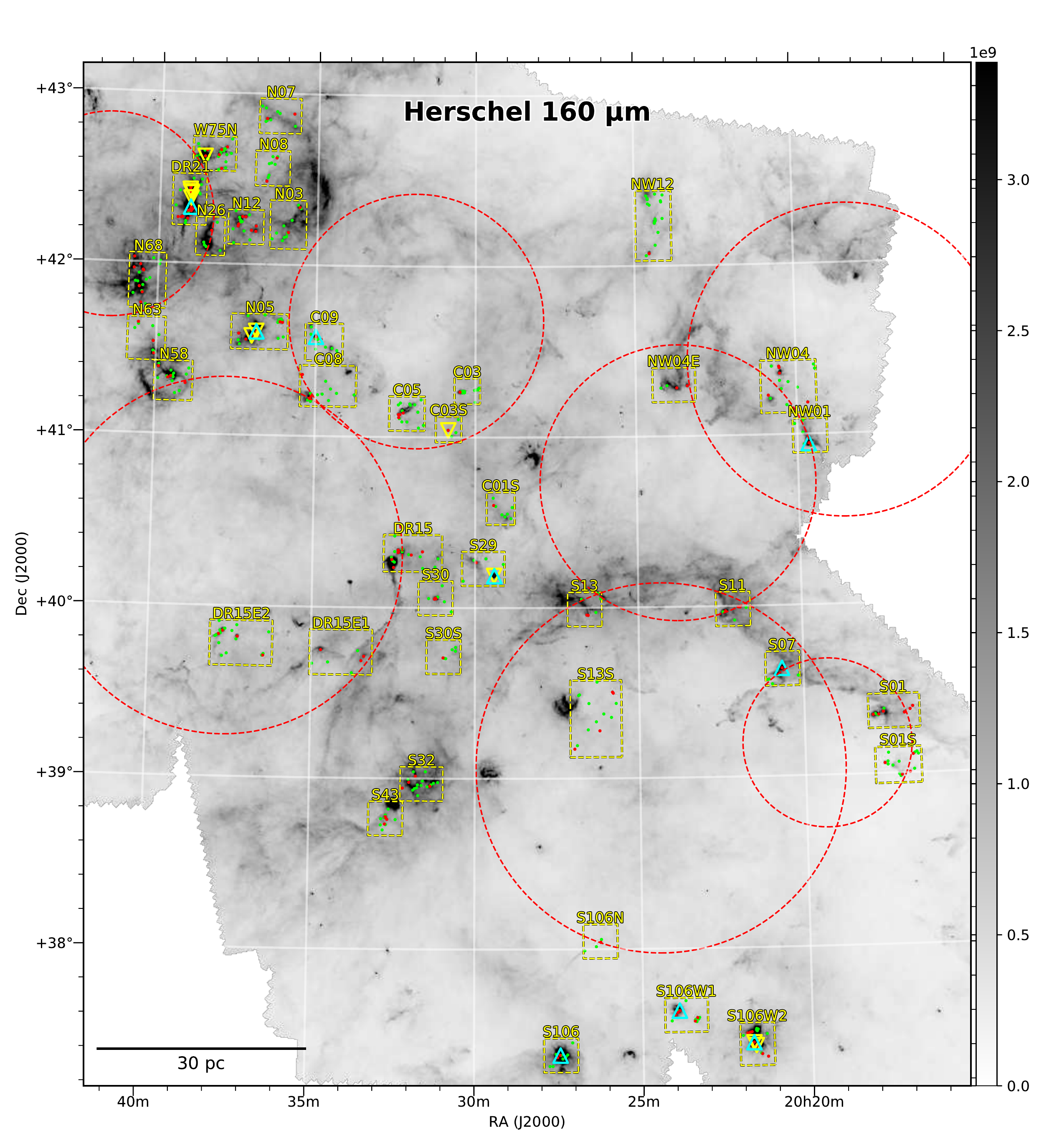}
\nfcaption{\emph{Herschel} 160 \um\ map ($\rm Jy\ sr^{-1}$) with the same legends as in Figure \ref{fig:panorama}. This map was created using the images from the Herschel Science Archive. \label{fig:H160}}
\end{minipage}

\clearpage
\begin{minipage}{\textwidth}
\figurenum{F5}\addtocounter{figure}{1}
\centering\vspace{0.2in}
\includegraphics[width=\scaleG in]{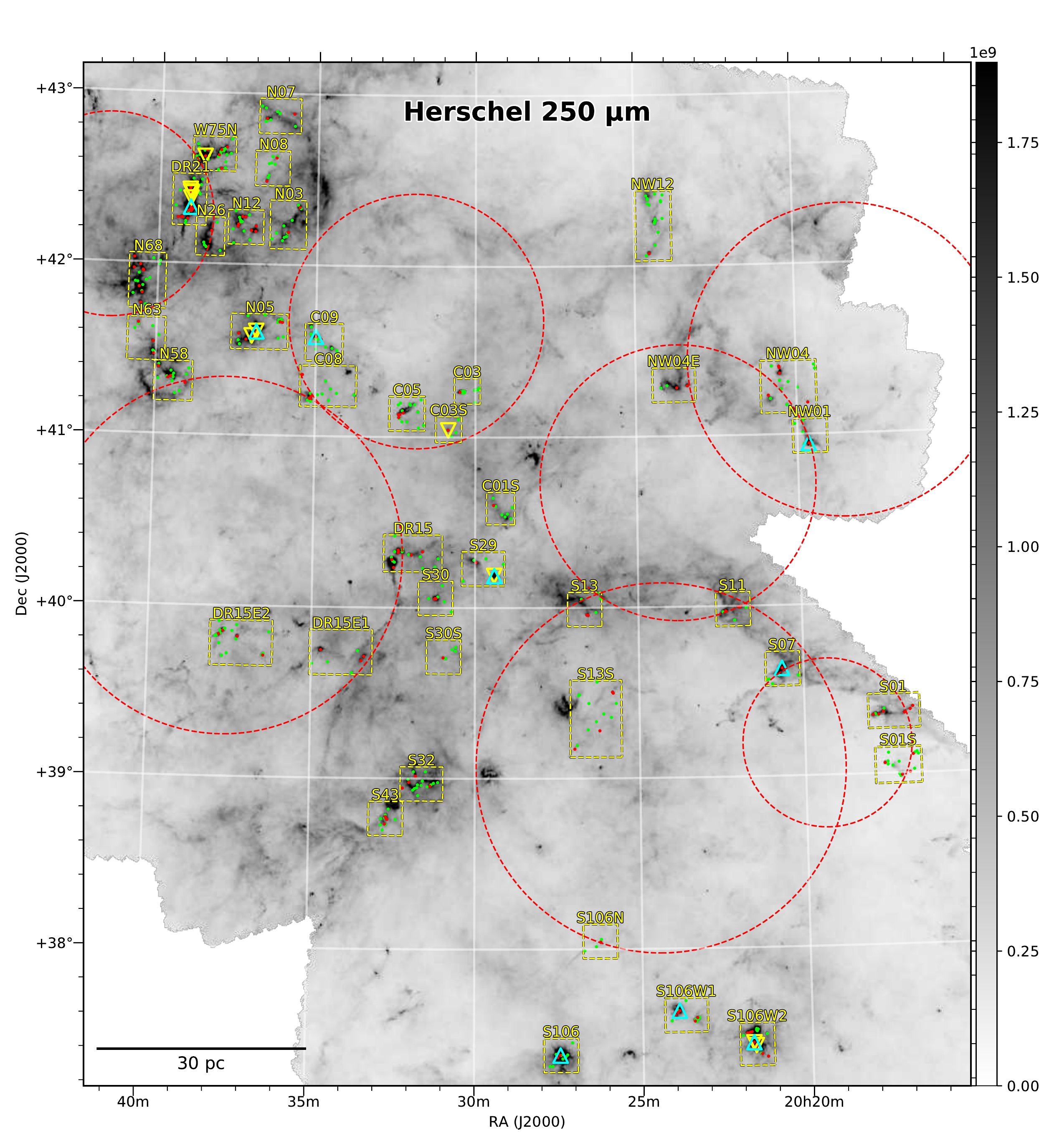}
\nfcaption{\emph{Herschel} 250 \um\ map ($\rm Jy\ sr^{-1}$) with the same legends as in Figure \ref{fig:panorama}. This map was created using the images from the Herschel Science Archive. \label{fig:H250}}
\end{minipage}

\clearpage
\begin{minipage}{\textwidth}
\figurenum{F6}\addtocounter{figure}{1}
\centering\vspace{0.2in}
\includegraphics[width=\scaleG in]{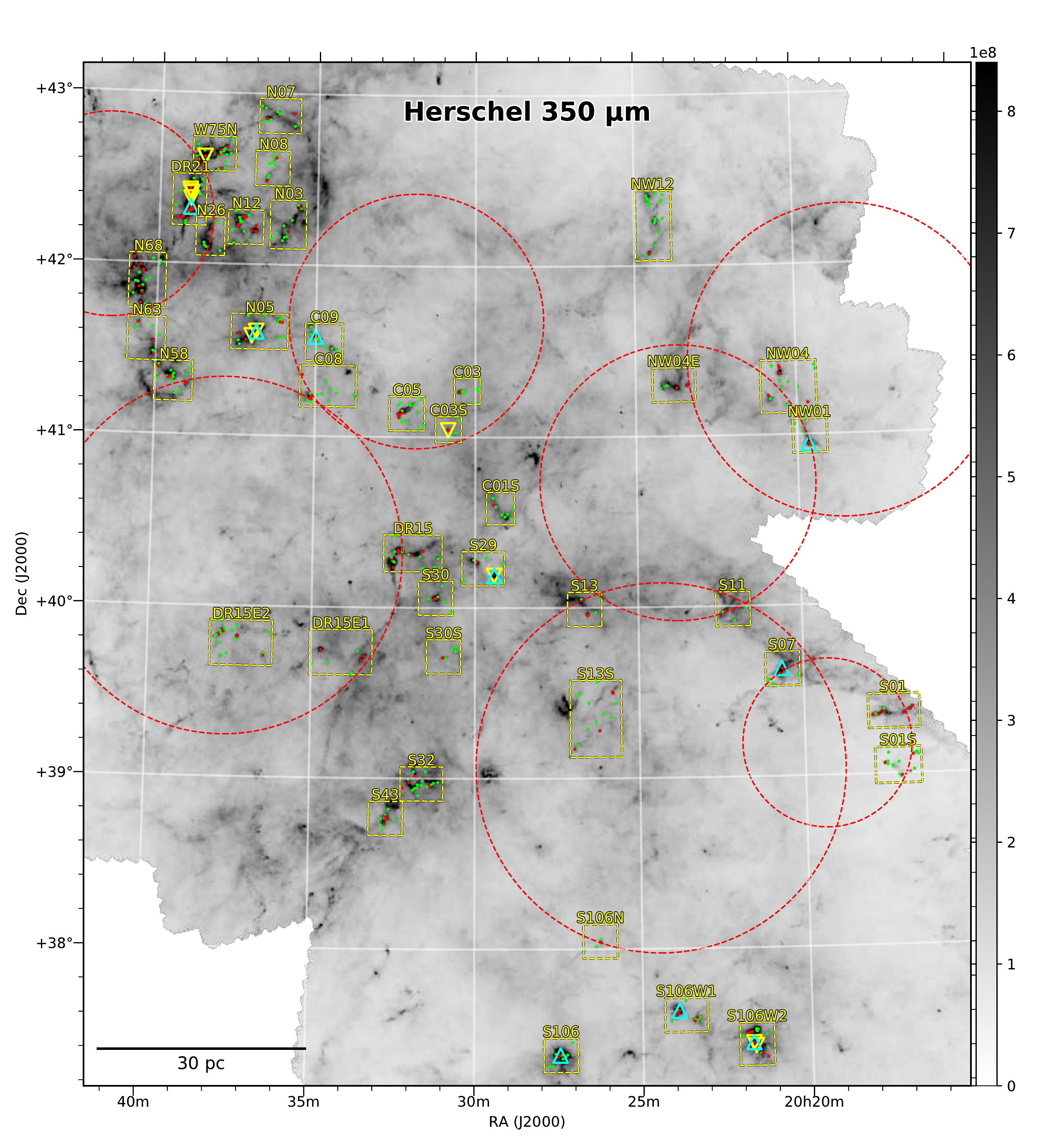}
\nfcaption{\emph{Herschel} 350 \um\ map ($\rm Jy\ sr^{-1}$) with the same legends as in Figure \ref{fig:panorama}. This map was created using the images from the Herschel Science Archive. \label{fig:H350}}
\end{minipage}

\clearpage
\begin{minipage}{\textwidth}
\figurenum{F7}\addtocounter{figure}{1}
\centering\vspace{0.2in}
\includegraphics[width=\scaleG in]{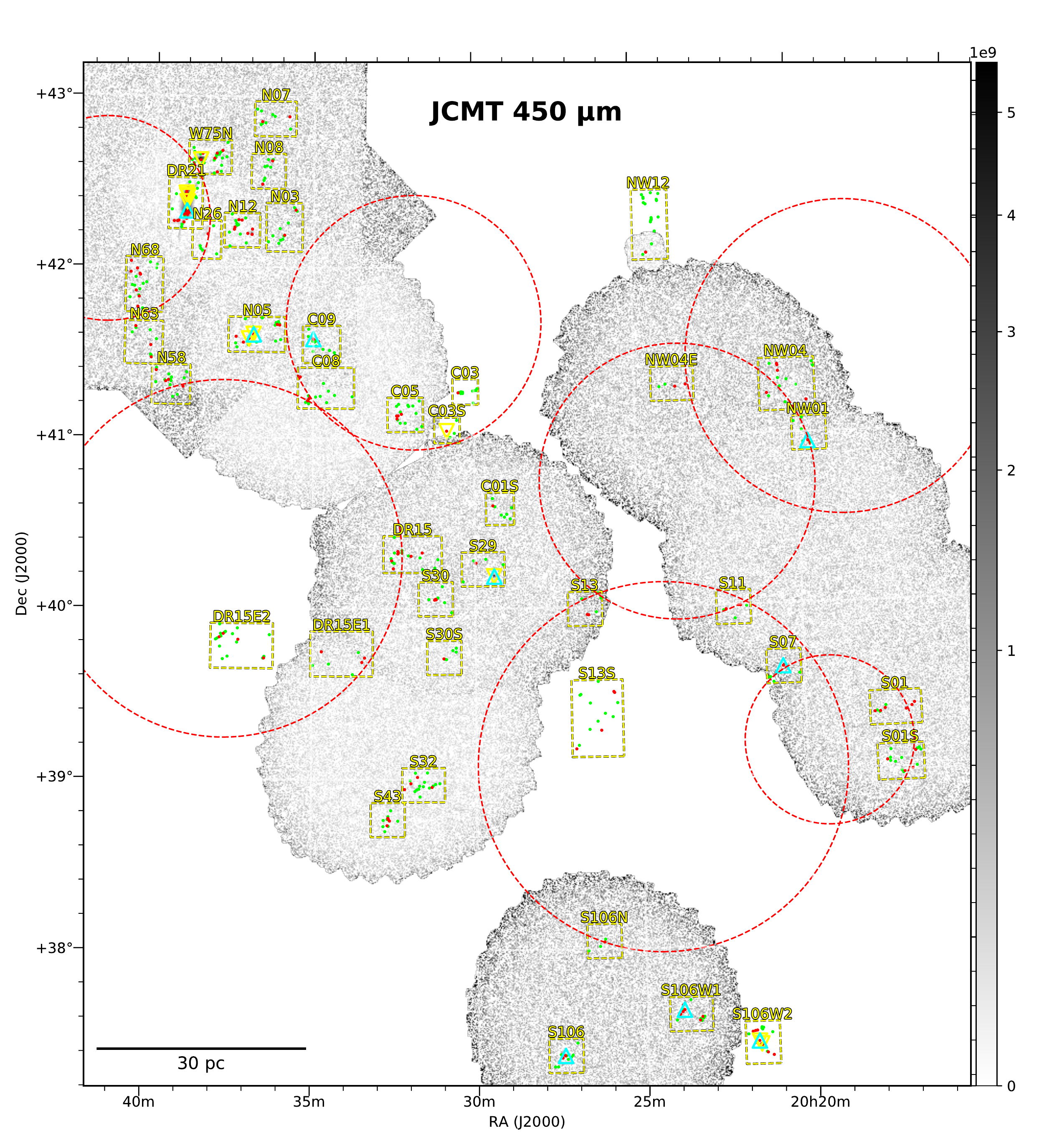}
\nfcaption{JCMT 450 \um\ map ($\rm Jy\ sr^{-1}$) with the same legends as in Figure \ref{fig:panorama}. This map was created using the data from the the Canadian Astronomy Data Centre. \label{fig:J450}}
\end{minipage}

\clearpage
\begin{minipage}{\textwidth}
\figurenum{F8}\addtocounter{figure}{1}
\centering\vspace{0.2in}
\includegraphics[width=\scaleG in]{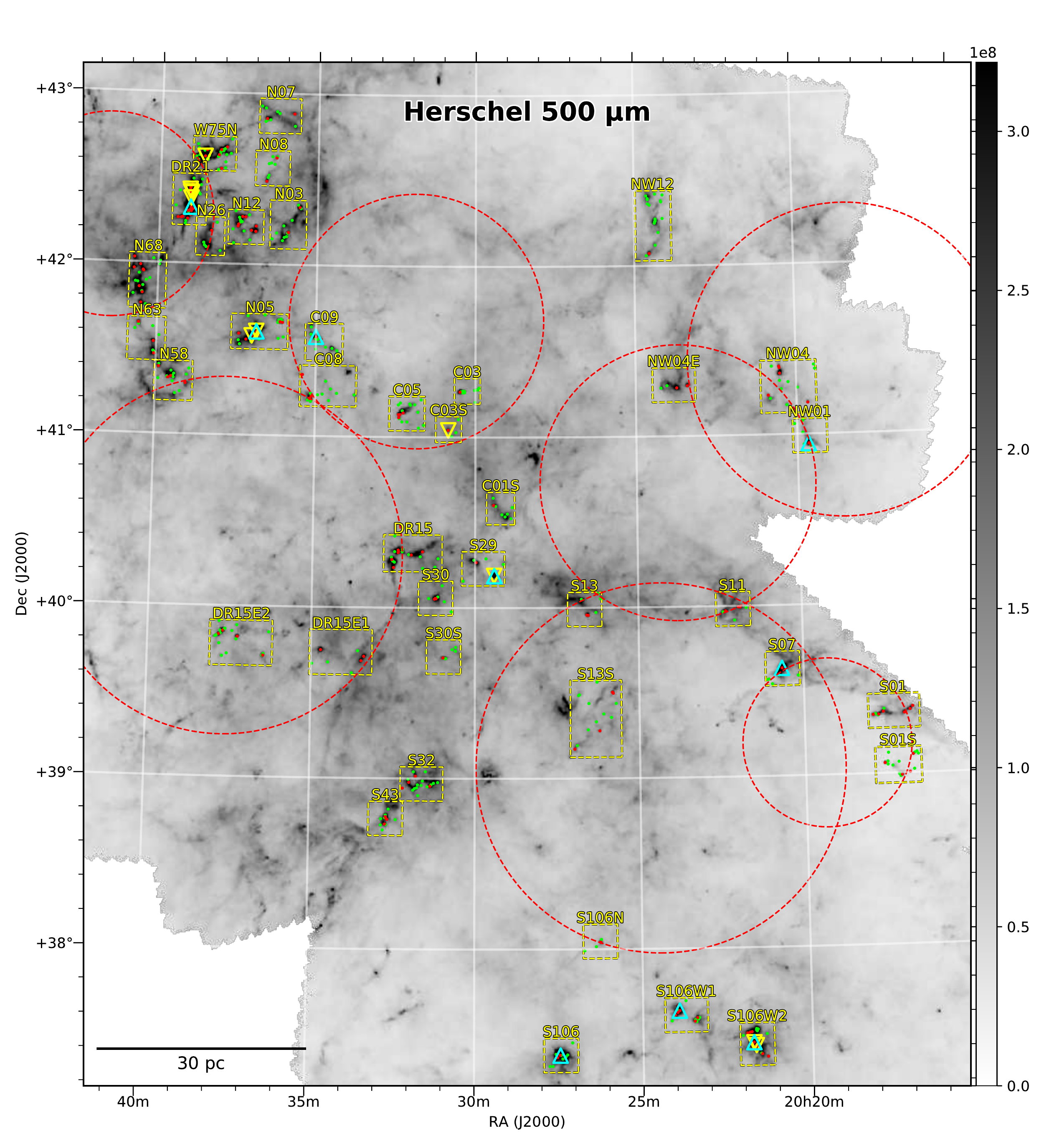}
\nfcaption{\emph{Herschel} 500 \um\ map ($\rm Jy\ sr^{-1}$) with the same legends as in Figure \ref{fig:panorama}.  This map was created using the images from the Herschel Science Archive.\label{fig:H500}}
\end{minipage}

\clearpage
\begin{minipage}{\textwidth}
\figurenum{F9}\addtocounter{figure}{1}
\centering\vspace{0.2in}
\includegraphics[width=\scaleG in]{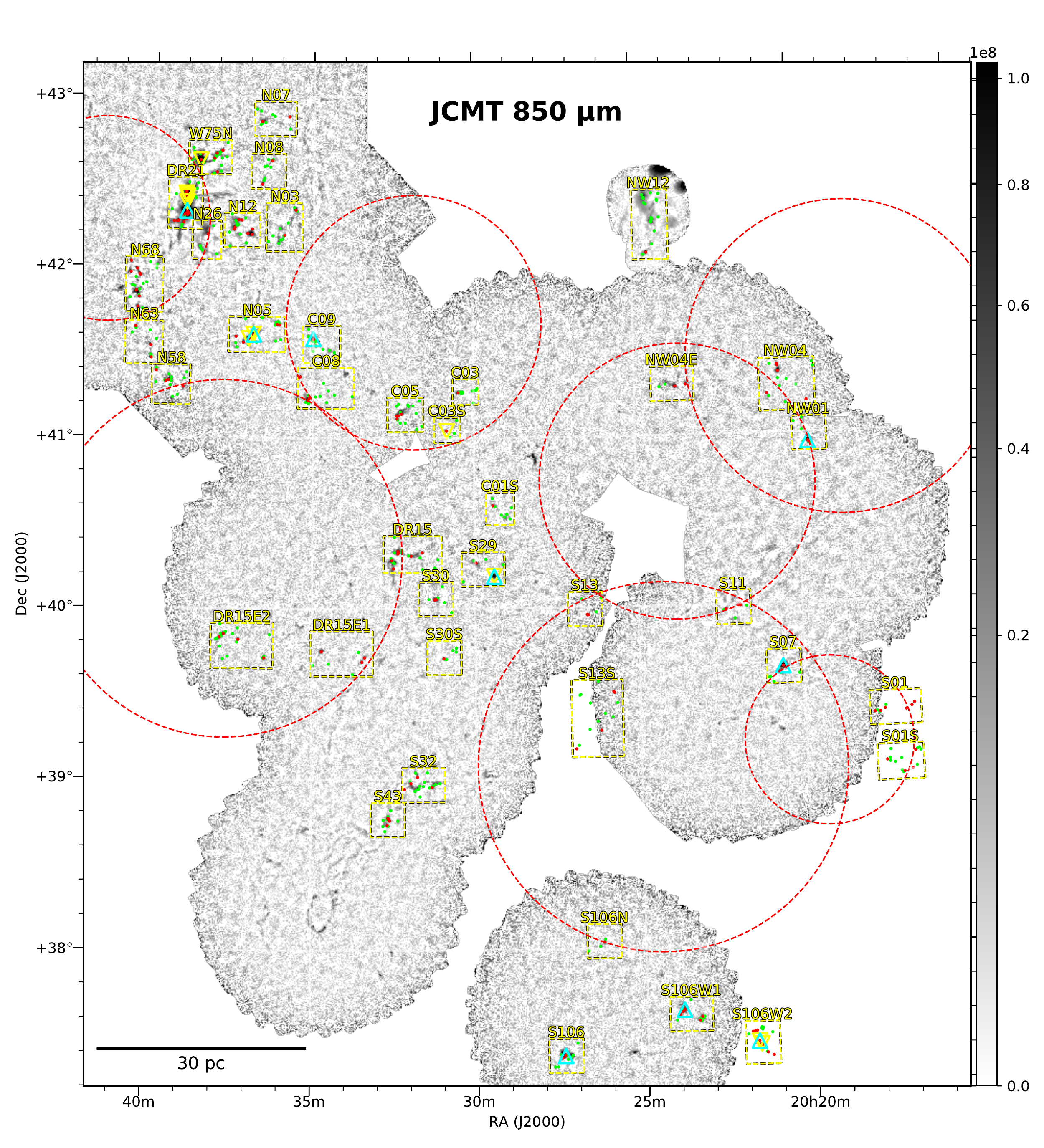}
\nfcaption{JCMT 850 \um\ map ($\rm Jy\ sr^{-1}$) with the same legends as in Figure \ref{fig:panorama}. This map was created using the data from the the Canadian Astronomy Data Centre.\label{fig:J850}}
\end{minipage}

\clearpage
\begin{minipage}{\textwidth}
\figurenum{F10}\addtocounter{figure}{1}
\centering\vspace{0.2in}
\includegraphics[width=\scaleG in]{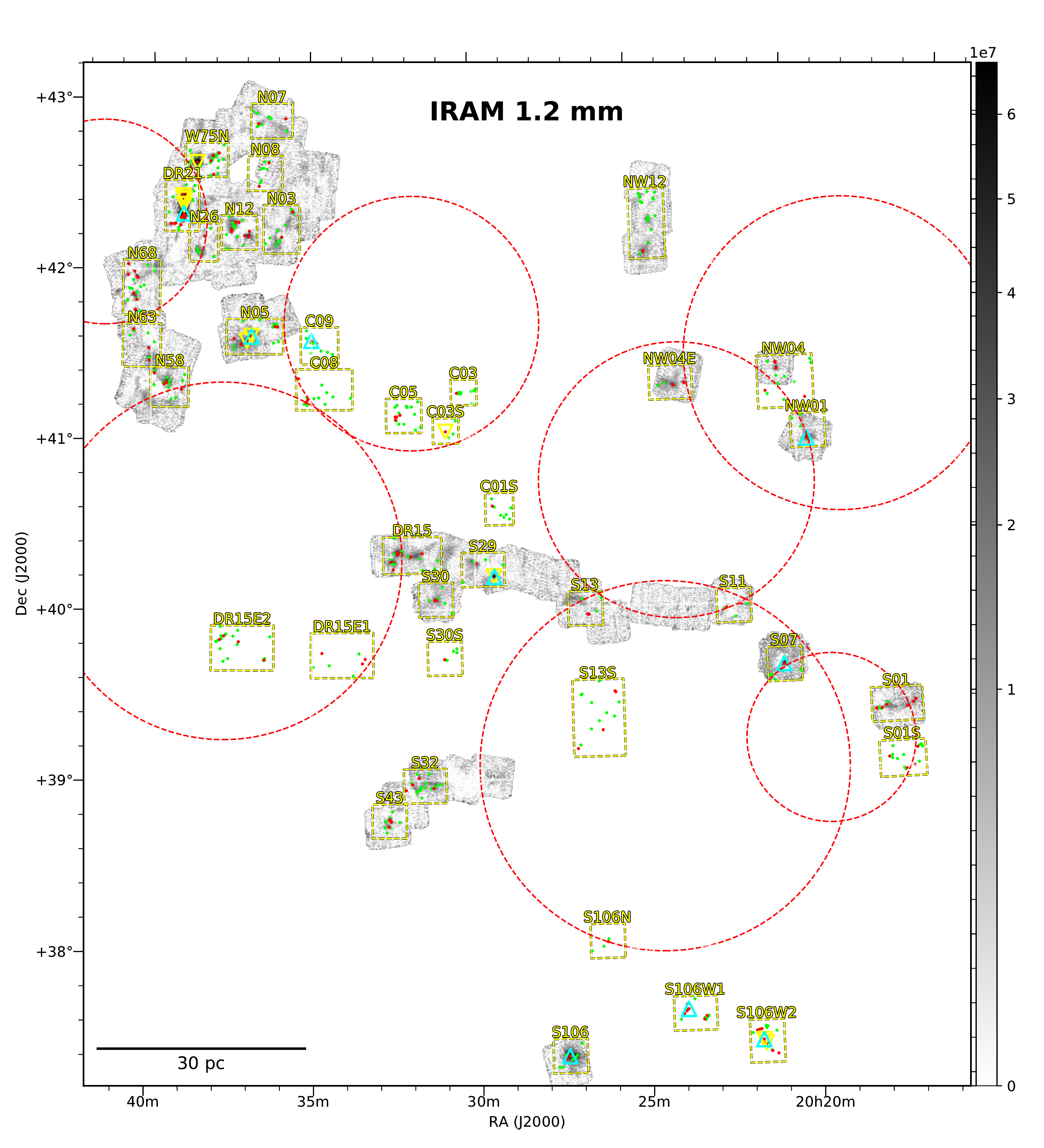}
\nfcaption{IRAM 1.2 mm map ($\rm Jy\ sr^{-1}$) with the same legends as in Figure \ref{fig:panorama}.  This map was created using the images in \motte.\label{fig:I1200}}
\end{minipage}

\clearpage
\begin{minipage}{\textwidth}
\figurenum{F11}\addtocounter{figure}{1}
\centering\vspace{0.2in}
\includegraphics[width=\scaleG in]{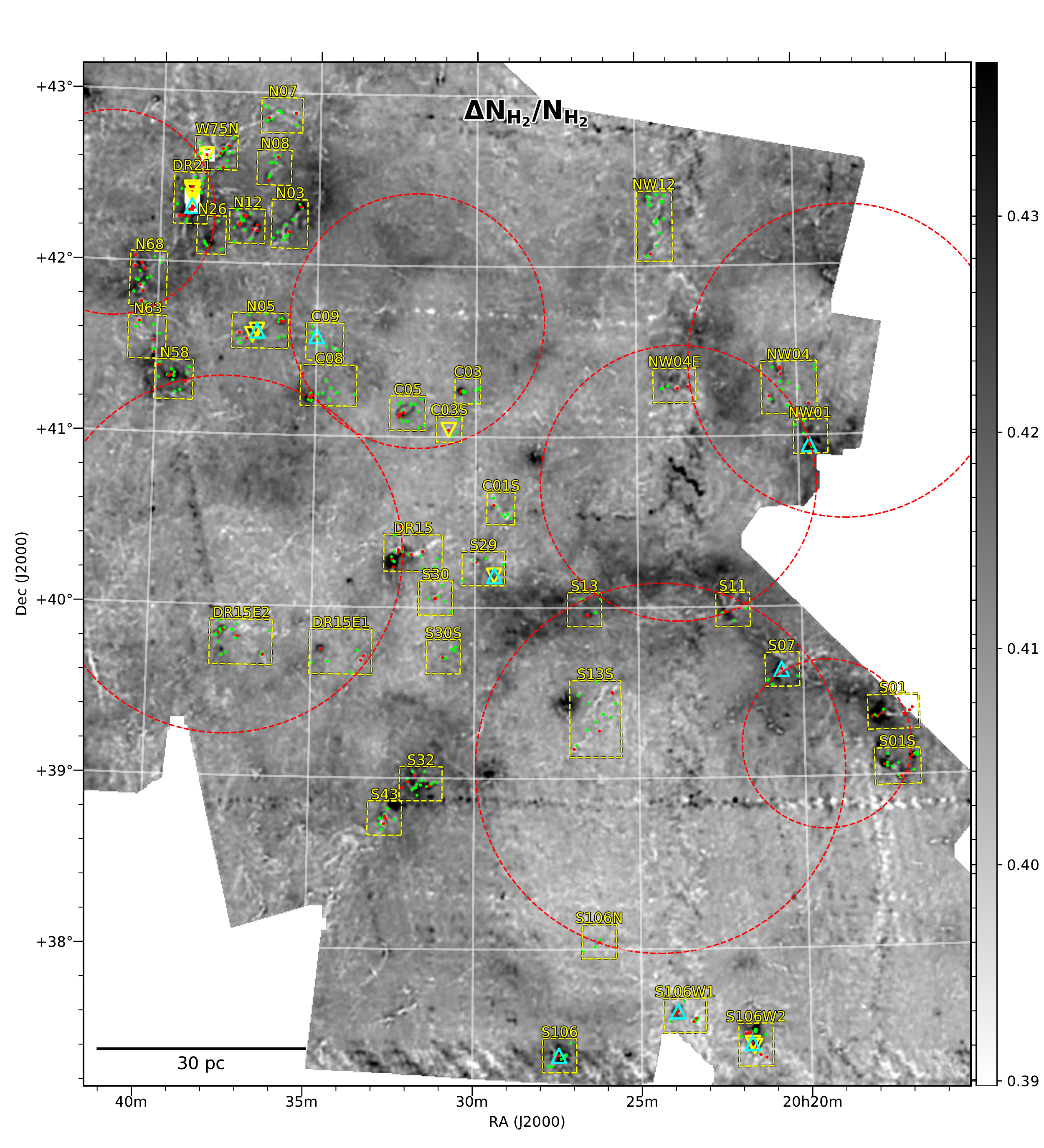}
\nfcaption{Relative uncertainty map of $\rm H_2$ column density with the same legends as in Figure \ref{fig:panorama}. This map was derived from the outputs of the \emph{hirescoldens} command. \label{fig:Ne2N}}
\end{minipage}

\clearpage
\begin{minipage}{\textwidth}
\figurenum{F12}\addtocounter{figure}{1}
\centering\vspace{0.2in}
\includegraphics[width=\scaleG in]{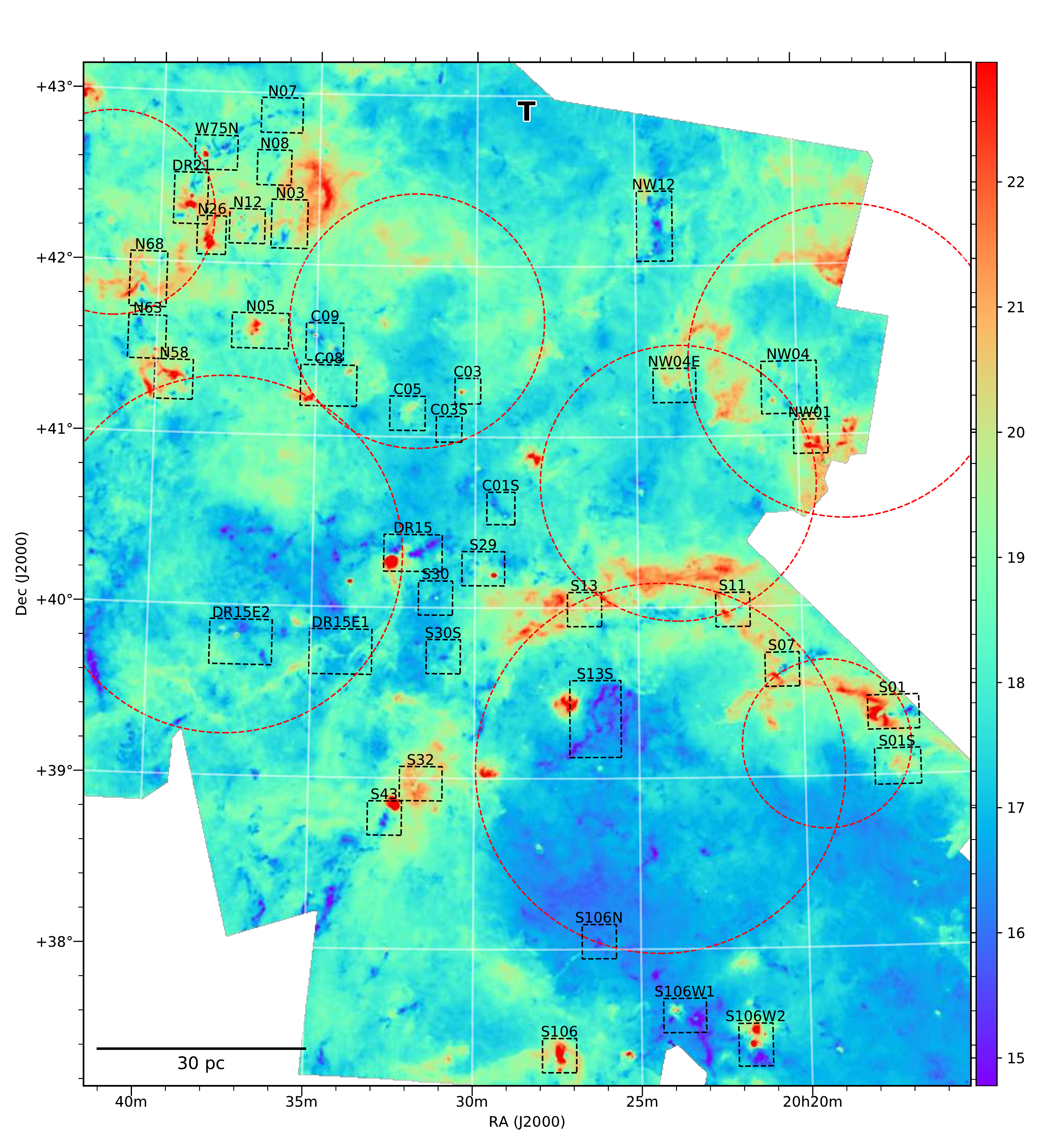}
\nfcaption{Temperature map (K). This map was derived from the outputs of the \emph{hirescoldens} command.\label{fig:T}}
\end{minipage}

\clearpage
\begin{minipage}{\textwidth}
\figurenum{F13}\addtocounter{figure}{1}
\centering\vspace{0.2in}
\includegraphics[width=\scaleG in]{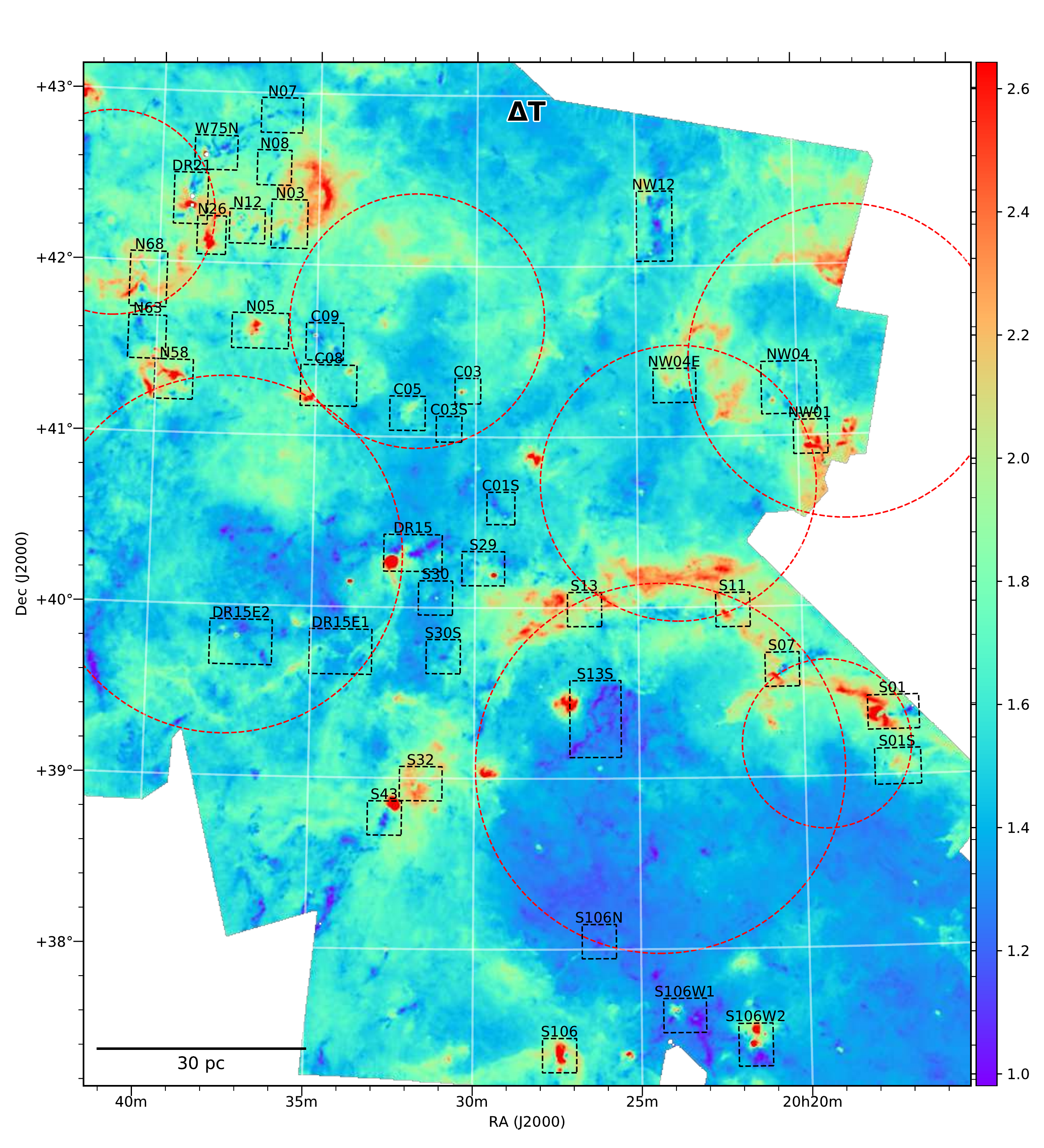}
\nfcaption{Map of temperature uncertainty (K). This map was derived from the outputs of the \emph{hirescoldens} command.\label{fig:T_err}}
\end{minipage}

\clearpage
\begin{minipage}{\textwidth}
\figurenum{F14}\addtocounter{figure}{1}
\centering\vspace{0.2in}
\includegraphics[width=\scaleG in]{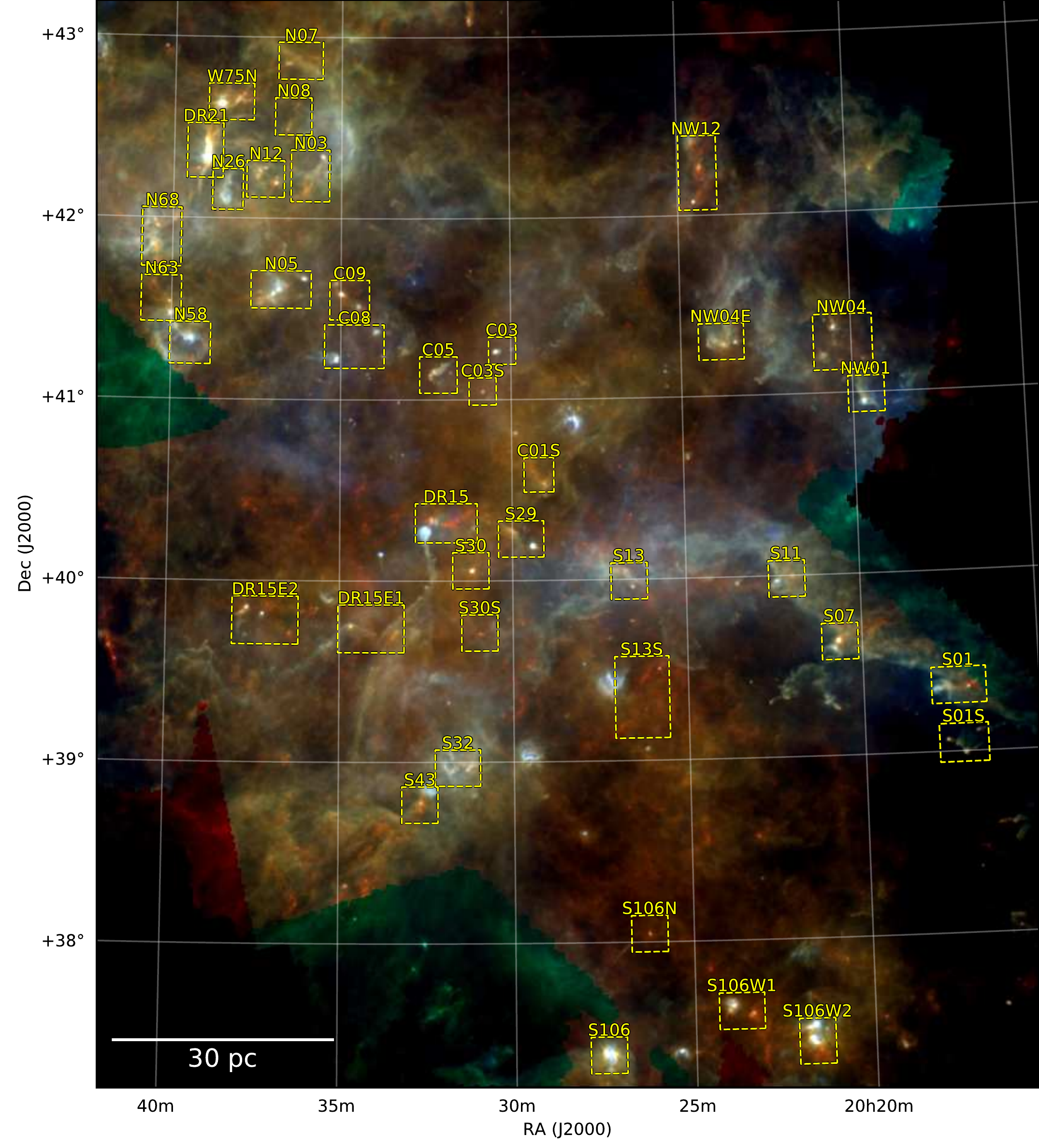}
\nfcaption{RGB map. R=\emph{Herschel} 500 \um, G=\emph{Herschel} 160 \um, B=\emph{Herschel} 70 \um. Color scales are logarithmically stretched. This map was created using the images from the Herschel Science Archive. \label{fig:rgb}}
\end{minipage}

\end{CJK*}
\end{document}